\documentclass[twocolumn,twocolappendix]{aastex631}

\usepackage{xspace}
\usepackage{mathtools}
\usepackage{float}
\usepackage{amsmath}	% Advanced maths commands
\usepackage{multirow}
\usepackage{enumitem}

\newcommand{\Omegam}{\xspace{\ensuremath{\Omega_{\mathrm{m}}}}\xspace}

\newcommand{\lcdm}{\xspace{\ensuremath{\Lambda\mathrm{CDM}}}\xspace}
\newcommand{\hmpc}{\xspace{$h^{-1}\mathrm{Mpc}$}\xspace}
\newcommand{\hkpc}{\xspace{$h^{-1}\mathrm{kpc}$}\xspace}
\newcommand{\hMsun}{\xspace{$h^{-1}\mathrm{M}_{\odot}$}\xspace}

\newcommand{\cemcee}{\xspace{\textsc{emcee\_in\_c}}\xspace}
\newcommand{\emcee}{\xspace{\textsc{emcee}}\xspace}

\newcommand{\mmin}{\xspace{$M_\mathrm{min}$}\xspace}
\newcommand{\logmmin}{\xspace{$\mathrm{log}{M_\mathrm{min}}$}\xspace}
\newcommand{\siglogm}{\xspace{$\sigma_{\mathrm{log} M}$}\xspace}
\newcommand{\logmzero}{\xspace{$\mathrm{log}{M_0}$}\xspace}
\newcommand{\logmone}{\xspace{$\mathrm{log}{M_1}$}\xspace}
\newcommand{\Acen}{\xspace{$A_\mathrm{cen}$}\xspace}
\newcommand{\Asat}{\xspace{$A_\mathrm{sat}$}\xspace}
\newcommand{\Bvel}{\xspace{$B_\mathrm{vel}$}\xspace}

\newcommand{\ngal}{\xspace{$n_{\mathrm{gal}}$}\xspace}
\newcommand{\wprp}{\xspace{$w_\mathrm{p}(r_\mathrm{p})$}\xspace}
\newcommand{\zxi}{\xspace{$\xi(s)$}\xspace}
\newcommand{\gmf}{\xspace{$n(N)$}\xspace}
\newcommand{\sigN}{\xspace{$\sigma_v(N)$}\xspace}
\newcommand{\mcf}{\xspace{$\mathrm{mcf}(s)$}\xspace}
\newcommand{\cic}{\xspace{$P_N(R)$}\xspace}
\newcommand{\Pzero}{\xspace{$P_0$}\xspace}
\newcommand{\Pone}{\xspace{$P_0$}\xspace}
\newcommand{\vpf}{\xspace{$\mathrm{VPF}(R)$}\xspace}
\newcommand{\spf}{\xspace{$\mathrm{SPF}(R)$}\xspace}

\newcommand{\wptwo}{\xspace{$w_\mathrm{p}(r_\mathrm{p} \sim 0.3 \ h^{-1}\mathrm{Mpc})$}\xspace}

\usepackage[normalem]{ulem} % Added by MS for \sout

\received{January 3, 2023}
\revised{March 15, 2023}
\accepted{March 16, 2023}
\submitjournal{ApJ}

\shorttitle{Extending the Halo Model}
\shortauthors{Beltz-Mohrmann et al.}
\graphicspath{{./}{figures/}}
\begin{document}

\title{Toward Accurate Modeling of Galaxy Clustering on Small Scales: Halo Model Extensions and Lingering Tension}

\correspondingauthor{Gillian D. Beltz-Mohrmann}
\email{gbeltzmohrmann@anl.gov}

\author[0000-0002-4392-8920]{Gillian D. Beltz-Mohrmann}
\affiliation{Department of Physics and Astronomy, Vanderbilt University, 2201 West End Ave, Nashville, TN 37235, USA}
\affiliation{High Energy Physics Division, Argonne National Laboratory, 9700 South Cass Avenue, Lemont, IL 60439, USA}

\author[0000-0001-9094-8433]{Adam O. Szewciw}
\affiliation{Department of Physics and Astronomy, Vanderbilt University, 2201 West End Ave, Nashville, TN 37235, USA}

\author[0000-0002-1814-2002]{Andreas A. Berlind}
\affiliation{Department of Physics and Astronomy, Vanderbilt University, 2201 West End Ave, Nashville, TN 37235, USA}
\affiliation{National Science Foundation, Division of Astronomical Sciences, Alexandria, VA 22314, USA}

\author[0000-0002-4845-1228]{Manodeep Sinha}
\affiliation{Department of Physics and Astronomy, Vanderbilt University, 2201 West End Ave, Nashville, TN 37235, USA}
\affiliation{SA 118, Center for Astrophysics \& Supercomputing, Swinburne University of Technology, 1 Alfred St., Hawthorn, VIC 3122, Australia}
\affiliation{ARC Centre of Excellence for All Sky Astrophysics in 3 Dimensions (ASTRO 3D), Australia}



\begin{abstract}
This paper represents an effort to provide robust constraints on the galaxy-halo connection and simultaneously test the Planck \lcdm cosmology using a fully numerical model of small-scale galaxy clustering.
We explore two extensions to the standard Halo Occupation Distribution model: assembly bias, whereby halo occupation depends on both halo mass and the larger environment, and velocity bias, whereby galaxy velocities do not perfectly trace the velocity of the dark matter within the halo.
Moreover, we incorporate halo mass corrections to account for the impact of baryonic physics on the halo population.
We identify an optimal set of clustering measurements to constrain this ``decorated” HOD model for both low- and high-luminosity galaxies in SDSS DR7.
We find that, for low-luminosity galaxies, a model with both assembly bias and velocity bias provides the best fit to the clustering measurements, with no tension remaining in the fit.
In this model we find evidence for both central and satellite galaxy assembly bias at the 99\% and 95\% confidence levels, respectively. 
In addition, we find evidence for satellite galaxy velocity bias at the 99.9\% confidence level.
For high luminosity galaxies, we find no evidence for either assembly bias or velocity bias, but our model exhibits significant tension with SDSS measurements. 
We find that all of these conclusions still stand when we include the effects of baryonic physics on the halo mass function, suggesting that the tension we find for high luminosity galaxies may be due to a problem with our assumed cosmological model.
\end{abstract}



\section{Introduction} \label{sec:intro}
Small-scale galaxy clustering contains a wealth of cosmological information.
However, harnessing the constraining power of small scales requires highly accurate models of both dark matter structure formation and the galaxy-halo connection.
Halo models are motivated by our understanding that galaxies form and reside in gravitationally bound, virialized regions of dark matter known as halos \citep[e.g.,][]{Neyman1952,Peebles1974,McClelland1977,Scherrer1991,Kauffmann1997,Jing1998,Baugh1999,Kauffmann1999,Benson2000b,Ma2000,Peacock2000,Seljak2000,Scoccimarro2001,Sheth2001,White2001,Cooray2002}. 
These models assume that the clustering of galaxies can be fully described by (i) the clustering of their host halos and (ii) the way in which galaxies occupy these halos. 

A key ingredient of the halo model is the Halo Occupation Distribution (HOD), which specifies via a few parameters the probability that a halo of mass $M$ contains $N$ galaxies (above some luminosity threshold) as well as how the galaxies are distributed within their halo \citep{Berlind2002,Berlind2003}. 
The standard form of the HOD \citep{Zheng2005} contains at most five free parameters that specify the mean occupation number of central and satellite galaxies.
This HOD formulation assumes that central galaxies reside at the center of their halo and move at the mean halo velocity, while satellite galaxies trace the spatial and velocity distribution of dark matter inside halos.
Constraining these parameters when fitting to observational data provides a useful empirical measurement against which we can test competing theories of galaxy formation and evolution.
Moreover, this framework provides a useful avenue for testing cosmology on small scales, assuming that the HOD model is sufficiently flexible to marginalize over the uncertainty of galaxy formation.
Compared to other methods like subhalo abundance matching, the HOD model is more flexible and does not rely on having well-resolved subhalos from simulations, making it a more robust framework for probing cosmology. 

Many works have used the standard HOD to model the clustering of galaxies in recent redshift surveys \citep[e.g.,][]{Zehavi2011,Guo2016}.
Several of these studies yield fits which would rule out the \lcdm+ HOD model if taken at face value. 
However, these studies typically rely on analytic approximations for calculating clustering statistics, which can introduce unknown systematic uncertainties.
Additionally, the errors and covariances used in these studies are typically derived via the jackknife method, which has been shown to produce biased results \citep{Norberg2009}. 

\citet{Sinha2018} (S18 hereafter) developed a fully numerical mock-based forward modeling procedure, whereby the standard HOD model is applied to halo catalogs from cosmological N-body simulations and observational survey systematics are added to create realistic mock catalogs. These mocks are then used both for model parameter exploration and for calculating covariance matrices.
This significantly improved the accuracy of the HOD modeling framework and allowed for the use of arbitrary clustering statistics that could be \textit{directly} measured on mocks (as opposed to calculated analytically).
Using the projected correlation function, group multiplicity function, and galaxy number density, S18 compared the clustering of galaxies in the Sloan Digital Sky Survey \citep[SDSS,][]{York2000} to a \lcdm cosmology \citep{Planck2014} + standard HOD model.
Carefully controlling for systematic errors allowed them to reliably interpret the goodness of fit of their model.
They found that their best-fit HOD model was unable to jointly fit the clustering statistics, revealing moderate tension with SDSS.
Because this tension did not exist when they considered only measurements of the projected correlation function (as is done in many studies), S18 demonstrated the value of adding additional statistics in small-scale clustering analyses.

\citet{Szewciw2022} (S22 hereafter) enhanced the procedure used in S18 in order to maximize the return from spectroscopic surveys.
They included the same clustering statistics used in S18 (galaxy number density, projected correlation function, and group multiplicity function) as well as four additional clustering statistics: redshift-space correlation function, group velocity dispersion, mark correlation function, and counts-in-cells statistics.
Additionally, they developed an algorithm to identify an optimal set of clustering measurements at a variety of different scales in order to maximize constraining power and minimize noise.
With these optimal observables, as well as several other improvements to the modeling procedure, they were able to significantly tighten their HOD parameter constraints, as well as dramatically increase the tension found in S18.

The tension found in S22 may be indicative of an issue with the cosmological model used, but it also may be the case that the standard HOD model is simply not flexible enough to accurately encompass the galaxy-halo connection. 
For example, the standard HOD model assigns galaxies to halos based solely on the halo's mass, but it is possible that halo occupation depends on additional (secondary) features of the halo (e.g., concentration) that correlate with the halo's larger scale environment, a phenomenon known as assembly bias \citep[e.g.,][]{Sheth2004, Gao2005, Wechsler2006, Gao2007, Croton2007, Salcedo2018, Wechsler2018}.
Evidence for galaxy assembly bias has been found in multiple hydrodynamic simulations \citep[e.g.,][]{Artale2018, Bose2019, Beltz-Mohrmann2020, Xu2020, Hadzhiyska2020, Hadzhiyska2021, Hadzhiyska2021b, Hadzhiyska2021c, Contreras2021}, indicating that it is an expected feature in a \lcdm universe.
Additionally, the standard HOD model assumes that galaxies trace the positions and velocities of the dark matter distribution within their host halo, but, as has been seen in hydrodynamic simulations, this may not be the case \citep[e.g.,][]{Berlind2003,Beltz-Mohrmann2020}.
Finally, the typical halo modeling framework relies on dark matter only simulations for creating halo catalogs.
These simulations lack baryonic physics, which has been shown to have a significant impact on the halo distribution itself \citep[e.g.,][]{Beltz-Mohrmann2021}. 
It is possible that failing to account for the impact of baryonic physics on the halo population is contributing to the tension between the halo model and the clustering of SDSS galaxies.

Several works have examined the potential for the presence of assembly bias to affect constraints on the galaxy-halo connection and cosmology.
For example, \citet{Zentner2014} examined the potential for assembly bias to induce systematic errors in inferred halo occupation statistics.
They built mock galaxy catalogs that exhibited assembly bias as well as companion mock catalogs with identical HODs but no assembly bias.
They fit HOD models to the galaxy clustering in each catalog, and found that the inferred HODs described the true HODs well in the mocks without assembly bias, but in the mocks with assembly bias the inferred HODs exhibited significant systematic errors.

In a later study, \citet{McCarthy2019} used a mock galaxy catalog with assembly bias to study how assembly bias might affect cosmological constraints.
Specifically, they used the large-scale redshift-space distortions to probe $f\sigma_8$.
They found that on small scales (a few to tens of \hmpc) galaxy assembly bias can introduce systematic uncertainties in cosmological constraints if unaccounted for.
They concluded that galaxy assembly bias can only be ignored when modeling scales above 8 \hmpc, where clustering is determined purely by the large scale bias. Similarly, \citet{Lange2019c} explored how galaxy assembly bias affects cosmological inference and found a degeneracy between assembly bias and $f\sigma_8$. 
Ultimately, they found that not including galaxy assembly bias in the model leads to a small shift in the posterior of $f\sigma_8$, indicating that it is important to account for galaxy assembly bias to obtain unbiased cosmological constraints.

Several recent works have attempted to constrain the galaxy-halo connection and/or cosmology in observational surveys using an extended HOD model that includes assembly bias \citep[e.g.,][]{Zentner2019, Vakili2019, Salcedo2022, Wang2022}.
Many of these works use the \citet{Hearin2016} ``decorated" HOD model, which adds two free parameters to the standard HOD model to control the strength of central and satellite occupation on a secondary property. Other works have extended the HOD model to include galaxy velocity bias \citep{Guo2015,Guo2015b}, while a few recent works have utilized an extended HOD that includes both assembly bias \textit{and} velocity bias to constrain the galaxy-halo connection and/or cosmology \citep[e.g.,][]{McCarthy2022, Lange2022, Zhai2022}.
\begin{deluxetable}{cccccc}
\tablenum{1}
\tablecaption{SDSS Volume-limited Sample Parameters\label{tab:sdss}}
\tablewidth{0pt}
\tablehead{
\colhead{$M_r^\mathrm{lim}$} & \colhead{$z_\mathrm{min}$} & \colhead{$z_\mathrm{max}$} & \colhead{$z_\mathrm{median}$} &
\colhead{$V_\mathrm{eff}$} & \colhead{$n_g$} \\
\colhead{} & \colhead{} & \colhead{} & \colhead{} &
\colhead{$(h^{-3}\mathrm{Mpc}^3)$} & \colhead{$(h^3\mathrm{Mpc}^{-3})$}
}
\startdata
$-19$ & 0.02 & 0.07 & 0.0562 & 6,087,119 & 0.01453 \\
$-21$ & 0.02 & 0.158 & 0.1285 & 67,174,396 & 0.00123
\enddata
\tablecomments{The columns list (from left to right): the absolute magnitude threshold of each sample at $z = 0.1$; the minimum, maximum, and median redshifts; the effective volume; and the galaxy number density of each sample. The volumes and number densities of the samples are corrected for survey incompleteness.}
\end{deluxetable}
In this work, we build on the procedure established in \citet{Sinha2018} and \citet{Szewciw2022}, extending the HOD model to include both assembly bias \textit{and} velocity bias.
We explore two different halo properties for implementing assembly bias, and identify an optimal set of clustering measurements to constrain our model. 
We also implement corrections to our halo masses to account for the impact of baryonic physics on the halo mass function. 
We use this framework to model the small-scale clustering of both low- and high-luminosity galaxies in SDSS. 
By using an optimal set of statistics, adding flexibility to our HOD model and accounting for the potential impact of baryonic physics, our goal is to make the most robust test to-date of our assumed \lcdm cosmological model using small-scale galaxy clustering.

In Section~\ref{sec:data} we describe our data, and in Section~\ref{sec:sims} we describe our simulations and halo catalogs.
In Section~\ref{sec:hod} we describe our halo model, and in Section~\ref{sec:modeling} we describe our full modeling procedure (including our mock galaxy catalogs, covariance matrices, clustering measurements, and MCMC framework). 
In Section~\ref{sec:opt} we describe our selection of optimal observables for constraining our HOD model, and in Section~\ref{sec:results} we describe our results.
We summarize our findings in Section~\ref{sec:conclusions}.

\begin{deluxetable*}{ccccccccc}
\tablenum{2}
\tablecaption{Simulation Parameters\label{tab:simulations}}
\tablewidth{0pt}
\tablehead{
\colhead{Use} & \colhead{Sample} & \colhead{Simulation} & \colhead{Seeds} &
\colhead{$L_\mathrm{box}$} & \colhead{$N_\mathrm{part}$} & \colhead{$m_\mathrm{part}$} & \colhead{$\epsilon$} & \colhead{Number} \\
\colhead{} & \colhead{} & \colhead{} & \colhead{} &
\colhead{(\hmpc)} & \colhead{} & \colhead{(\hMsun)} & \colhead{(\hkpc)} & \colhead{}
}
\startdata
Covariance matrix & $-19$ & Consuelo & 4001 - 4100 & 420 & $1400^3$ & $2.26 \times 10^9$ & 8 & 100 \\
Covariance matrix & $-21$ & Carmen & 2001 - 2100 & 1000 & $1120^3$ & $5.97 \times 10^{10}$ & 25 & 100 \\
MCMC & $-19$ & ConsueloHD & 4002, 4022 & 420 & $2240^3$ & $5.53 \times 10^8$ & 5 & 2 \\
MCMC & $-21$ & CarmenHD & 2007, 2023 & 1000 & $2240^3$ & $7.46 \times 10^9$ & 12 & 2
\enddata
\tablecomments{The columns list (from left to right): what each simulation is used for, the absolute magnitude threshold of the corresponding SDSS sample, the name of the simulation, the seeds used, the (comoving) boxsize, number of particles, mass resolution, (comoving) force softening, and the number of simulations.}
\end{deluxetable*}

\section{Observational Data} \label{sec:data}
In this work, we use the same observational dataset as that used in S22. 
We utilize the large scale structure samples from the NYU Value Added Galaxy Catalog \citep[NYU-VAGC;][]{Blanton2005} from the seventh data release \citep[DR7;][]{Abazajian2009} of the Sloan Digital Sky Survey \citep[SDSS;][]{York2000}.
The absolute magnitudes of the galaxies in this sample have been k-corrected to rest-frame magnitudes at redshift $z=0.1$ but have not been corrected for passive luminosity evolution.

From this sample, we construct two volume-limited subsamples, each complete down to a specified r-band absolute magnitude threshold ($M_r < -19$ and $M_r < -21$).
We refer to these samples as the $-19$ and $-21$ samples throughout this paper.
The luminosity thresholds, redshift limits, median redshifts, effective volumes, and number densities of our samples are listed in Table~\ref{tab:sdss}. 
The co-moving distances of the SDSS galaxies in our samples are determined using a flat \lcdm cosmological model with \Omegam = 0.302 and $h=1$.
Our distances are in units of \hmpc, and our absolute magnitudes are actually $M_r + 5\mathrm{log}h$\footnote{Throughout this paper, $\mathrm{log}$ refers to $\mathrm{log}_{10}$.}.

Fiber collisions are handled in the same way as in S22.
Briefly, we first adopt the nearest neighbor correction and then, informed by SDSS plate overlap regions, we apply additional corrections to our galaxy clustering measurements in order to account for errors in the nearest neighbor correction.
This method was applied in S22, and was recently further validated using the Uchuu-SDSS galaxy lightcones \citep{Uchuu2022}.
For more details on our observational data and our treatment of fiber collisions, see S22.

\section{Simulations and Halo Catalogs} \label{sec:sims}
In our modeling procedure, we make use of the same dark matter only cosmological N-body simulations as those used in S22. These simulations are from the Large Suite of Dark Matter Simulations project \citep[LasDamas;][]{McBride2009} and were run on the Texas Advanced Computing Center's Stampede supercomputer using the public code \textsc{gadget-2} \citep{Springel2005}. 
Power spectra were generated with \textsc{cmbfast} \citep{Seljak1996,Zaldarriaga1998,Zaldarriaga2000}, and initial conditions were generated with \textsc{2lptic} \citep{Scoccimarro1998, Crocce2006, Crocce2012}. 
All simulations were run with the following cosmological parameters, based on results from the Planck experiment \citep{Planck2014}: $\Omega_\mathrm{m}=0.302$, $\Omega_{\Lambda}=0.698$, $\Omega_\mathrm{b}=0.048$, $h=0.681$, $\sigma_8=0.828$, and $n_s=0.96$. 
For each observational sample of interest (i.e., -19 and -21), we run two sets of simulations: 100 low resolution simulations to build a covariance matrix, and 2 high resolution simulations for model parameter exploration.
The details of these simulations are given in Table~\ref{tab:simulations}.
We identify halos with a spherical over-density algorithm \citep[SO;][]{Lacey1994} using the \textsc{rockstar} phase-space temporal halo finder \citep{Behroozi2013}. We adopt a virial mass definition with a density threshold given by \citep{Bryan1998}.
Finally, for computational purposes, we randomly downsample to keep only 5\% of the dark matter particles in each halo, with no loss of accuracy (see S22).

\section{Halo Model} \label{sec:hod}
\subsection{The Standard HOD Model}
The Halo Occupation Distribution framework governs the number, positions, and velocities of galaxies within dark matter halos.
The standard HOD model assigns galaxies to halos based on five free parameters, which depend only on the halo's mass \citep{Zheng2007a}.
Galaxies are split into centrals and satellites within their halos \citep{Kravtsov2004,Zheng2005}.
In this model, the mean number of central galaxies in a halo of mass $M$ is described by
\begin{equation}
\langle N_\mathrm{cen} \rangle = \frac{1}{2}\bigg[1 + \mathrm{erf} \bigg(\frac{\mathrm{log} M - \mathrm{log}M_\mathrm{min}}{\sigma_{\mathrm{log} M}}\bigg)\bigg],
\end{equation}
where \mmin is the mass at which half of halos host a central galaxy, \siglogm is the scatter around this halo mass, and $\mathrm{erf}(x)$ is the error function, $\mathrm{erf}(x)=\frac{2}{\sqrt{\pi}}\int_0^x \mathrm{exp}(-y^2)dy$.
For a specific halo of mass $M$, we draw a random number $R$ from a uniform distribution on the interval $[0,1)$. 
If $R < \langle N_\mathrm{cen} \rangle$, then a central galaxy is assigned to the halo.
The central galaxy is always placed at the center of the halo and given the mean velocity of the halo.

The number of satellite galaxies in a given halo is drawn from a Poisson distribution with mean
\begin{equation}
\langle N_\mathrm{sat} \rangle = \langle N_\mathrm{cen} \rangle \times \bigg(\frac{M - M_0}{M_1}\bigg)^\alpha,
\end{equation}
where $M_0$ is the halo mass below which there are no satellite galaxies, $M_1$ is the mass where halos contain one satellite galaxy on average, and $\alpha$ is the slope of the power-law occupation function at high masses. Each satellite galaxy is given the position and velocity of a randomly selected dark matter particle within the halo.

\subsection{Assembly Bias}
One way in which we can extend the standard HOD model is to relax the assumption that halo occupation depends solely on the halo's mass.
In other words, we can allow for halo occupation to depend on both halo mass \textit{and} a secondary halo property, a phenomenon known as assembly bias \citep{Gao2005, Croton2007}.
To implement assembly bias, we use the decorated HOD (dHOD) model of \citet{Hearin2016}.
In order to apply this decorated HOD model, we first split halos by mass into bins of width 0.05 dex. 
Then, within each mass bin, we split halos into two groups based on the median value of the secondary property $s$ in each bin.
We then assign galaxies to halos based on the following conditional relations
\begin{equation}
\langle N_\mathrm{cen} | M, s_\mathrm{high} \rangle = \langle N_\mathrm{cen} | M \rangle + \delta N_\mathrm{cen},
\end{equation}
\begin{equation}
\langle N_\mathrm{cen} | M, s_\mathrm{low} \rangle = \langle N_\mathrm{cen} | M \rangle - \delta N_\mathrm{cen},
\end{equation}
\begin{equation}
\langle N_\mathrm{sat} | M, s_\mathrm{high} \rangle = \langle N_\mathrm{sat} | M \rangle + \delta N_\mathrm{sat},
\end{equation}
\begin{equation}
\langle N_\mathrm{sat} | M, s_\mathrm{low} \rangle = \langle N_\mathrm{sat} | M \rangle - \delta N_\mathrm{sat},
\end{equation}
where
\begin{equation}
\delta N_\mathrm{cen} = A_\mathrm{cen} \mathrm{MIN} [\langle N_\mathrm{cen} | M \rangle, 1 - \langle N_\mathrm{cen} | M \rangle]
\end{equation}

for central galaxies and
\begin{equation}
\delta N_\mathrm{sat} = A_\mathrm{sat} \langle N_\mathrm{sat} | M \rangle
\end{equation}
for satellite galaxies.
$A_\mathrm{cen}$ and $A_\mathrm{sat}$ have values between $-1$ and $1$; values of $0$ indicate no assembly bias. 
A key feature of this dHOD model is that, regardless of the strength of the assembly bias, $\langle N_\mathrm{cen} \rangle$ and $\langle N_\mathrm{sat} \rangle$ are preserved for a given halo mass. In other words, at fixed mass, for the same 5-parameter standard HOD model, the decorated HOD has the same halo occupation distribution when averaged over all halos.

Several works have explored the variety of different halo properties that can be used to model assembly bias.
\citet{Salcedo2018} explored halo assembly bias in the LasDamas simulations and found that a clustering bias exists if halos are binned by mass or by any other halo property, indicating that no single halo property encompasses all the spatial clustering information of the halo population. 
They also found that the mean values of some halo properties depend on their halo's distance to a more massive neighbor and concluded that this ``neighbor bias" largely accounts for the secondary bias seen in halos binned by mass and split by concentration or age. 
However, they also found that halos binned by other mass-like properties still show a secondary bias even when the neighbor bias is removed.

Meanwhile, \citet{Mao2018} presented a summary of secondary halo biases of high-mass halos due to various halo properties (e.g., concentration, spin, several proxies of assembly history, and subhalo properties) in the MultiDark Planck 2 simulation. 
They found that, while concentration, spin, and the abundance and radial distribution of subhalos exhibit significant secondary biases, properties that directly quantify halo assembly history do not.

Finally, \citet{Behroozi2022} examined the correlation of different properties of dark matter halos (e.g., growth rate, spin, concentration) with environment in the Small MultiDark Planck simulation and demonstrated that these halo properties imprint distinct signatures in the galaxy two-point correlation function and in the distribution of distances to galaxies' $k$th nearest neighbors. 
They demonstrated that the agreement with observed clustering can be improved with a simple empirical model in which galaxy size correlates with halo growth.

In this work, the first secondary halo property that we use to model assembly bias is halo concentration, $c$, defined as the ratio of the virial radius $R_\mathrm{vir}$ of the halo to the scale radius $R_s$ \citep{Navarro1997}.
The dependence of the galaxy-halo connenction on concentration or circular velocity has been explored in a number of previous works \citep[e.g.,][]{Lehmann2017,Xu2020}.
For a given halo, concentration can be found using the relationship between virial mass, maximum circular velocity, and concentration at $z=0$:
\begin{equation}
v_\mathrm{circ} (M_\mathrm{vir}) = \frac{6.72 \times 10^{-3} M_\mathrm{vir}^{1/3} \sqrt{c}} {\sqrt{ln(1+c) - c/(1+c)}}
\end{equation}
where $M_\mathrm{vir}$ is the virial mass of the halo in units of $h^{-1} M_\odot$ and $v_\mathrm{circ}$ is the maximum circular velocity of the halo in units of km/s \citep{Klypin2011}.
In our case, when implementing halo concentration as our secondary bias property, we determined that the normalization is irrelevant and it is only the halo ranking that matters; thus, we use $v_\mathrm{circ} / M_\mathrm{vir}^{1/3}$ as a proxy for concentration.
We refer to this assembly bias model using concentration as ``ABcon."

Another halo property that can be used to model assembly bias is its larger scale environment. The reason that conditioning the galaxy occupation on concentration has an impact on clustering statistics is that concentration is correlated with a halo's larger scale environment at fixed mass. Since we do not know a priori what secondary halo property to use in modeling assembly bias, it makes sense to skip this intermediate step and condition galaxy occupation directly on environment. 
Several works have explored the dependence of the galaxy-halo connection on environment \citep[e.g.,][]{Pujol2017,Shi2018,Hadzhiyska2020,Yuan2021}.
Motivated by these works, we choose to also explore the effects of using local halo environment to model assembly bias.
We define local environment as the total mass of \textit{halos} within a 5 \hmpc spheres centered on the halo of interest (excluding the mass of the halo of interest itself). 
We do not impose any lower mass limit on the halos included in this sum.
We refer to this assembly bias model using environment as ``ABenv."

\subsection{Velocity Bias}
Another way in which we can extend the HOD model is to relax the assumption that satellite galaxies trace the velocities of the dark matter particles within their host halo.
In other words, we can introduce satellite velocity bias (``VB") into our model.
We do this by introducing a new parameter, \Bvel, to the model.
\Bvel is defined as the ratio between the velocities of satellite galaxies and dark matter in the halo frame of reference:
\begin{equation}
B_\mathrm{vel} = \frac{v_g - v_h}{v_m - v_h}
\end{equation}
where $v_g$ is the velocity of the satellite galaxy, $v_h$ is the velocity of the halo, and $v_m$ is the velocity of the randomly chosen dark matter particle on which the satellite galaxy is placed.
A value of \Bvel less than $1$ indicates that satellite galaxies are moving with slower velocities than the dark matter, while a value of \Bvel greater than $1$ indicates that satellite galaxies are moving faster than the dark matter, and a value of \Bvel equal to $1$ indicates no velocity bias.

In this study we only model satellite velocity bias and not central velocity bias. In other words, we stick with the standard assumption that central galaxies inherit the same velocity as their host halo. In principle, central galaxies can move relative to their halo as predicted by some hydrodynamic simulations \citep{Berlind2003} and suggested for SDSS galaxies \citep[e.g.,][]{Guo2015,Guo2015b}. However, when comparing HOD modeling to hydrodynamic simulations \cite{Beltz-Mohrmann2020} found that the presence of central velocity bias is likely to have a negligible effect on the galaxy clustering statistics that we use in our analysis, unlike satellite velocity bias which is likely important for low luminosity galaxies.

\subsection{Accounting for Baryonic Effects}
While not strictly part of the HOD model, another way in which we can extend the standard halo modeling framework is to account for the effect of baryonic physics on the halo mass function.
The HOD model is typically applied to a halo catalog generated from a dark matter only (DMO) simulation, which does not account for the impact of baryonic physics on halo mass.
\citet{Beltz-Mohrmann2021} investigated the differences in halo mass functions between matched DMO and hydrodynamic simulations in EAGLE, Illustris, and IllustrisTNG, and found that, for halos at $z=0$, stellar feedback generally reduces the masses of low mass halos ($\lesssim 10^{11}$ \hMsun), while AGN feedback generally reduces the masses of high mass halos (between $10^{12}$ and $10^{13}$ \hMsun) compared to their DMO counterparts. 
However, the exact effect that feedback has on the halo masses differs dramatically from one hydrodynamic simulation to the next.
By matching halos according to mass between dark matter and hydrodynamic simulations, \citet{Beltz-Mohrmann2021} produced formulae which can be used to adjust the halo masses in a DMO simulation in order to reproduce the halo mass function from a given hydrodynamic simulation.
Additionally, they produced fits based on matching halos between dark matter and hydrodynamic simulations based on both mass \textit{and} local halo environment.
By taking halo environment into account, these fits can be used to adjust halo masses in a DMO simulation to not only reproduce the global halo mass function from a hydrodynamic simulation, but also to reproduce the conditional mass function, which then also reproduces the halo \textit{correlation} function.
In Section~\ref{subsec:hmf_corrections} we apply several of the halo mass corrections from \citet{Beltz-Mohrmann2021} to our halo catalogs, in order to assess the robustness of our results to changes in the mass function due to baryonic physics.
It should be noted that these mass corrections do not modify the halo profile, nor do they alter the velocity dispersion of dark matter within the halo, they adjust only the mass of each halo.

\subsection{Summary}
In this work, we extend the standard HOD model in several ways.
We first explore the effects of extending the standard HOD model to include concentration-based assembly bias.
We then explore the effects of instead using environment-based assembly bias.
Next we extend the model to include both assembly bias \textit{and} satellite velocity bias.
Finally, we implement halo mass corrections to account for the impact of baryonic physics, and we investigate the effects this has on the results from our most complete halo model (i.e., the model with both assembly bias and velocity bias).

\section{Modeling Procedure} \label{sec:modeling}
\subsection{Building mock galaxy catalogs}
We build mock galaxy catalogs to use as our model by populating the two high-resolution simulations for each sample (ConsueloHD and CarmenHD) with galaxies.
Once we populate our dark matter halos with galaxies, we build realistic mock galaxy catalogs that resemble our SDSS samples of interest.
To do this, we transpose the mock galaxies from Cartesian to spherical coordinates by positioning a virtual observer at the center of the box and converting the positions of the galaxies into RA, DEC, and comoving distances.
We then carve out four independent mock galaxy catalogs from each simulation box and incorporate the same systematic effects that plague our observational dataset, such as redshift-space distortions, sample geometry, and incompleteness.
For more information on the forward modeling details, see S22.

\subsection{Covariance Matrices}
If we wish to take advantage of the information present at small scales to constrain the galaxy-halo connection, it is essential that we construct accurate covariance matrices for our clustering measurements.
To do this, we run 100 low-resolution simulations for each sample (Consuelo and Carmen) which differ in the phases of the density modes of the power spectrum, which is controlled by a seed supplied to \textsc{2lptic}.
We populate these low-resolution simulations with galaxies using the same HOD parameters\footnote{The covariance matrices are built using a model that does not include assembly bias or velocity bias.} as were used to build the matrices in S22.
These parameters are listed in Table~\ref{tab:cov}.

We then build 400 \textit{independent} mock galaxy catalogs for each sample, from which we can construct a covariance matrix to represent cosmic variance.
The elements of the covariance matrix are given by
\begin{equation}
\label{eq:covariance}
C_{ij} = \frac{1}{N-1} \sum_{1}^{N}(y_i - \overline{y_i})(y_j - \overline{y_j})
\end{equation}
where the sum is taken over the $N=400$ mocks.
The values $y_i$ and $y_j$ are the $i$th and $j$th observables measured on each mock, while $\overline{y_i}$ and $\overline{y_j}$ are the mean values of the $i$th and $j$th observables, respectively.
Each diagonal element, $C_{ii}$, of the matrix is the variance across 400 mocks for observable $i$, and $\sqrt{C_{ii}}$ is the cosmic variance uncertainty of observable $i$.
For an arbitrary observable, we refer to this uncertainty as $\sigma_\mathrm{obs}$. S22 showed that the noise from using a finite number (400) of mock catalogs does not significantly affect our clustering analysis. 
Additionally, S22 examined the impact of resolution on the accuracy of our covariance matrices, and determined that the lower resolution of the Carmen and Consuelo simulations causes us to overestimate the error on the smallest scales of the correlation function by 10-20\%. However, not only is this a small effect, but larger cosmic variance uncertainties result in lower chi-square measurements and in general make it more difficult to rule out incorrect models, and we would rather have slightly broader constraints than artificially tight constraints.

\begin{deluxetable}{cccccccc}
\tablenum{3}
\tablecaption{Fiducial HOD parameters for covariance matrices\label{tab:cov}}
\tablewidth{0pt}
\tablehead{
\colhead{$M_r^\mathrm{lim}$} & \colhead{\logmmin} & \colhead{\siglogm} & \colhead{\logmzero} & \colhead{\logmone} & \colhead{$\alpha$}
}
\startdata
$-19$ & 11.54 & 0.22 & 12.01 & 12.74 & 0.92 \\
\hline
$-21$ & 12.72 & 0.46 & 7.87 & 13.95 & 1.17
\enddata
\tablecomments{The HOD parameters used to construct the covariance matrices in our analysis. Note that the matrices were constructed assuming zero assembly bias and velocity bias.}
\end{deluxetable}

\subsection{Clustering Statistics}
Several works have demonstrated the power of using a variety of different clustering statistics to constrain both the galaxy-halo connection as well as cosmology \citep{Berlind2002, Sinha2018, Hadzhiyska2021, Szewciw2022, Storey-Fisher2022}.
In our analysis, we employ the following clustering statistics: the projected correlation function \wprp \citep[e.g.,][]{Zehavi2002, Zehavi2004, Zheng2004, Zehavi2005, Zheng2007a, Zehavi2011, Leauthaud2012, Zentner2014, Coupon2015}, the redshift-space correlation function \zxi \citep[e.g.,][]{Tinker2006b, Parejko2013, Guo2015b, Padilla2019, Beltz-Mohrmann2020, Tonegawa2020}, the group multiplicity function \gmf \citep[e.g.,][]{Berlind2006a, Zheng2007b, Sinha2018, Beltz-Mohrmann2020}, the average group velocity dispersion \sigN, the mark correlation function \mcf \citep[e.g.,][]{Zu2018,Storey-Fisher2022}, and two special cases of counts-in-cells \cic: the void probability function \Pzero (\vpf) and the singular probability function \Pone (\spf) \citep[e.g.,][]{Tinker2006a, Tinker2008, McCullagh2017, Walsh2019, Wang2019, Beltz-Mohrmann2020, Perez2021}.
A detailed description of each of these clustering statistics is given in S22.
To calculate \wprp, \zxi, \mcf, \vpf, and \spf, we make use of the publicly available code \textsc{corrfunc} \citep{Sinha2019, Sinha2020}.
In our modeling procedure, we measure each clustering statistic in the same way (i.e., either on the full box/es or on the mock galaxy catalogs) as was done in S22.
It is important to note that our clustering statistics range in scale from about $0.1$ to $20$ \hmpc for both samples; thus, our analysis extends from the highly-nonlinear regime all way out to the ``quasi-linear" regime of clustering.

One of the main motivations for including so many higher-order statistics in our analysis is to ultimately obtain constraining power for both our model of the galaxy-halo connection and our cosmological model.
For example, redshift-space correlation function contains information about galaxy peculiar velocities due to redshift-space
distortions that change the apparent positions of galaxies
along the line of sight, which can help us constrain cosmic structure growth.
Additionally, \citet{Storey-Fisher2022} found that statistics beyond the standard galaxy clustering statistics (e.g. \wprp) significantly increase the constraining power on cosmological parameters of interest. Specifically, they found that including counts in cells statistics and the mark correlation function improves the precision of constraints on $\sigma_8$ by 33\%, $\Omega_m$ by 28\%, and the growth of structure parameter, $f\sigma_8$, by 18\% compared to standard statistics.
While we do not vary our cosmological model in this work, and thus cannot comment on the specific ability of each clustering measurement to constrain cosmological parameters, we include such a wide variety of clustering statistics in this work with the goal of ultimately constraining both HOD and cosmological parameters.

\subsection{MCMC}
We explore the HOD parameter space with a Markov Chain Monte Carlo (MCMC) algorithm, using a privately developed C-implementation of the popular affine-invariant sampler \emcee \citep{EMCEE2013}, which we call \cemcee.\footnote{https://github.com/aszewciw/emcee\_in\_c} 
We impose flat priors on the same parameter ranges given in S18, as well as flat priors of [-1.0,1.0] on \Acen and \Asat, and a flat prior of [0.5,1.5] on \Bvel.

At each point in the chain, we evaluate the likelihood that a particular HOD model could have generated a dataset with the same clustering as SDSS.
This likelihood is given by
\begin{equation}
    \label{eq:likelihood}
    \mathcal{L}(\mathbf{D}|\mathbf{M}) = \frac{\exp(-\frac{1}{2}{(\mathbf{D} - \mathbf{M}) \mathbf{C}^{-1} (\mathbf{D} - \mathbf{M})^{T}})}{\sqrt{(2\pi)^K\mathrm{det}(\mathbf{C})}},
\end{equation}
where \textbf{D} is the K-dimensional vector of observables measured on the SDSS dataset, \textbf{M} is the corresponding vector of observables measured on the HOD model, and \textbf{C} is the K-dimensional covariance matrix of these observables representing cosmic variance (see Equation \ref{eq:covariance}).
(The the term within the exponential is essentially $\chi^2$, multiplied by a factor of $-1/2$.)

It is worth noting that our likelihood calculation assumes that all of our observables are Gaussian.
However, \citet{Hahn2019} found that assuming a Gaussian likelihood in a group multiplicity function analysis slightly underestimates the uncertainties and biases HOD parameter constraints by $\sim 0.5\sigma$.
We have examined all of our observables and determined that the vast majority of them (including the group multiplicity function) appear to follow Gaussian distributions and pass typical tests of Gaussianity, so we have proceeded with assuming a Gaussian likelihood.

In the HOD framework, the process of populating halos with galaxies is stochastic, and is controlled with a ``population seed." 
For a fixed HOD model, changes in this population seed can lead to significant differences in clustering statistics.
To minimize the noise in our results due to this random variation, at each point in the chain we populate halos four times, using four fixed population seeds. 
Thus the clustering measurements for a given point in HOD parameter space are the average measurements over these four population seeds.
(See S22 for details.)

\begin{deluxetable}{ccccc}
\tablenum{4}
\tablecaption{Optimal Observable Order\label{tab:order}}
\tablewidth{\columnwidth}
\tablehead{
\colhead{Index} & \colhead{-19 sHOD} & \colhead{-19 dHOD} & \colhead{-21 sHOD} & \colhead{-21 dHOD}}
\startdata
1  & \ngal   & \ngal        & \ngal     & \ngal         \\
2  & \wprp 2 & \wprp 2      & \wprp 2   & \wprp 2       \\
3  & \wprp 4 & \sigN 3      & \zxi 8    & \bf{\sigN 1}  \\ 
4  & \vpf 3  & \bf{\zxi 8}  & \wprp 4   & \zxi 9        \\ 
5  & \wprp 8 & \gmf 3       & \mcf 9    & \zxi 3        \\ 
6  & \zxi 1  & \bf{\spf 1}  & \wprp 1   & \mcf 10       \\ 
7  & \gmf 3  & \wprp 3      & \zxi 9    & \bf{\wprp 5}  \\ 
8  & \zxi 5  & \gmf 2       & \mcf 7    & \gmf 1        \\ 
9  & \gmf 2  & \wprp 8      & \zxi 4    & \sigN 3       \\ 
10 & \gmf 4  & \zxi 1       & \zxi 7    & \mcf 3        \\ 
11 & \gmf 1  & \wprp 4      & \mcf 10   & \zxi 1        \\ 
12 & \spf 4  & \bf{\vpf 2}  & \zxi 1    & \zxi 8        \\ 
13 & \zxi 13 & \mcf 1       & \wprp 14  & \zxi 5        \\ 
14 & \mcf 14 & \zxi 10      & \gmf 1    & \wprp 1       \\ 
15 & \zxi 6  & \spf 2       & \spf 4    & \gmf 2        \\
16 & \gmf 5  & \bf{\zxi 4}  & \mcf 3    & \spf 4        \\
17 & \zxi 2  & \gmf 1       & \zxi 6    & \mcf 5        \\
18 & \spf 2  & \gmf 5       & \sigN 4   & \sigN 4       \\
19 & \zxi 10 & \wprp 1      & \zxi 5    & \bf{\mcf 14}  \\
20 & \mcf 2  & \spf 4       & \zxi 3    & \spf 3        \\
21 & \mcf 3  & \bf{\mcf 7}  & \gmf 4    & \wprp 3       \\
22 & \sigN 1 & \bf{\mcf 11} & \wprp 7   & \bf{\sigN 5}  \\
23 & \sigN 3 & \sigN 5      & \wprp 3   & \sigN 2       \\
24 & \zxi 9  & \bf{\spf 3}  & \mcf 8    & \zxi 7        \\
25 & \sigN 4 & \bf{\zxi 3}  & \vpf 3    & \bf{\gmf 3}   \\
26 & \mcf 1  & \gmf 4       & \zxi 2    & \gmf 4        \\
27 & \sigN 2 & \mcf 2       & \gmf 5    & \zxi 4        \\
28 & \gmf 6  & \sigN 2      & \gmf 2    & \zxi 2        \\
29 & \vpf 1  & \bf{\vpf 4}  & \zxi 11   & \gmf 5        \\
30 & \wprp 1 & \bf{\mcf 8}  & \sigN 3   & \bf{\wprp 8}  \\
31 & \wprp 6 & \wprp 6      & \sigN 2   & \wprp 4       \\
32 & \wprp 5 & \zxi 9       & \spf 2    & \zxi 6        \\
33 & \sigN 5 & \gmf 6       & \mcf 5    & \mcf 8        \\
34 & \wprp 3 & \mcf 14      & \spf 3    & \bf{\zxi 10}  \\
35 & \sigN 7 & \bf{\vpf 5}  & \mcf 4    & \zxi 11       \\
36 & \gmf 7  & \bf{\mcf 12} & \spf 1    & \mcf 7        \\
37 & --      & --           & --        & \bf{\mcf 12}  \\
38 & --      & --           & --        & \bf{\zxi 14}  \\
39 & --      & --           & --        & \bf{\vpf 5}   \\
40 & --      & --           & --        & \bf{\vpf 2}   \\
41 & --      & --           & --        & \bf{\mcf 1}
\enddata
\tablecomments{The type of clustering statistic and the bin number (1-indexing) for the observables chosen (in order) for each sample. ``sHOD" refers to the observables chosen for each sample using the standard HOD model in S22. ``dHOD" refers to the observables chosen in this work using the decorated HOD model. The observables chosen in this work that were \textit{not} chosen in S22 are shown in bold.}
\end{deluxetable}

\section{Choosing Optimal Observables} \label{sec:opt}
In order to constrain the dHOD when fit to SDSS, we must first choose a set of observables to use in our MCMC.
We cannot arbitrarily continue to increase the number of observables we use, because doing so increases the noise in our modeling and degrades our final constraints.
Noise is introduced into the covariance matrix due to the fact that we are constructing it from only 400 mocks.
This noise propagates into the likelihood function and ultimately into our posterior results.
Thus, we need to choose our observables wisely.
We seek a subset of observables that produce the tightest constraints on our HOD parameters, at the cost of little noise.

To choose an ``optimal" set of high-information, low-noise observables, we employ the importance sampling algorithm described in S22.
In this algorithm, we first create four mock SDSS catalogs for which we will determine optimal statistics. We use four mocks instead of the actual SDSS data in order to minimize the impact of cosmic variance on the selection of optimal statistics. We build these four mocks using a fiducial dHOD (with concentration) model with parameters that we obtain by fitting this model to the S22 set of clustering statistics (listed in Table~\ref{tab:order} under ``sHOD") measured on the SDSS. 

We run an initial MCMC on each of the four mock galaxy catalogs, fitting the dHOD (with concentration) model to only two observables: \ngal and \wptwo. This results in a fairly broad MCMC non-uniform grid of points in parameter space for each mock. We then use importance sampling on these grids to explore the constraining power of different combinations of clustering statistics.
The algorithm chooses observables one by one, each time selecting the observable that, when combined with all previously chosen observables, produces the tightest projected constraints on all HOD parameters of interest.
When choosing an observable, we consider how it performs on across all four grids, minimizing any bias due to cosmic variance.
Thus, at the end of running this algorithm, we have a list of observables (ordered in terms of cumulative constraining power) and a corresponding list of cumulative projected constraints for each sample.
(We refer the reader to S22 for a more complete description of this procedure.)

\begin{figure*}
\centering
\includegraphics[width=6in]{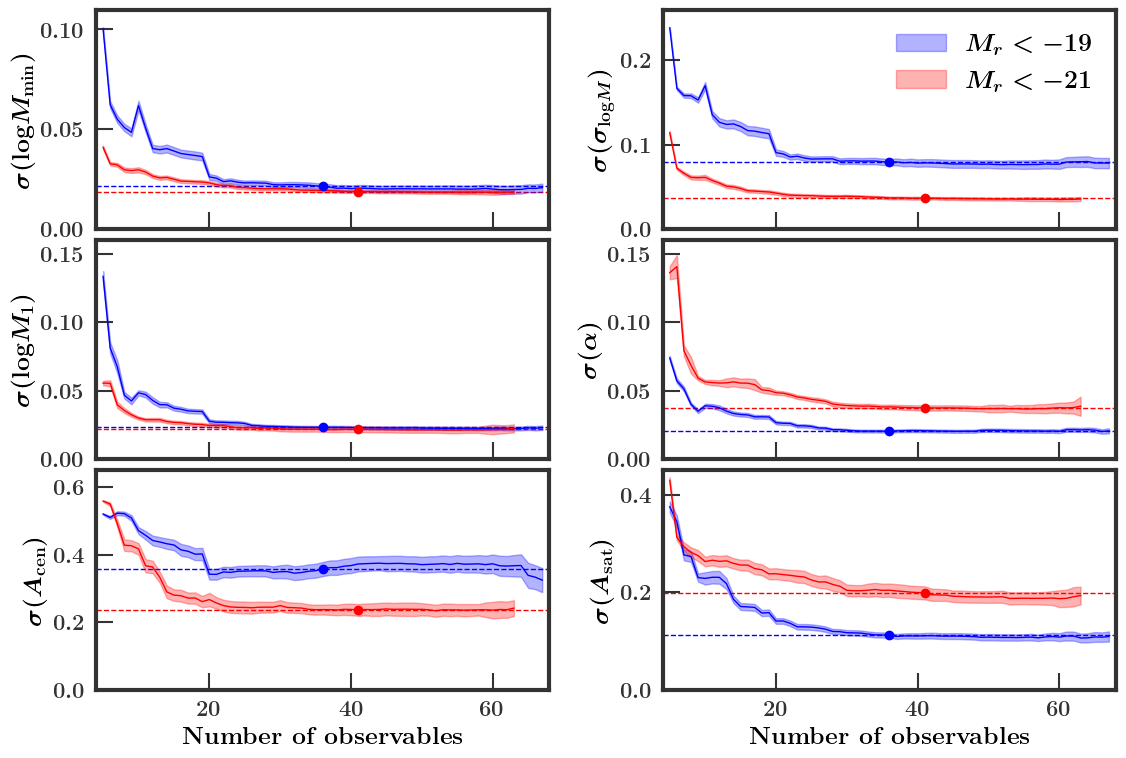}
\caption{\small Constraints on each HOD parameter as we increase the number of observables, for the $-19$ sample (blue) and the $-21$ sample (red). The solid line in each panel shows the average mock constraint (1-$\sigma$) across four mocks, and the shaded region is an estimate of the uncertainty (inner 68\%) in these constraints due to the noise present in the covariance matrix. The dot indicates the optimal number of observables for each sample, and the dashed line indicates the corresponding constraining power for each parameter.}
\label{fig:posterior_error}
\end{figure*}

There are two key differences in our implementation of this algorithm compared to S22.
First, when choosing the third observable for each sample, we only attempt to constrain \Acen and \Asat.
This is because these parameters are entirely unconstrained when using only \ngal and \wptwo, which causes the MCMC to explore unrealistic HOD models; thus, it is essential to choose an observable early on that provides constraining power for these parameters.
After the third observable is chosen, we make all successive choices by attempting to jointly constrain all HOD parameters (excluding \logmzero for the $-21$ sample).
Second, in the S22 algorithm, new grids are created (by running new MCMCs using the already chosen observables) whenever the old grids become insufficiently dense for importance sampling.
S22 creates these new grids after choosing five observables for each sample, and again for the $-19$ sample after choosing eight observables.
In our case, because we are trying to constrain two additional parameters, our grids become insufficiently dense more quickly, and so we ultimately build denser grids after choosing three, five, ten, and twenty observables for each sample.

In Figure~\ref{fig:posterior_error}, we show our estimated constraint for each HOD parameter (excluding \logmzero) as we choose successive observables.
The results for the $-19$ sample are shown in blue, and the results for the $-21$ sample are shown in red.
The solid lines show the average constraint across the four mocks used in the algorithm described above.
In Table~\ref{tab:order}, we list the observables chosen (in order) that we use for each sample (labeled ``dHOD").
We also list the observables chosen in S22 using a standard HOD model (i.e., no assembly bias, labeled ``sHOD").
The observables chosen in this work that were \textit{not} chosen in S22 are shown in bold.

\begin{figure*}
\centering
\includegraphics[width=6in]{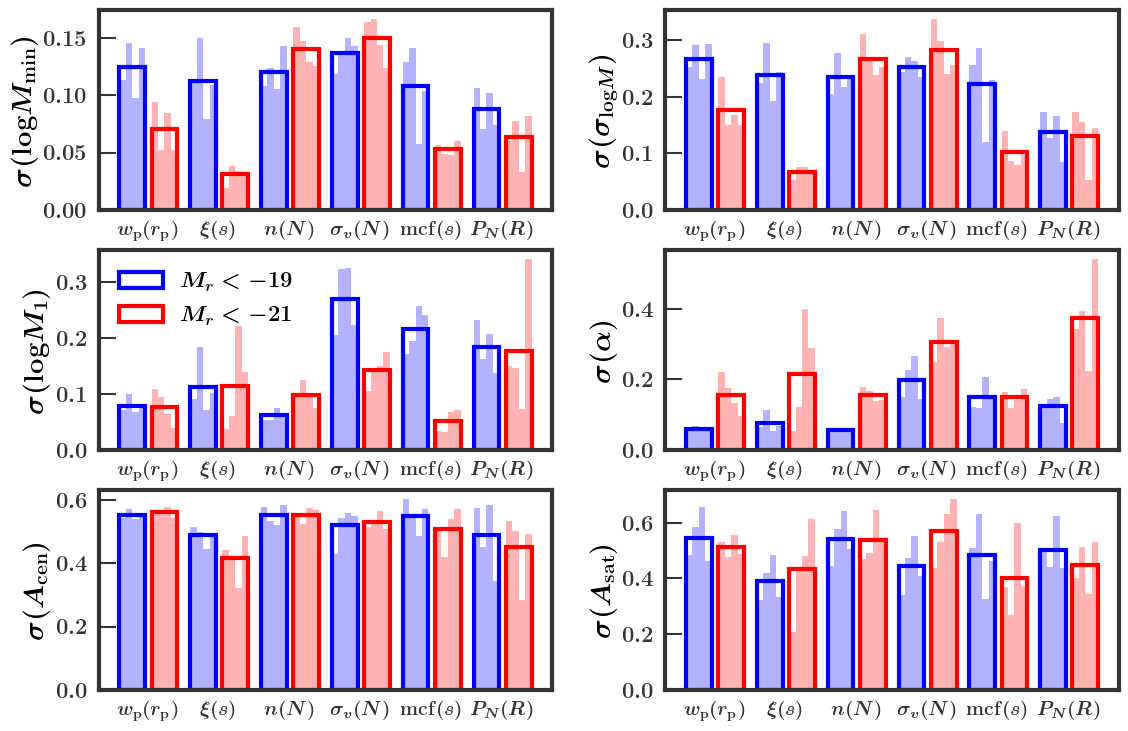}
\caption{\small Projected constraints (1-$\sigma$) of each clustering statistic (combined with \ngal) for each HOD parameter. The constraints for the $-19$ and $-21$ mocks are shown in blue and red, respectively. The height of each smaller vertical bar shows the projected constraints on one mock, while the larger open bar shows the average constraint across four mocks.}
\label{fig:statpower}
\end{figure*}

After ordering the observables from greatest to least constraining power, we need to choose the total number of observables to use in our analysis to maximize our constraining power and minimize the noise in our procedure due to building our covariance matrix from a finite number of mocks (i.e., 400).
To do this, we employ the same procedure as S22.
Briefly, we estimate an uncertainty associated with each projected constraint (for a given number of observables $K$) by resampling the covariance matrix 100 times and then importance sampling the chain with each of these resampled matrices.
Doing so lets us approximate the uncertainty in our constraints due to noise in the covariance matrix for each combination of observables that we consider.
The shaded regions in each panel of Figure~\ref{fig:posterior_error} show this uncertainty for each HOD parameter as we increase $K$.
We choose the lowest value of $K$ such that the constraint at this value is within one standard error of the constraint at all higher values of $K$.
We require that this condition is met for all HOD parameters (except \logmzero for the $-21$ sample).
The optimal number of observables for each sample is indicated with a dot in each panel, and the corresponding constraining power is shown with a dashed line.
For the $-19$ sample, the optimal number of observables is 36.
For the $-21$ sample, the optimal number of observables is 41.
Thus, the size of our data vector for each sample is 36 and 41, respectively.

Using these observables, we confirm that we can recover the truth when running chains on mocks created with different HOD parameters (i.e., different amounts of assembly bias) for each sample.
In all of our validation tests with mocks, all parameters for the -21 sample are always recovered within $1\sigma$, and in the -19 sample all parameters are always recovered within or just outside the $1\sigma$ region.
Additionally, for both samples, the best-fit result is always a good fit (i.e., it always has a p-value greater than 0.85).

Looking at the optimal observables in Table~\ref{tab:order}, it is noteworthy that for both the $-19$ and $-21$ samples, the third observable (chosen to constrain only $A_\mathrm{cen}$ and $A_\mathrm{sat}$) is a small bin of the average group velocity dispersion function (\sigN 3 for $-19$ and \sigN 1 for $-21$).
It is also noteworthy that for both samples, the majority of the first twenty observables chosen in this analysis (16/20 or 17/20) were also chosen in S22 to constrain an HOD model without assembly bias.
Meanwhile, about half of the observables chosen beyond the initial twenty (8/16 or 9/21) in this analysis are unique to the model with assembly bias (i.e., they were not chosen in S22). 
This possibly indicates that the initial observables are chosen for their ability to constrain the standard HOD parameters, while the later observables are selected for their ability to constrain the assembly bias parameters.
This may also indicate that it is difficult to constrain assembly bias until the standard HOD parameters are constrained, or that assembly bias is a smaller signal on top of the global clustering signal.

For the $-19$ sample, the unique observables chosen for this analysis include a large and small scale of \zxi, five scales of \cic, and four large scales of \mcf.
For the $-21$ sample, the unique observables chosen for this analysis include two bins of \sigN, two intermediate scales of \wprp, one small scale and two large scales of \mcf, one intermediate bin of \gmf, two large scales of \zxi, and two bins of \vpf.
It is worth mentioning that for the $-19$ sample, it is difficult to accurately constrain the decorated HOD model until the parameter \logmzero is constrained.
This occurs by about 15 observables, particularly after \zxi 1 and \wprp 4 are included.
In the $-21$ sample, the parameter \logmzero remains unconstrained.
This is consistent with the results of S22, which found that constraining \logmzero is important for obtaining accurate results in the $-19$ sample, but not in the $-21$ sample.

\begin{figure*}
\centering
\includegraphics[width=.33\linewidth]{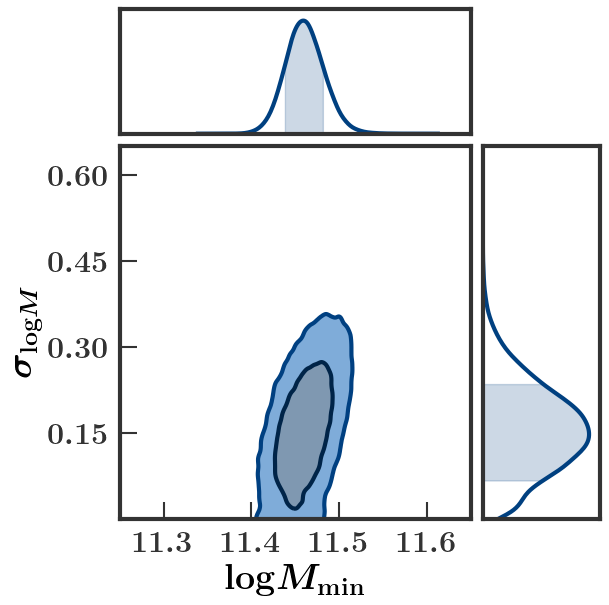}%
\includegraphics[width=.33\linewidth]{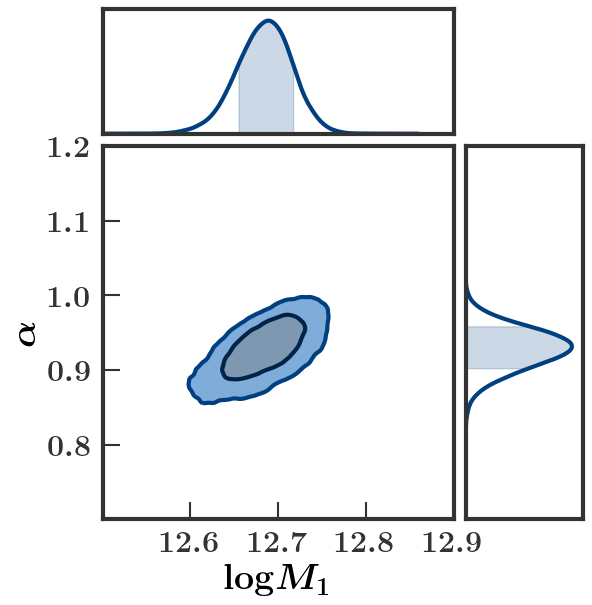}%
\includegraphics[width=.33\linewidth]{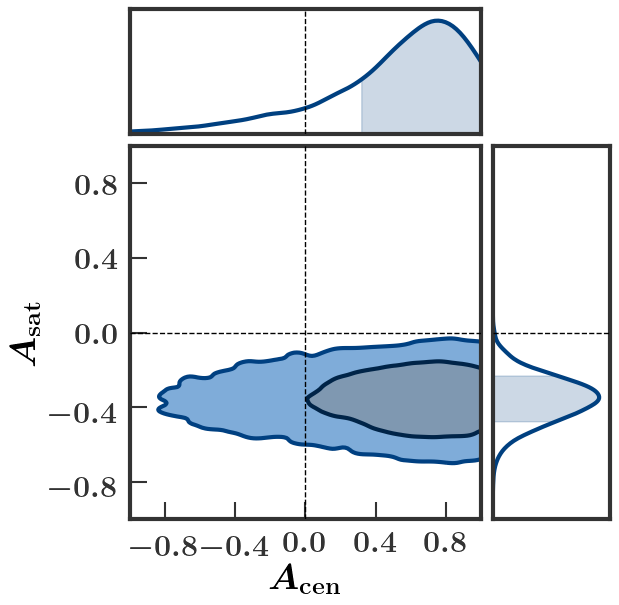}
\caption{Parameter constraints for the SDSS $-19$ sample, using concentration as the secondary halo property and the ``dHOD" optimal observables (listed in Table~\ref{tab:order}). The crosshairs in the third panel indicate $A_\mathrm{cen} = A_\mathrm{sat} = 0$ (i.e., no assembly bias).}
\label{fig:sdss19con}
\end{figure*}

\begin{figure*}
\centering
\includegraphics[width=.33\linewidth]{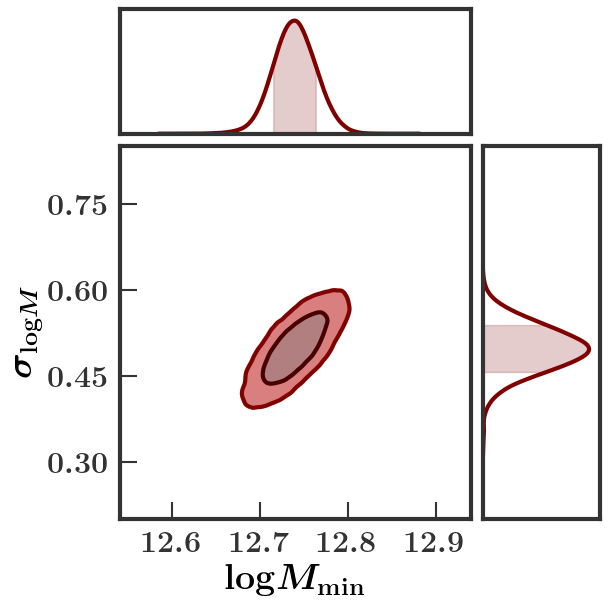}%
\includegraphics[width=.33\linewidth]{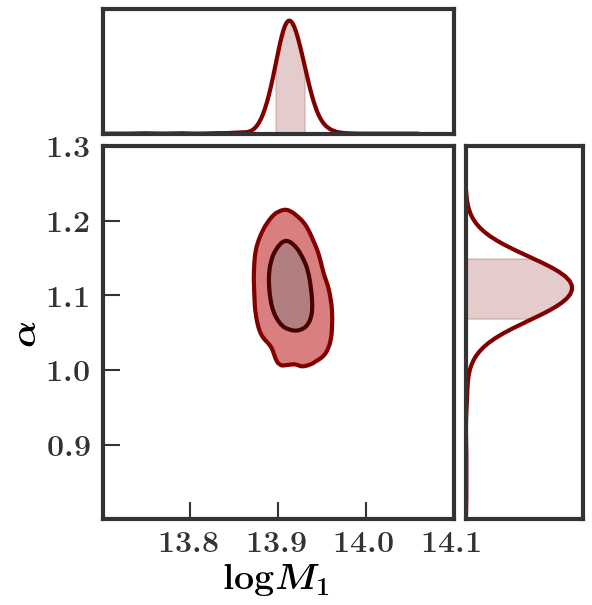}%
\includegraphics[width=.33\linewidth]{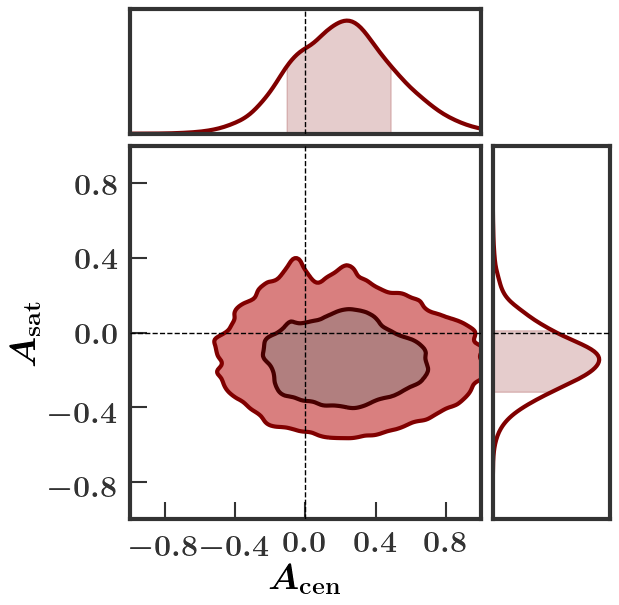}
\caption{Parameter constraints for the SDSS $-21$ sample, using concentration as the secondary halo property and the 41 ``dHOD" optimal observables (listed in Table~\ref{tab:order}). The crosshairs in the third panel indicate $A_\mathrm{cen} = A_\mathrm{sat} = 0$ (i.e., no assembly bias).}
\label{fig:sdss21con}
\end{figure*}

Given the MCMC grids that we obtained from the first three observables in our optimal selection algorithm, we can use importance sampling to estimate the constraining power we would achieve for each HOD parameter had we run a chain using only one clustering statistic (e.g., \wprp) plus \ngal.
We display the results of this exercise in Figure~\ref{fig:statpower}.
In each panel, the y-axis shows the projected constraint (1-$\sigma$) for a particular HOD parameter as we use different clustering statistics. 
The constraints for the $-19$ and $-21$ mocks are shown in blue and red, respectively. 
The height of each smaller vertical bar shows the projected constraints on one mock, while the larger open bar shows the average constraint across four mocks.

For the central and satellite parameters, our results are similar (though not identical) to the results from S22.
For the assembly bias parameters, it is interesting to note that for both samples, no single clustering statistic provides significant constraining power for either \Acen or \Asat.
\zxi seems to have the most constraining power for both \Acen and \Asat in both samples, but it performs only slightly better than the other clustering statistics.

Due to the nature of importance sampling, these results should be interpreted as \textit{estimates}, purely for visual purposes.
However, this figure illustrates that while no single clustering statistic provides significant constraining power for assembly bias, the \textit{combination} of different scales of each clustering statistic is able to produce tighter constraints on the assembly bias parameters than any one statistic.

\begin{figure*}
\centering
\includegraphics[width=6in]{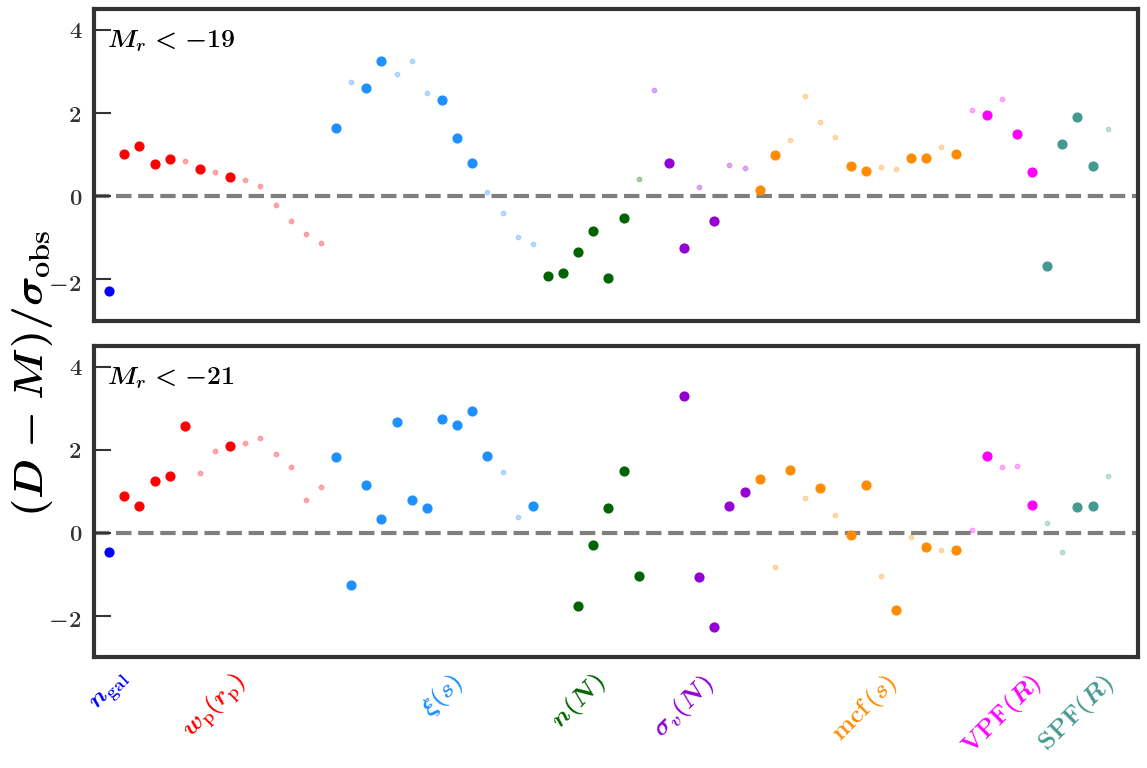}%
\caption{Residuals between the best-fit model and the SDSS measurements for the $-19$ (top) and $-21$ (bottom) samples. For each sample, the model includes concentration-based assembly bias. We show residuals for all observables, but the model was constrained using the ``dHOD" optimal observables for each sample (listed in Table~\ref{tab:order}), which are displayed with larger points.}
\label{fig:con_residuals}
\end{figure*}

\section{Results} \label{sec:results}
\subsection{Concentration-based Assembly Bias}
Here we present the results from using the optimal observables identified in the previous section to constrain the galaxy-halo connection in SDSS using a decorated HOD model with concentration-based assembly bias.
The results for the $-19$ sample are shown in Figure~\ref{fig:sdss19con}, while the results for the $-21$ sample are shown in Figure~\ref{fig:sdss21con}.
Dark and light regions depict the 1- and 2-$\sigma$ regions, respectively.
The best-fit parameters are listed in Table~\ref{tab:bestfit}, along with their corresponding p-values, as well as the results from the previous S22 analysis for comparison.
The constraints for each parameter are listed in Table~\ref{tab:constraints}.

For the $-19$ sample, our best-fit results suggest positive central galaxy assembly bias (\Acen = 0.793) and negative satellite galaxy assembly bias (\Asat = -0.368). 
In other words, central galaxies preferentially reside in halos with higher concentrations, while satellite galaxies preferentially reside in halos with lower concentrations, at fixed mass.
This is consistent with previous results \citep[e.g.,][]{Lange2022, Wang2022} which also found positive central galaxy assembly bias and negative satellite galaxy assembly bias when using concentration as the secondary halo property.
Additionally, this best-fit model yields a significant decrease in tension compared to the results of S22 ($2.0\sigma$ compared to $4.5\sigma$).
Unfortunately, even for our optimal combination of observables, it is difficult to tightly constrain central galaxy assembly bias for this sample (see the third panel in Figure~\ref{fig:sdss19con}).
Despite the lack of tight constraints on \Acen, we are able to rule out a model with zero assembly bias (i.e., the standard HOD model).

\begin{figure*}
\centering
\includegraphics[width=.25\linewidth]{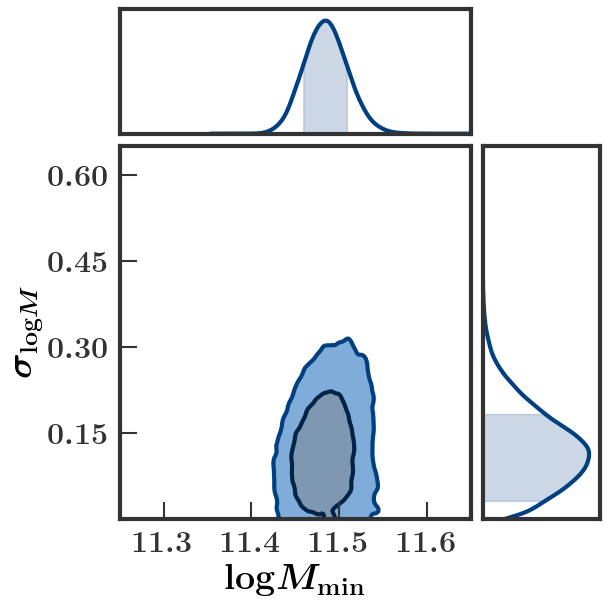}%
\includegraphics[width=.25\linewidth]{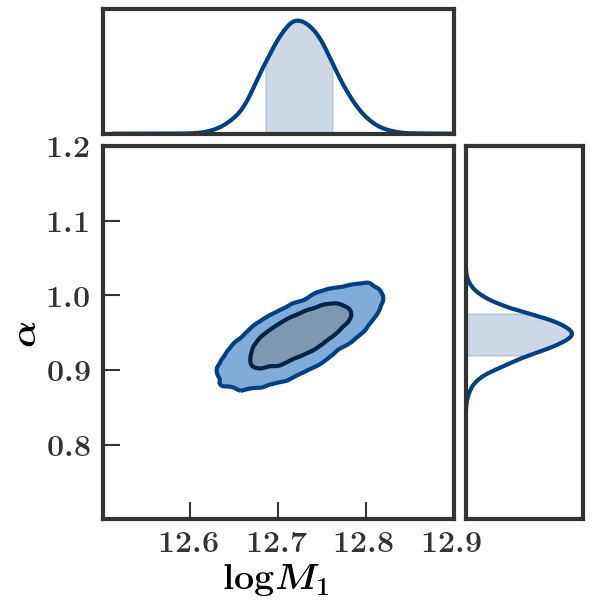}%
\includegraphics[width=.25\linewidth]{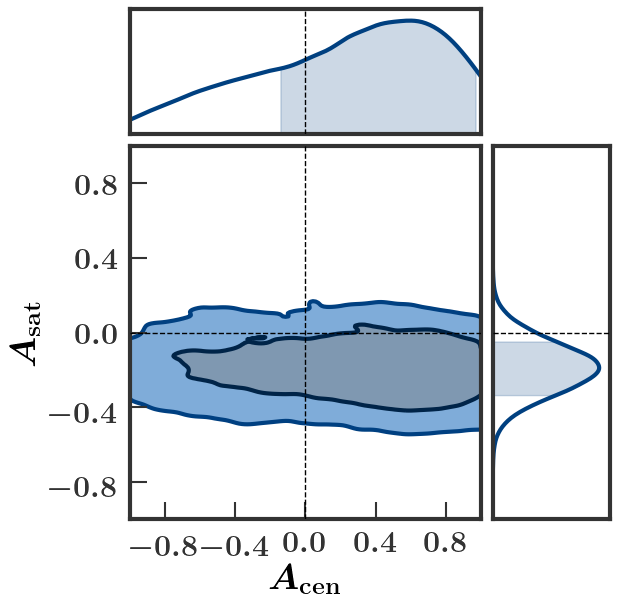}%
\includegraphics[width=.25\linewidth]{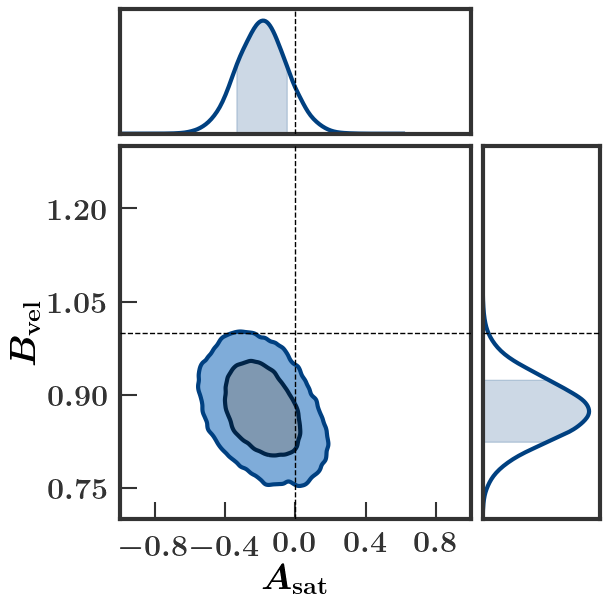}
\caption{Parameter constraints for the SDSS $-19$ sample, from a model with assembly bias (with concentration as the secondary halo property) and satellite velocity bias, using the ``dHOD" optimal observables (listed in Table~\ref{tab:order}). The crosshairs in the third and fourth panels indicate no assembly bias and no velocity bias ($A_\mathrm{cen} = A_\mathrm{sat} = B_\mathrm{vel} =  0$).}
\label{fig:sdss19con_vbias}
\end{figure*}

\begin{figure*}
\centering
\includegraphics[width=.25\linewidth]{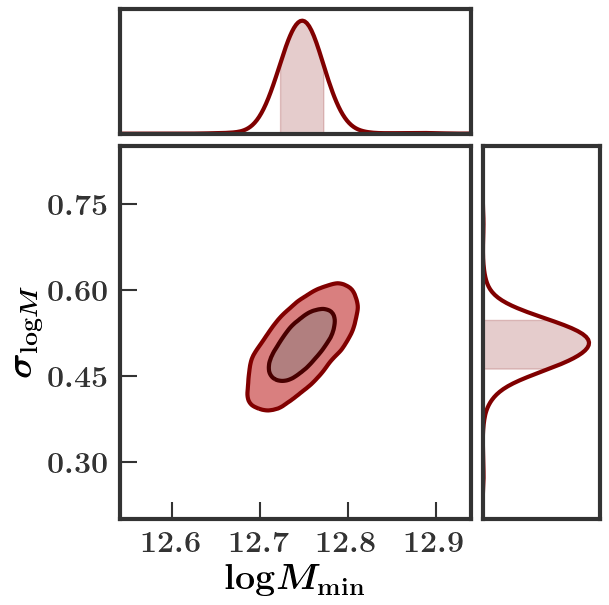}%
\includegraphics[width=.25\linewidth]{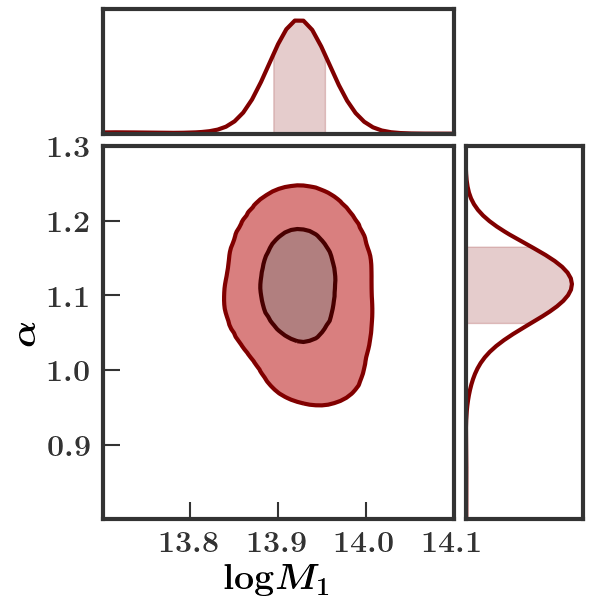}%
\includegraphics[width=.25\linewidth]{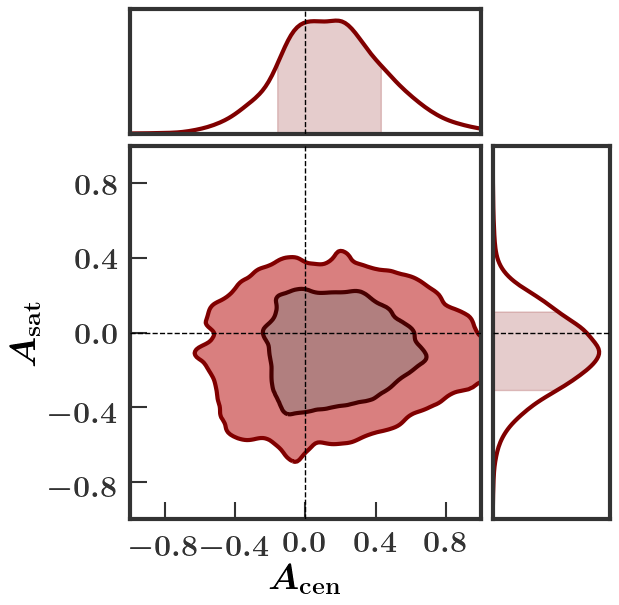}%
\includegraphics[width=.25\linewidth]{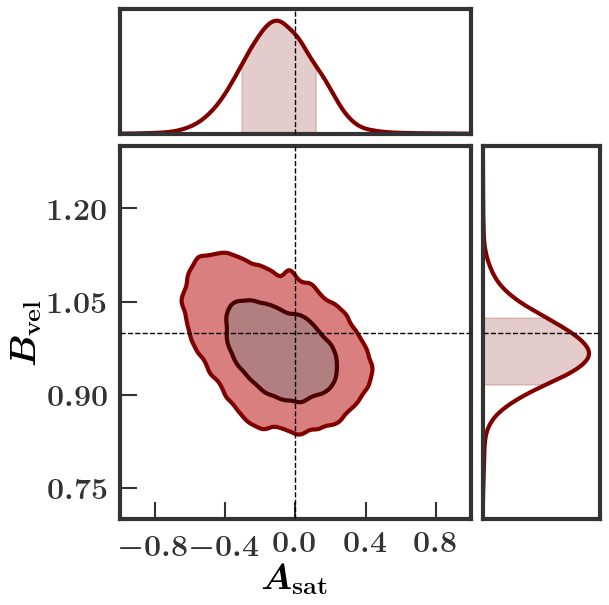}
\caption{Parameter constraints for the SDSS $-21$ sample, from a model with assembly bias (with concentration as the secondary halo property) and satellite velocity bias, using the ``dHOD" optimal observables (listed in Table~\ref{tab:order}). The crosshairs in the third and fourth panels indicate no assembly bias and no velocity bias ($A_\mathrm{cen} = A_\mathrm{sat} = B_\mathrm{vel} =  0$).}
\label{fig:sdss21con_vbias}
\end{figure*}

For the $-21$ sample, we obtain slightly tighter constraints on \Acen than we are able to achieve in the $-19$ sample.
However, our best-fit results are consistent with zero assembly bias.
Additionally, this model does not result in any decrease in tension compared to the results from S22.\footnote{In fact, the tension actually increased slightly compared to the previous analysis, from $4.1\sigma$ up to $4.7\sigma$. This slight increase in tension can be attributed to the change in observables between this work and the previous work.}
This finding is consistent with the results of \citet{Beltz-Mohrmann2020}, which found assembly bias to be present in hydrodynamic simulations for lower luminosity galaxies but not significant for higher luminosity galaxies.
It is thus to be expected that for the $-21$ sample, the addition of assembly bias parameters to the model did not result in any relief of tension.
Furthermore, the constraints on the standard HOD parameters in the $-21$ sample do not change considerably compared to what they were in S22, indicating that the addition of assembly bias has very little affect on the outcome of the model.

In Figure~\ref{fig:con_residuals} we show the deviation between each observable as measured on SDSS (\textbf{D}) and on our best-fit model (\textbf{M}) for each sample.
This deviation is shown as a factor of the cosmic variance uncertainty, $\sigma_\mathrm{obs}$, for each observable.
This quantity is shown for all observables, where each point is a different scale or bin of a given clustering statistic.
Each clustering statistic is plotted in a different color and is labeled on the x-axis.
The specific observables actually used in our analysis are plotted as larger bold points.
The results for the $-19$ sample are shown in the top panel, and the results for the $-21$ panel are shown in the bottom panel.
In the $-19$ sample, much of the remaining tension seems to be coming from several scales of \zxi and, to a lesser extent, \ngal, \gmf, \vpf, and \spf. 
In the $-21$ sample, most of the clustering statistics exhibit a high degree of tension at various scales.
Overall, the $-19$ sample exhibits a greater improvement in the observable residuals compared to the S22 results than the $-21$ sample does, which explains the greater overall reduction in tension seen in this sample.

The remaining tension found for both the $-19$ and $-21$ samples could indicate that the HOD model needs to be made even more flexible with the inclusion of spatial and velocity bias parameters \citep[e.g.,][]{Beltz-Mohrmann2020}.
Additionally, it is possible that a different secondary halo property other than concentration could be more strongly correlated with galaxy occupation and is thus a more appropriate choice for our assembly bias model.
It is also possible that accounting for the impact of baryonic physics on the halo mass function could relieve some of the remaining tension.
Finally, these results are for a fixed cosmology sample; it is possible that a slight change in cosmological parameter values could also result in a further relief of this tension.
We explore some of these possibilities in the remaining sections.

\begin{figure*}
\centering
\includegraphics[width=6in]{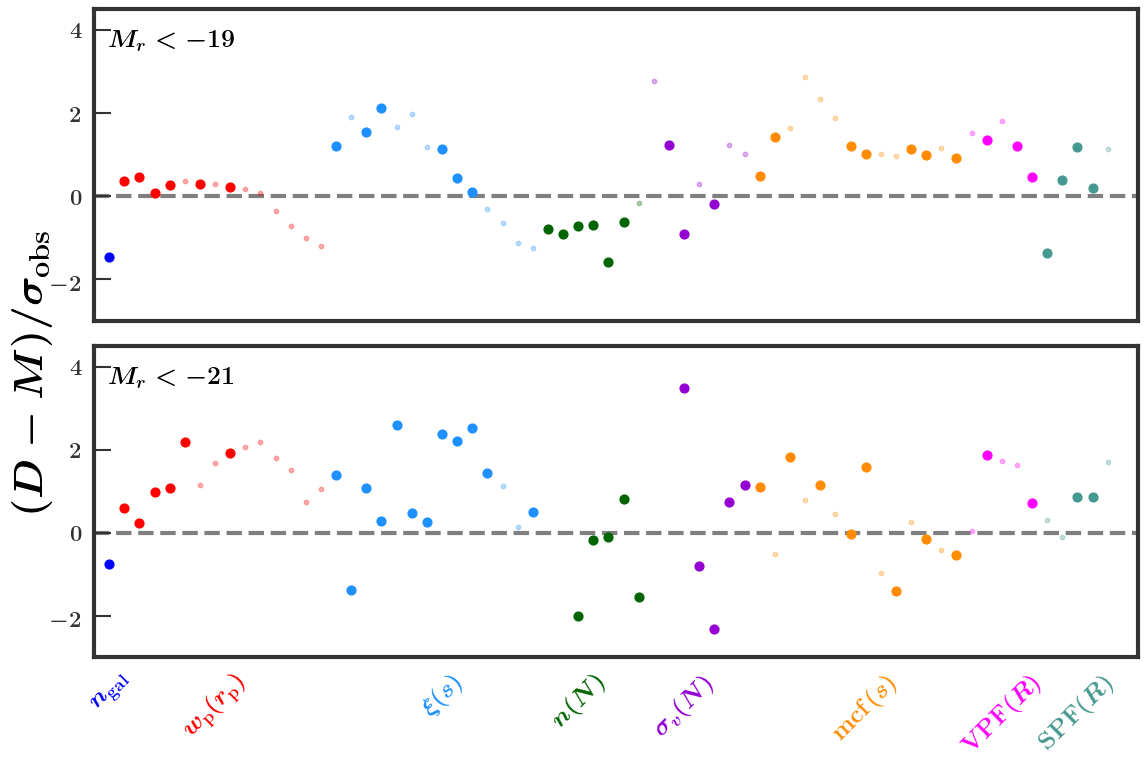}%
\caption{Residuals between the best-fit model and the SDSS measurements for the $-19$ (top) and $-21$ (bottom) samples. For each sample, the model includes assembly bias (with concentration as the secondary halo property) as well as velocity bias. We show residuals for all observables, but the model was constrained using the ``dHOD" optimal observables for each sample (listed in Table~\ref{tab:order}), which are displayed with larger points.}
\label{fig:con_vbias_residuals}
\end{figure*}

\subsection{Satellite Velocity Bias} \label{subsec:vbias}
Here we investigate whether adding satellite velocity bias to our HOD model results in better agreement with SDSS.
Using the same set of optimal observables for each sample listed in Table~\ref{tab:order}, we run chains on our SDSS samples using an HOD model with both concentration-based central and satellite galaxy assembly bias (i.e., \Acen and \Asat) and additionally with satellite galaxy velocity bias (\Bvel).
The results for the $-19$ sample are shown in Figure~\ref{fig:sdss19con_vbias}, while the results for the $-21$ sample are shown in Figure~\ref{fig:sdss21con_vbias}.
The best-fit parameter values and constraints are listed in Tables~\ref{tab:bestfit} and \ref{tab:constraints}.
Additionally, in Figure~\ref{fig:con_vbias_residuals} we show the deviation between each observable as measured on SDSS (\textbf{D}) and on our best-fit model (\textbf{M}) for each sample, with the same layout as in Figure~\ref{fig:con_residuals}.

We do not identify a new set of optimal observables for constraining this new model, but rather run chains for each of our SDSS samples using the same optimal observables listed in Table~\ref{tab:order} (``dHOD").
While choosing a new set of optimal observables could potentially lead to tighter constraints on \Bvel, we emphasize that our goal is not to optimally constrain satellite velocity bias, but rather to allow \Bvel to vary with the hope of alleviating the lingering tension with SDSS.
Additionally, we note that our optimal set of observables already includes many measurements that are sensitive to galaxy velocities (e.g., \zxi, \gmf, \sigN) and so should contain constraining power for \Bvel. 

For the $-19$ sample, our best-fit results for this model indicate moderate satellite galaxy velocity bias, with satellite galaxies moving slightly slower than the dark matter distribution (\Bvel = 0.898).
However, when velocity bias is included in the model, the strength of the assembly bias signal is significantly reduced. 
This is in part due to the anti-correlation between \Bvel and (concentration-based) \Asat, which can be seen in the fourth panel of Figure~\ref{fig:sdss19con_vbias}: when $B_\mathrm{vel} = 1$, lower values of \Asat are preferred by the model, but when \Bvel is allowed to be less than 1, \Asat increases.
While the best-fit parameter values still suggest positive central assembly bias and negative satellite assembly bias (\Acen = 0.825 and \Asat = -0.251), the constraints on \Acen and \Asat are much weaker, and we can no longer rule out a model with zero assembly bias (though we can rule out a model with zero velocity bias).

\begin{figure*}
\centering
\includegraphics[width=.25\linewidth]{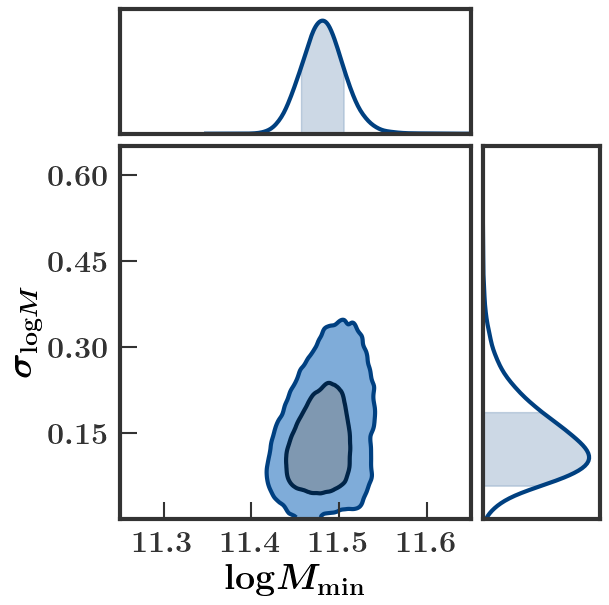}%
\includegraphics[width=.25\linewidth]{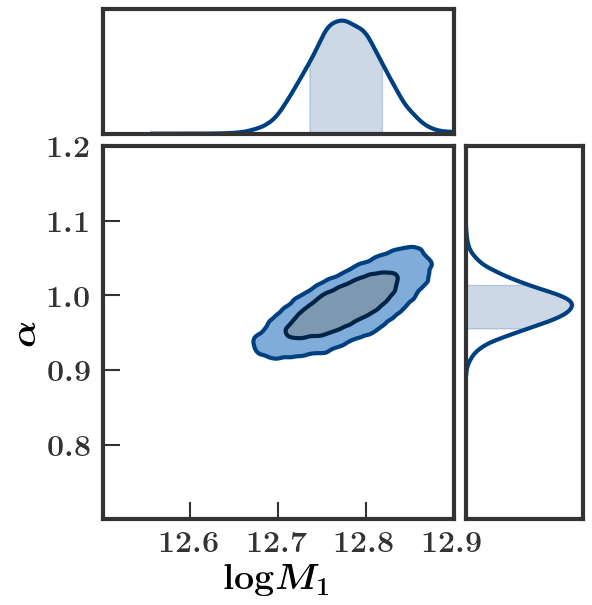}%
\includegraphics[width=.25\linewidth]{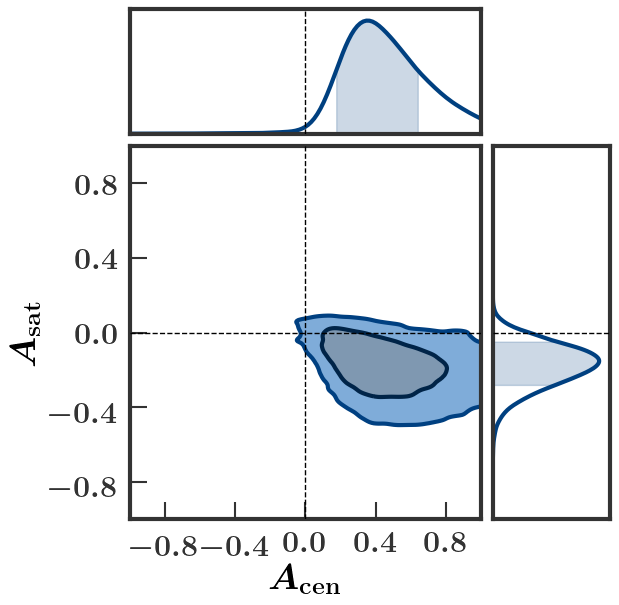}%
\includegraphics[width=.25\linewidth]{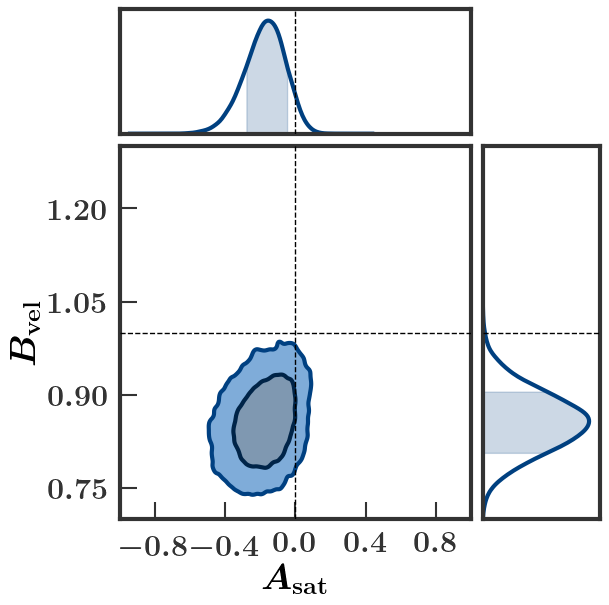}
\caption{Parameter constraints for the SDSS $-19$ sample, from a model with assembly bias (with environment as the secondary halo property) and satellite velocity bias, using the ``dHOD" optimal observables (listed in Table~\ref{tab:order}). The crosshairs in the third and fourth panels indicate no assembly bias and no velocity bias ($A_\mathrm{cen} = A_\mathrm{sat} = B_\mathrm{vel} =  0$).}
\label{fig:sdss19env_vbias}
\end{figure*}

\begin{figure*}
\centering
\includegraphics[width=.25\linewidth]{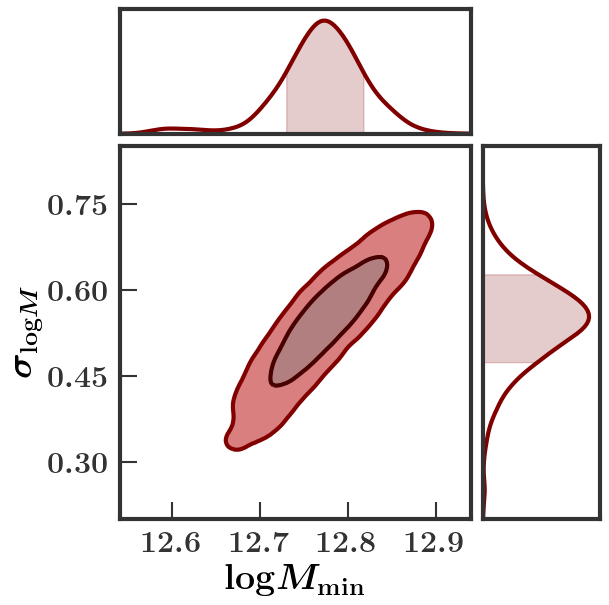}%
\includegraphics[width=.25\linewidth]{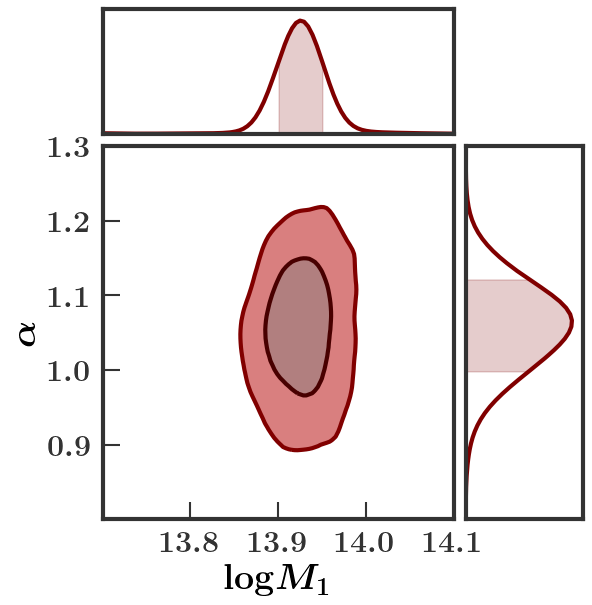}%
\includegraphics[width=.25\linewidth]{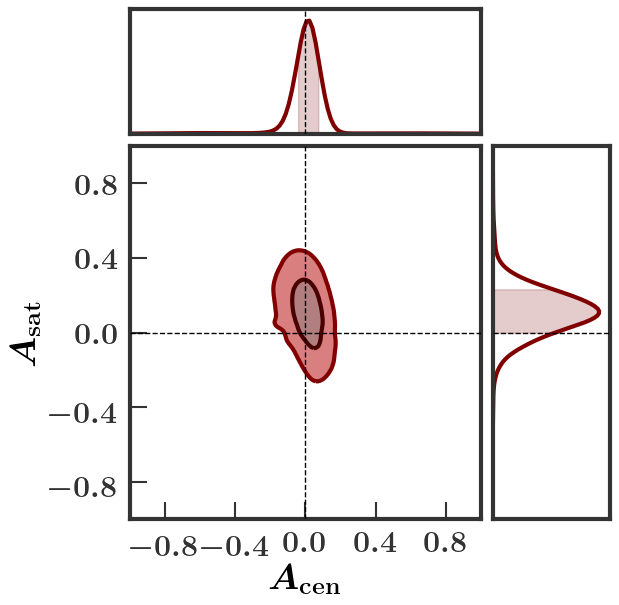}%
\includegraphics[width=.25\linewidth]{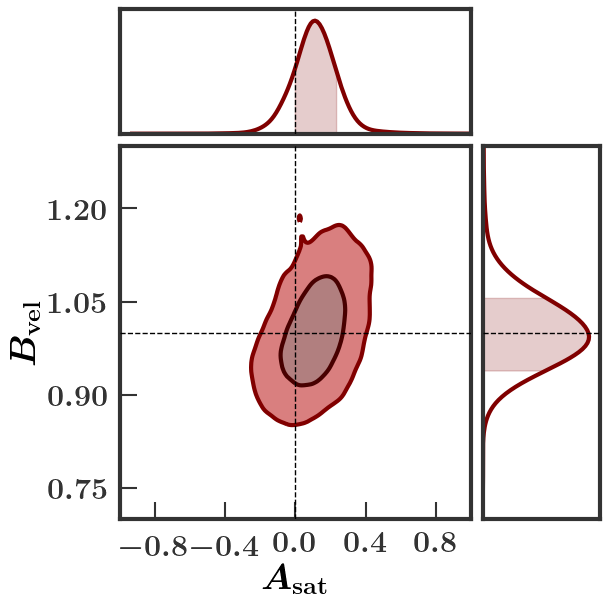}
\caption{Parameter constraints for the SDSS $-21$ sample, from a model with assembly bias (with environment as the secondary halo property) and satellite velocity bias, using the ``dHOD" optimal observables (listed in Table~\ref{tab:order}). The crosshairs in the third and fourth panels indicate no assembly bias and no velocity bias ($A_\mathrm{cen} = A_\mathrm{sat} = B_\mathrm{vel} =  0$).}
\label{fig:sdss21env_vbias}
\end{figure*}

The fact that adding velocity bias to our model reduces the strength of the assembly bias signature is strong evidence in favor of having a sufficiently flexible HOD model, without which we cannot claim to have made a robust detection of assembly bias, nor can we hope to reliably constrain cosmology.
It is likely that our analysis is sensitive to \Bvel because of our use of clustering measurements that are particularly affected by galaxy velocities (e.g., \zxi). 
It is possible that an analysis that does not include these statistics would have less constraining power for \Bvel and therefore might not affirm the presence of velocity bias.

After including \Bvel as well as \Acen and \Asat in our HOD model, the best-fit result for the $-19$ sample exhibits a substantial decrease in tension, down from $2.0\sigma$ to $1.4\sigma$.
The $\chi^2/\mathrm{d.o.f}$ also decreased, from 1.53 to 1.27. 
Thus, when we include both assembly bias and velocity bias in our model, our clustering results are in agreement with SDSS.
Looking at the top panel of Figure~\ref{fig:con_vbias_residuals} and comparing it to Figure~\ref{fig:con_residuals}, we can see that this relief in tension comes from the improvement in observables across the board, but particularly in \ngal, \wprp\footnote{We note that the improvement in \ngal and \wprp comes not from their relationship with satellite velocity, but rather from the changes in \textit{other} HOD parameters as a result of including velocity bias in the model.}, \zxi, and \gmf.

For the $-21$ sample, our best-fit results for this model yield minimal satellite galaxy velocity bias (\Bvel = 0.976), as well as minimal central and satellite galaxy assembly bias (\Acen = 0.144 and \Asat = -0.198).
For this sample, the constraints on \Acen and \Asat do not significantly degrade after adding velocity bias to the model, and the constraints on \Bvel comparable to what they are in the $-19$ sample.
In spite of this, we cannot claim a significant detection of assembly or satellite galaxy velocity bias in the SDSS $-21$ sample.
Once again, the constraints on the standard HOD parameters remain roughly the same, suggesting that the further addition of velocity bias to the model has negligible impact.
Additionally, we are still unable to relieve the tension present in the $-21$ sample even after adding this new flexibility to the HOD model.

\begin{figure*}
\centering
\includegraphics[width=6in]{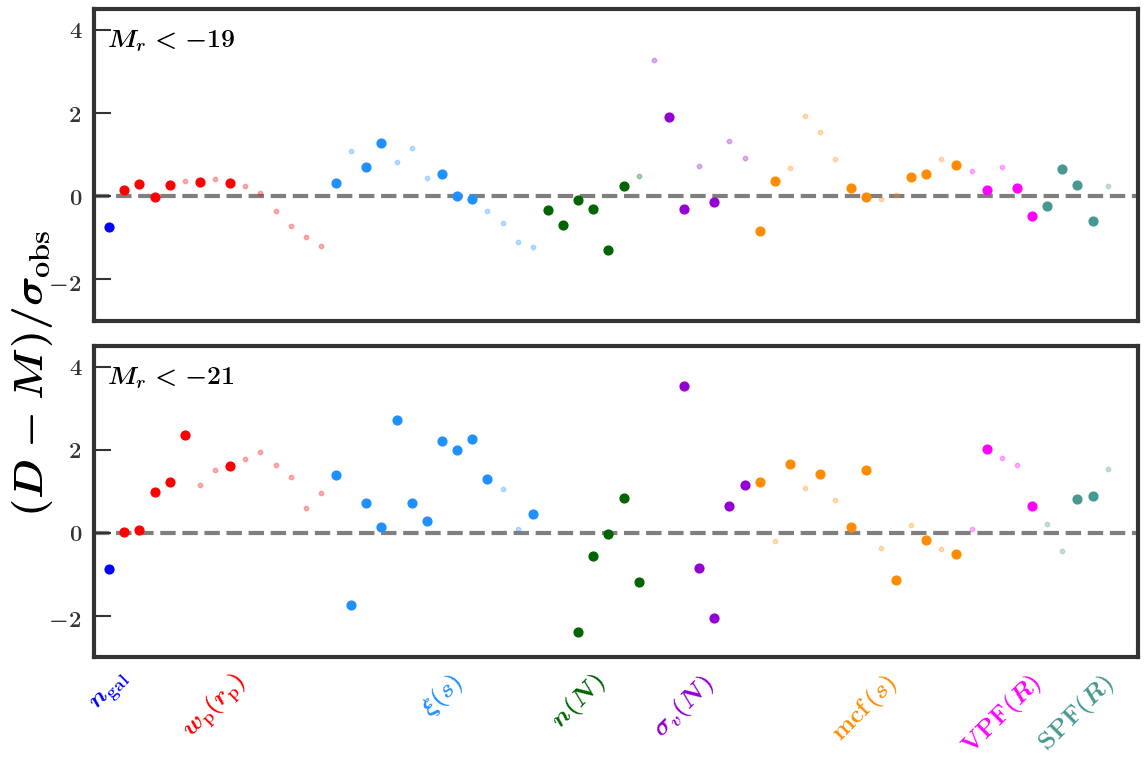}%
\caption{Residuals between the best-fit model and the SDSS measurements for the $-19$ (top) and $-21$ (bottom) samples. For each sample, the model includes assembly bias (with environment as the secondary halo property) as well as velocity bias. We show residuals for all observables, but the model was constrained using the ``dHOD" optimal observables for each sample (listed in Table~\ref{tab:order}), which are displayed with larger points.}
\label{fig:env_vbias_residuals}
\end{figure*}

Looking at the lower panel of Figure~\ref{fig:con_vbias_residuals}, we see very little improvement in our residuals compared to Figure~\ref{fig:con_residuals}, which illustrates why the addition of velocity bias to the model results in no relief in tension for this sample.
It is particularly noteworthy that even the statistics that contain velocity information (like \zxi and \sigN) do not show any substantial improvement after adding velocity bias to the model.
While it is possible that a different velocity bias prescription could lead to more improvement, it is also possible that these dynamical clustering measurements are sensitive to an issue that exists not within our HOD model but within our cosmological model. 

\subsection{Environment-based Assembly Bias}
\label{subsec:environment}
The previous section shows that central galaxy occupation is, at most, only loosely tied to the concentration of the host halo.
We next investigate whether modeling assembly bias with a different halo property can improve the goodness of fit of our model.
Given that we do not know which halo property other than mass most strongly affects the presence of a central galaxy in a given halo, we choose to use local halo environment as our new assembly bias property.
This allows us to investigate the general assumption that central galaxy occupation is tied to some halo property that is \textit{correlated} with the local environment.
For this reason, environment is a useful property for probing assembly bias and gives our model the flexibility that it needs to model galaxy clustering without knowing the ``true" halo property that leads to assembly bias.
We define local environment as the total mass in \textit{halos} within a 5 \hmpc radius.

Once again, we do not identify a new set of optimal observables for constraining this environment-based assembly bias model, but rather run chains for each of our SDSS samples using the same optimal observables listed in Table~\ref{tab:order} (``dHOD").
The results for the $-19$ sample are shown in Figure~\ref{fig:sdss19env_vbias}, while the results for the $-21$ sample are shown in Figure~\ref{fig:sdss21env_vbias}.
The best-fit parameter values and constraints are listed in Tables~\ref{tab:bestfit} and \ref{tab:constraints}.

For the $-19$ sample, our best-fit results once again indicate positive central galaxy assembly bias (\Acen = 0.533), and negative satellite galaxy assembly bias (\Asat = -0.224).
In other words, low-luminosity central galaxies preferentially reside in halos with denser environments, while satellite galaxies preferentially reside in halos with less dense environments, at fixed mass. We can rule out a model with no central galaxy assembly bias at the 99\% confidence level and a model with no satellite galaxy assembly bias at the 95\% confidence level.
We again find that satellite galaxies have velocities that are slightly slower than the dark matter distribution (\Bvel = 0.826). We can rule out a model with no satellite velocity bias at the 99.9\% confidence level.
It is noteworthy that when environment is used to model assembly bias instead of concentration, \Bvel and \Asat are correlated rather than anti-correlated.

Additionally, the constraints on \Acen are much improved compared to the constraints when using concentration, and the constraints on \Asat are slightly improved.
This improvement in constraints is seen in spite of the fact that the observables used were chosen to optimally constrain a \textit{concentration-based} assembly bias model. 
We attribute this improvement to the fact that environment is more directly associated with clustering than concentration.

\begin{deluxetable*}{ccccccccccccc}
\tablenum{5}
\tablecaption{SDSS best-fit results for different halo models\label{tab:bestfit}}
\tablewidth{0pt}
\tablehead{
\colhead{$M_r^\mathrm{lim}$} & \colhead{Model} & \colhead{Obs.} & \colhead{\logmmin} & \colhead{\siglogm} & \colhead{\logmzero} & \colhead{\logmone} & \colhead{$\alpha$} & \colhead{$A_\mathrm{cen}$} & \colhead{$A_\mathrm{sat}$} & \colhead{$B_\mathrm{vel}$} & \colhead{p-value} & \colhead{AIC}
}
\startdata
$-19$ & Standard HOD & S22 & 11.445 & 0.099 & 11.651 & 12.703 & 0.958 & -- & -- & -- & $6.8\cdot10^{-6}$ & 87.77 \\
& ABcon & BM22 & 11.455 & 0.141 & 11.757 & 12.685 & 0.925 & 0.793 & -0.368 & -- & 0.047 & 56.83 \\
& ABcon + VB & BM22 & 11.474 & 0.132 & 11.877 & 12.715 & 0.950 & 0.825 & -0.251 & 0.898 & 0.155 & 51.54 \\
& ABenv + VB & BM22 & 11.490 & 0.125 & 11.855 & 12.783 & 0.985 & 0.533 & -0.224 & 0.826 & 0.364 & 45.99 \\
\hline
$-21$ & Standard HOD & S22 & 12.728 & 0.467 & 9.015 & 13.929 & 1.112 & -- & -- & -- & $3.5\cdot10^{-5}$ &  \\
& ABcon & BM22 & 12.774 & 0.554 & 9.447 & 13.926 & 1.067 & -0.090 & -0.240 & -- & $2.6\cdot10^{-6}$ & 99.38 \\
& ABcon + VB & BM22 & 12.756 & 0.525 & 9.804 & 13.915 & 1.108 & 0.144 & -0.198 & 0.976 & $1.8\cdot10^{-6}$ & 100.96 \\
& ABenv + VB & BM22 & 12.740 & 0.495 & 9.984 & 13.917 & 1.079 & -0.025 & 0.165 & 1.011 & $4.1\cdot10^{-6}$ & 98.43
\enddata
\tablecomments{Best-fit HOD parameters for each SDSS sample using four different models: the standard 5-parameter model, a model with concentration-based assembly bias (``ABcon"), a model with concentration-based assembly bias plus satellite velocity bias (``ABcon + VB"), and a model with environment-based assembly bias plus satellite velocity bias (``ABenv + VB"). The Standard HOD results are taken from S22 and thus use the S22 observables, while the chains using extended HOD models use the optimal observables identified in this work (listed in Table~\ref{tab:order}). We indicate the goodness-of-fit of each parameter combination with a p-value, as well as assess the success of the model using the AIC.}
\end{deluxetable*}

Ultimately, this model is an even better fit to the data than the model with concentration-based assembly bias plus velocity bias ($0.9\sigma$ compared to $2.0\sigma$), and we can rule out a model with zero assembly bias and zero velocity bias.
Looking at the top panel of Figure~\ref{fig:env_vbias_residuals} and comparing it to Figure~\ref{fig:con_vbias_residuals}, we can see that switching the assembly bias property from concentration to environment leads to improvement in almost every observable that we measure, which explains the reduction in tension.
In particular, \ngal, \zxi, \mcf, \vpf, and \spf see sizeable improvement. 

For the $-21$ sample, our best-fit results indicate negligible central galaxy assembly bias (\Acen = -0.025) and minimal satellite galaxy assembly bias (\Asat = 0.165), as well as negligible velocity bias (\Bvel = 1.011). A model with with no assembly bias and no velocity bias is entirely consistent with the data.
This means that for high-luminosity galaxies, neither central nor satellite galaxies show any meaningful preference toward local halo environment, and satellite galaxies move with velocities similar to the dark matter within the halo.

Like in the $-19$ sample, the constraints on \Acen and \Asat are substantially improved for the $-21$ sample when using environment as the secondary halo property as opposed to concentration.
However, the constraints on \mmin and \siglogm are actually degraded when environment-based assembly bias is used.
Once again we see no improvement in tension between our model and SDSS ($4.6\sigma$).
Looking at the lower panel of Figure~\ref{fig:env_vbias_residuals}, we see little to no improvement in our residuals compared to Figure~\ref{fig:con_vbias_residuals}, which illustrates why switching the assembly bias parameter from concentration to environment fails to reduce the tension for this sample.

These results demonstrate that low-luminosity galaxies exhibit an assembly bias signature that is present in some capacity regardless of the secondary halo property used, although the exact strength of the central and satellite assembly bias may differ for different secondary properties.
Meanwhile, high-luminosity galaxies do not display any assembly bias for either secondary halo property.
Moreover, the tension found between our model and SDSS is not easily alleviated with a change in secondary halo property.
Investigating many different secondary halo properties for modeling assembly bias is beyond the scope of this paper; however, given our results using concentration and environment, we do not anticipate that some other secondary halo property would alleviate all of the tension that we find in the $-21$ sample.

\begin{deluxetable*}{cccccccccc}
\tablenum{6}
\tablecaption{SDSS Constraints\label{tab:constraints}}
\tablewidth{0pt}
\tablehead{
\colhead{$M_r^\mathrm{lim}$} & \colhead{Model} & \colhead{\logmmin} & \colhead{\siglogm} & \colhead{\logmzero} & \colhead{\logmone} & \colhead{$\alpha$} & \colhead{$A_\mathrm{cen}$} & \colhead{$A_\mathrm{sat}$} & \colhead{$B_\mathrm{vel}$}
}
\startdata
$-19$ & sHOD & $11.442^{+0.016}_{-0.015}$ & $0.106^{+0.074}_{-0.065}$ & $11.674^{+0.089}_{-0.094}$ & $12.691^{+0.028}_{-0.029}$ & $0.954^{+0.019}_{-0.019}$ & -- & -- & -- \\
& ABcon & $11.469^{+0.019}_{-0.017}$ & $0.159^{+0.074}_{-0.077}$ & $11.750^{+0.093}_{-0.095}$ &  $12.685^{+0.029}_{-0.031}$ & $0.930^{+0.025}_{-0.028}$ & $0.673^{+0.245}_{-0.529}$ & $-0.361^{+0.107}_{-0.103}$ & -- \\
& ABcon+VB & $11.484^{+0.022}_{-0.021}$ & $0.121^{+0.072}_{-0.067}$ & $11.864^{+0.100}_{-0.099}$ &  $12.724^{+0.037}_{-0.037}$ & $0.947^{+0.024}_{-0.028}$ & $0.338^{+0.472}_{-0.720}$ & $-0.194^{+0.145}_{-0.134}$ & $0.876^{+0.042}_{-0.041}$ \\
& ABenv+VB & $11.481^{+0.021}_{-0.021}$ & $0.132^{+0.073}_{-0.052}$ & $11.786^{+0.119}_{-0.140}$ & $12.776^{+0.041}_{-0.040}$ & $0.986^{+0.027}_{-0.027}$ & $0.444^{+0.265}_{-0.190}$ & $-0.173^{+0.108}_{-0.115}$ & $0.857^{+0.041}_{-0.042}$ \\
\hline
$-21$ & sHOD & $12.748^{+0.015}_{-0.015}$ & $0.517^{+0.029}_{-0.029}$ & $9.015^{+2.017}_{-2.036}$ & $13.919^{+0.014}_{-0.014}$ & $1.088^{+0.031}_{-0.033}$ & -- & -- & -- \\
& ABcon & $12.737^{+0.019}_{-0.020}$ & $0.494^{+0.038}_{-0.040}$ & $8.980^{+2.019}_{-2.013}$ & $13.914^{+0.015}_{-0.015}$ & $1.110^{+0.035}_{-0.039}$ & $-0.236^{+0.290}_{-0.297}$ & $-0.148^{+0.181}_{-0.156}$ & -- \\
& ABcon+VB & $12.747^{+0.020}_{-0.020}$ & $0.505^{+0.038}_{-0.040}$ & $9.525^{+1.676}_{-1.879}$ & $13.923^{+0.018}_{-0.017}$ & $1.111^{+0.043}_{-0.050}$ & $0.154^{+0.305}_{-0.283}$ & $-0.101^{+0.215}_{-0.208}$ & $0.971^{+0.047}_{-0.041}$ \\
& ABenv+VB & $12.773^{+0.039}_{-0.043}$ & $0.546^{+0.067}_{-0.081}$ & $9.984^{+1.161}_{-1.456}$ & $13.925^{+0.022}_{-0.022}$ & $1.059^{+0.052}_{-0.062}$ & $0.014^{+0.038}_{-0.045}$ & $0.115^{+0.108}_{-0.118}$ & $1.000^{+0.050}_{-0.044}$
\enddata
\tablecomments{Marginalized constraints on SDSS for both samples using four different models: the standard HOD model from S22 (using the optimal observables from S22), an HOD model with concentration-based assembly bias, a model with concentration-based assembly bias plus satellite velocity bias, and a model with environment-based assembly bias plus satellite velocity bias, using the optimal observables identified in this work. We present the median parameter values along with upper and lower limits corresponding to the 84 and 16 percentiles respectively. All of these chains were run using the optimal observables identified in this work  (listed in Table~\ref{tab:order}).}
\end{deluxetable*}

\subsection{Baryonic Effects}
\label{subsec:hmf_corrections}
In this section, we present the results from applying the halo mass corrections from \citet{Beltz-Mohrmann2021} to our halo catalogs and then repeating our analysis using an HOD model with both assembly bias and velocity bias.
Specifically, we utilize the mass corrections for $M_\mathrm{vir}$ halos at $z=0$ according to the IllustrisTNG and EAGLE simulations, as well as the environment-dependent mass correction from IllustrisTNG.
(The EAGLE mass correction shows very little environmental dependence, so we do not employ it here.)
The best-fit model parameters for these analyses are listed in Table~\ref{tab:masscorrections}.
The constraints on the model parameters remain similar in size after the different mass corrections, and thus are not listed separately; we refer the reader to Table~\ref{tab:constraints} for the constraints on the assembly bias + velocity bias (ABe+VB) model.

\begin{figure*}
\centering
\includegraphics[width=.25\linewidth]{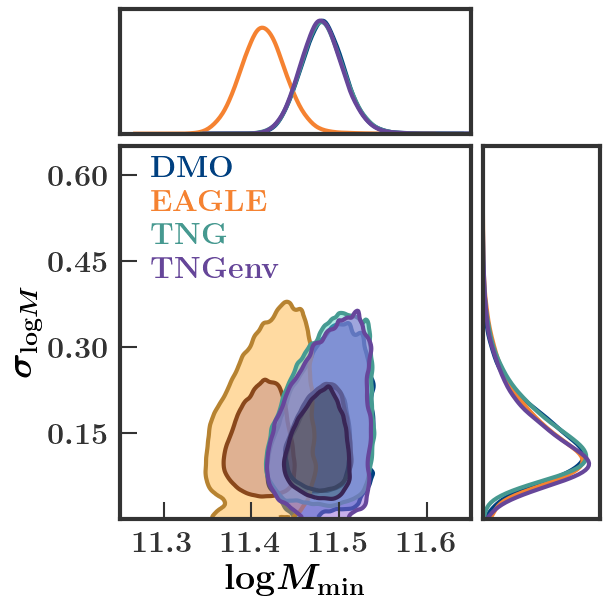}%
\includegraphics[width=.25\linewidth]{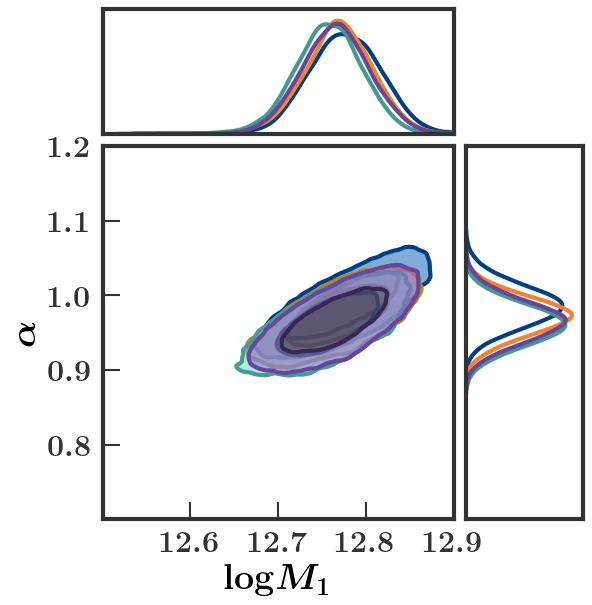}%
\includegraphics[width=.25\linewidth]{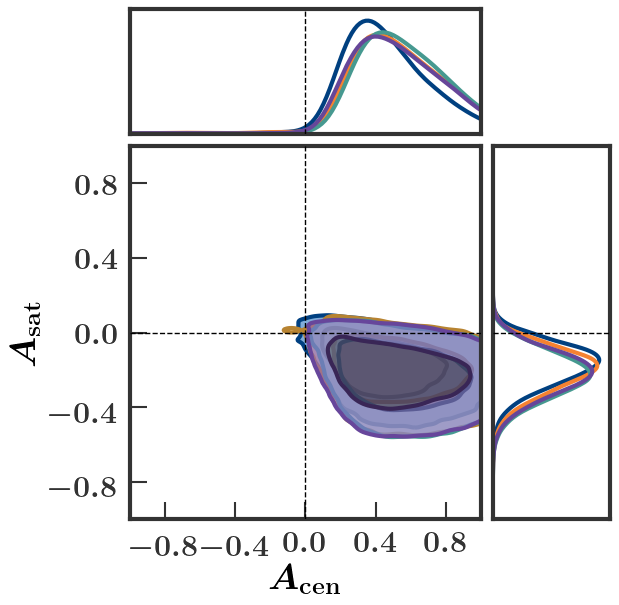}%
\includegraphics[width=.25\linewidth]{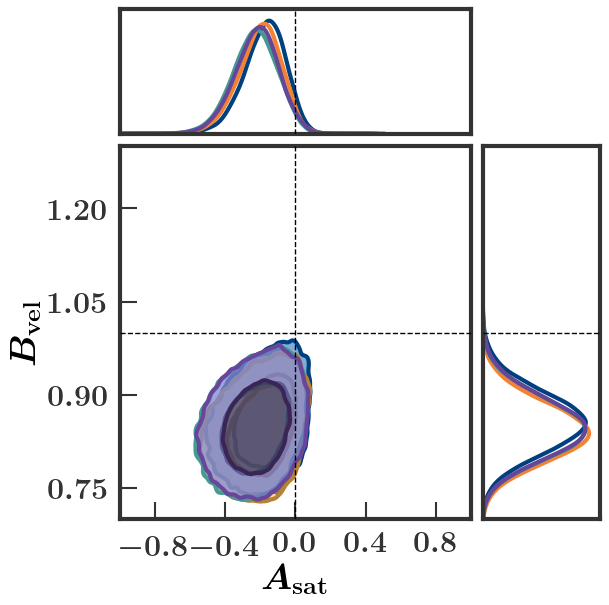}
\caption{HOD parameter constraints for the SDSS $-19$ sample, from a model with assembly bias (with environment as the secondary halo property) and velocity bias, after applying three different mass corrections. The crosshairs in the third and fourth panels indicate no assembly bias and no velocity bias ($A_\mathrm{cen} = A_\mathrm{sat} = B_\mathrm{vel} =  0$).}
\label{fig:sdss19_halomass}
\end{figure*}

\begin{figure*}
\centering
\includegraphics[width=.25\linewidth]{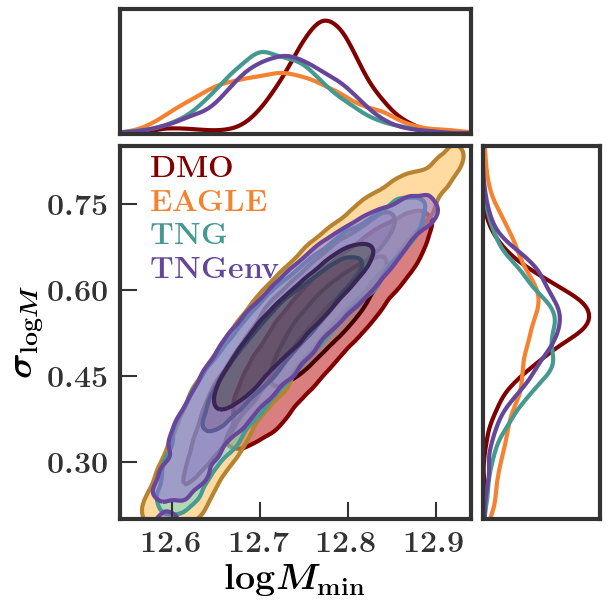}%
\includegraphics[width=.25\linewidth]{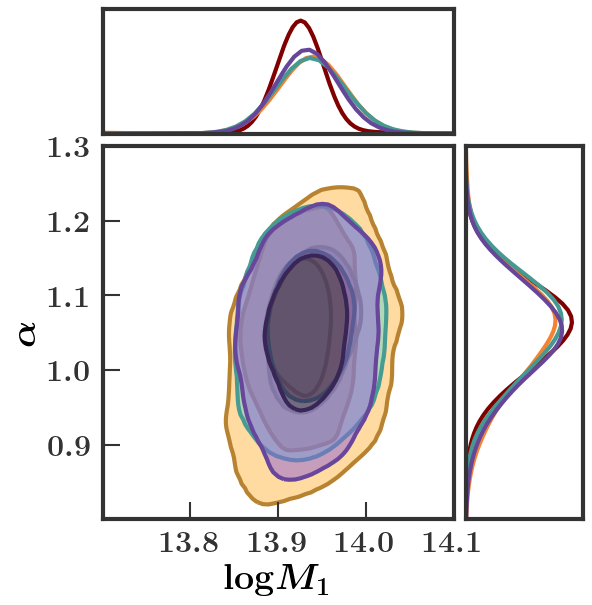}%
\includegraphics[width=.25\linewidth]{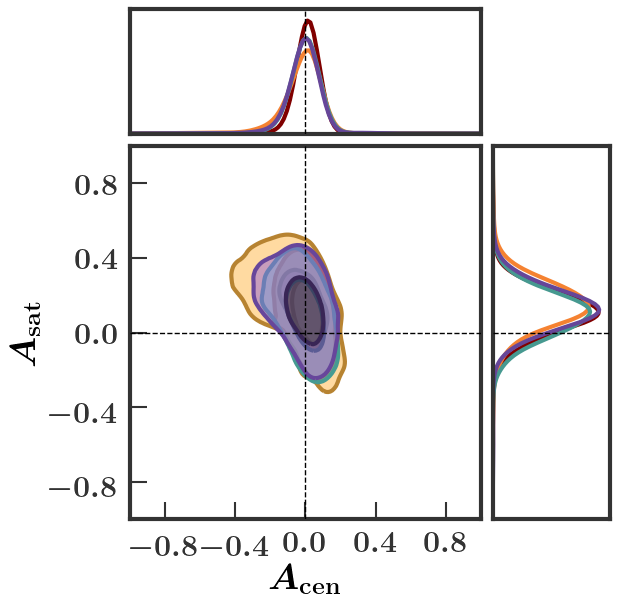}%
\includegraphics[width=.25\linewidth]{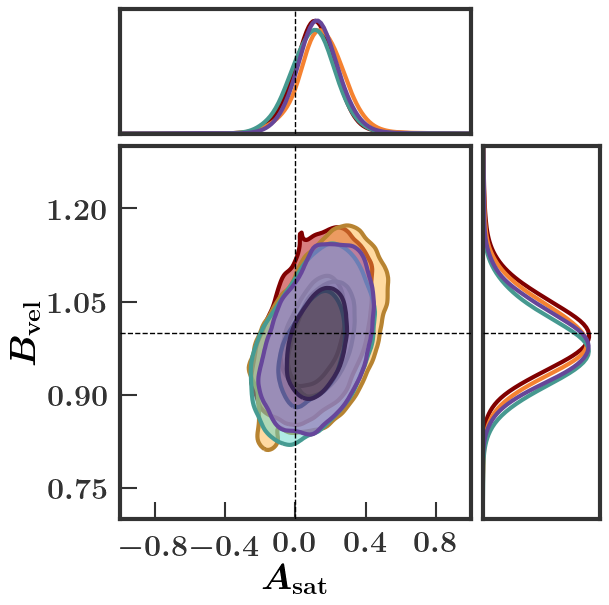}
\caption{HOD parameter constraints for the SDSS $-21$ sample, from a model with assembly bias (with environment as the secondary halo property) and velocity bias, after applying three different mass corrections. The crosshairs in the third and fourth panels indicate no assembly bias and no velocity bias ($A_\mathrm{cen} = A_\mathrm{sat} = B_\mathrm{vel} =  0$).}
\label{fig:sdss21_halomass}
\end{figure*}

In Figure~\ref{fig:sdss19_halomass}, we show the results of applying these halo mass corrections to the $-19$ halo catalogs and performing our analysis.
The panels show the model parameters, using the same layout as in Figure~\ref{fig:sdss19con_vbias}.
The original results (i.e., with no mass correction) are depicted in blue.
The results from the EAGLE mass correction are shown in yellow, the results from the TNG mass correction are shown in green, and the results from the environment-dependent TNG mass correction (``TNG,env") are shown in purple.
In Figure~\ref{fig:sdss21_halomass}, we show the results of applying these halo mass corrections to the $-21$ sample.
The original results (i.e., with no mass correction) are depicted in red, and the mass-corrected results are shown with the same colors as in Figure~\ref{fig:sdss19_halomass}.

For both samples, the mass corrections produce minimal changes to our HOD parameter constraints (with the exception of \logmmin, which does experience significant shifts in each sample).
Additionally, our conclusions about the presence of assembly bias and velocity bias remain the same for both samples after the mass corrections: the $-19$ samples exhibits significant postive central assembly bias and negative satellite assembly bias, as well as significant velocity bias, while the $-21$ sample exhibits no such biases.
Furthermore, the goodness-of-fit of the model remains roughly the same after each of the mass corrections: an HOD model with environment-based assembly bias and satellite velocity bias produces good agreement with the clustering of low-luminosity SDSS galaxies, while the same model yields significant tension with the clustering of high-luminosity SDSS galaxies.
None of the mass corrections are able to alleviate this tension.

It is unsurprising that the mass corrections mainly affect the best-fit value of \logmmin in each sample, have a slight impact on the other standard HOD parameters, and have a negligible affect on the assembly bias and velocity bias parameters. 
This is because the mass corrections shift the masses of our halos (albeit in a non-trivial way), and so the parameter that governs the minimum halo mass that can host a galaxy shifts to compensate.
To a lesser extent, the parameter that governs the scatter in this minimum halo mass (\siglogm), and the parameters that determine the number of satellite galaxies in a halo of a given mass (\logmone and $\alpha$) also shift to compensate for the halo mass correction.
Meanwhile, the parameters that govern the dependence of halo occupation on a halo property \textit{other} than mass (\Acen and \Asat) and the parameter that governs the relative velocities of the satellite galaxies to the dark matter (\Bvel) are unaffected by changes to the halo mass function.

Overall, it is difficult to distinguish between any of the mass corrections based on their agreement with the clustering of SDSS, nor would we rule out any of these models of baryonic physics based on our results. 
It is possible that with different clustering statistics, we could tighten some of our constraints and thus differentiate between the results of the different mass corrections.
It is also possible that with a better fitting model for the $-21$ sample, the effect of different mass corrections on the overall tension could be seen.
Ideally, we would be able to vary HOD parameters, cosmology parameters, and mass correction prescriptions simultaneously and use our results to rule out certain baryonic physics models; however, this is a challenge left for future work.

\begin{deluxetable*}{cccccccccccc}
\tablenum{7}
\tablecaption{SDSS best-fit results with different mass corrections\label{tab:masscorrections}}
\tablewidth{0pt}
\tablehead{
\colhead{$M_r^\mathrm{lim}$} & \colhead{Model} & \colhead{Mass Correction} & \colhead{\logmmin} & \colhead{\siglogm} & \colhead{\logmzero} & \colhead{\logmone} & \colhead{$\alpha$} & \colhead{$A_\mathrm{cen}$} & \colhead{$A_\mathrm{sat}$} & \colhead{$B_\mathrm{vel}$} & \colhead{p-value}
}
\startdata
$-19$ & ABenv + VB & -- & 11.490 & 0.125 & 11.855 & 12.783 & 0.985 & 0.533 & -0.224 & 0.826 & 0.364 \\
& ABenv + VB & TNG & 11.478 & 0.140 & 11.780 & 12.760 & 0.963 & 0.606 & -0.286 & 0.844 & 0.250 \\
& ABenv + VB & EAGLE & 11.412 & 0.146 & 11.758 & 12.766 & 0.963 & 0.575 & -0.273 & 0.825 & 0.311 \\
& ABenv + VB & TNG, env. & 11.481 & 0.122 & 11.777 & 12.777 & 0.971 & 0.506 & -0.228 & 0.816 & 0.301 \\
\hline
$-21$ &  ABenv + VB & -- & 12.740 & 0.495 & 9.984 & 13.917 & 1.079 & -0.025 & 0.165 & 1.011 & $4.1\cdot10^{-6}$ \\
& ABenv + VB & TNG & 12.671 & 0.455 & 9.258 & 13.924 & 1.122 & -0.035 & 0.145 & 1.001 & $1.1\cdot10^{-6}$ \\
& ABenv + VB & EAGLE & 12.720 & 0.539 & 11.028 & 13.925 & 1.059 & -0.026 & 0.177 & 0.970 & $1.5\cdot10^{-6}$ \\
& ABenv + VB & TNG, env. & 12.745 & 0.575 & 8.594 & 13.914 & 1.053 & 0.023 & 0.074 & 0.998 & $8.5\cdot10^{-7}$
\enddata
\tablecomments{Best-fit HOD parameters for each SDSS sample using different HOD models as well as different mass corrections from \citet{Beltz-Mohrmann2021}. The model includes concentration-based assembly bias plus velocity bias. All of these chains were run using the optimal observables identified in this work  (listed in Table~\ref{tab:order}). We list the best-fit values of each parameter and indicate the goodness of fit of each parameter combination with a p-value.}
\end{deluxetable*}

\section{Conclusions} \label{sec:conclusions}
In this work we have explored several extensions to the standard HOD model and employed an optimal set of galaxy clustering measurements to constrain this model for both high- and low-luminosity galaxies in SDSS.
We first extended the standard HOD model to include parameters for central and satellite galaxy assembly bias, using halo concentration as the secondary halo property for implementing this assembly bias.
We identified a set of observables to best constrain this model, using the algorithm laid out in \citet{Szewciw2022}.
We then further extended our model to include an additional parameter for satellite galaxy velocity bias and repeated our analysis for both SDSS samples using this new model with both concentration-based assembly bias \textit{and} satellite velocity bias.
We then repeated this analysis using local halo environment as the assembly bias property instead of concentration.
Lastly, we applied three different halo mass corrections to our dark matter halos to account for the impact of baryonic physics on the halo mass function; we repeated our analysis by applying our extended halo model (with environment-based assembly bias \textit{and} satellite velocity bias) to these corrected halo masses.
This is the first time that an extended HOD modeling framework, with assembly bias \textit{and} velocity bias, a prescription to account for the impact of baryonic physics on the halo mass function, and a variety of galaxy clustering statistics measured on a wide range of scales, has been used to constrain the galaxy-halo connection in the SDSS $-19$ and $-21$ samples. Our conclusions are listed below:

\begin{itemize}[noitemsep]
\item Low-luminosity galaxies in SDSS exhibit both central and satellite galaxy assembly bias when fit with an HOD model that includes concentration-based assembly bias, with satellite galaxies displaying a negative dependence of occupation on concentration and central galaxies displaying a positive dependence on concentration (although central galaxy assembly bias is difficult to constrain). With this model, we find evidence for satellite assembly bias at the 99.8\% confidence level. Additionally, this model is a substantially better fit to the clustering of low-luminosity galaxies than the standard HOD model (i.e., a model with no assembly bias).
\item When fitting the clustering of low-luminosity galaxies with an HOD model that includes both concentration-based assembly bias \textit{and} satellite galaxy velocity bias, we find evidence for satellite velocity bias at the 99.8\% confidence level, with satellite galaxies moving $\sim10-15\%$ slower than the dark matter. The assembly bias is quite unconstrained, making it difficult to rule out a model with zero assembly bias. However, this model does further reduce the tension with SDSS.
\item When fit with an HOD model that instead uses environment-based assembly bias, low-luminosity galaxies exhibit significant negative satellite assembly bias and significant positive central assembly bias. Using environment also helps to tighten the constraints on the assembly bias parameters. This model also results in significant satellite velocity bias. We find evidence for satellite assembly bias, central assembly bias, and satellite velocity bias at the 95\%, 99\%, and 99.9\% confidence levels, respectively. This model ultimately results in the tightest constraints on assembly bias and velocity bias, as well as the best agreement with SDSS, with essentially no remaining tension.
\item[$\blacklozenge$] High-luminosity galaxies exhibit negligible assembly bias when using either concentration or local environment as the assembly bias property (although the constraints are once again tighter using environment.) They also exhibit negligible satellite velocity bias when fit with a model that includes both assembly bias and velocity bias. Additionally, none of these models yield good agreement with SDSS ($>4\sigma$ tension).
\item[$\bigstar$] While each different treatment of baryonic physics leads to a slight change in best-fit HOD parameters, none of them significantly change our conclusions about the presence of assembly bias and velocity bias in each sample, nor do they change the goodness-of-fit of the HOD model used. Thus, we cannot draw any conclusions on the accuracy of our baryonic physics models based on this analysis, nor can we use baryonic physics to explain the tension we find in the $-21$ sample.
\end{itemize}

For low-luminosity galaxies, our results using either concentration or environment are consistent with recent results from semi-analytic models and hydrodynamic simulations \citep[e.g.,][]{Artale2018, Zehavi2018,Bose2019}.
Additionally, the presence of assembly bias and velocity bias among low-luminosity galaxies but not among high-luminosity galaxies is consistent with recent findings from semi-analytic models and hydrodynamic simulations \citep[e.g.,][]{Contreras2019, Contreras2021, Contreras2023, Beltz-Mohrmann2020}.

Our findings are also consistent with several recent observational studies.
For example, \citet{Zentner2019} used concentration to model assembly bias in SDSS and found evidence for satellite assembly bias among faint galaxies ($M_r < -19$) but found no evidence for assembly bias in the $M_r < -21$ sample.
Similarly, \citet{Vakili2019} used concentration to model assembly bias in SDSS and detected moderate central assembly bias among faint galaxies ($M_r < -20.5, -20, -19.5$) but did not detect central galaxy assembly bias among bright galaxies ($M_r < -21.5, -21$).
Meanwhile, \citet{Salcedo2022} instead used environment to model assembly bias in SDSS and similarly found no evidence for assembly bias among bright galaxies.
\citet{Wang2022} also used concentration to model assembly bias in SDSS, and detected positive central assembly bias for faint galaxies ($M_r < -20.5, -20, -19.5, -19$), and marginal negative satellite galaxy assembly bias in the $M_r < -20$ and $M_r < -19$ samples, but did not detect assembly bias in the $M_r < -21$ sample. 

The assembly bias signature among low-luminosity galaxies can be understood as follows:
Early forming halos ultimately contain fewer satellites, because they acquired their satellites earlier, and thus these satellites were subject to the destructive processes of the host halo (i.e., merging) for a longer period of time \citep{Zentner2005}.
Thus, it is reasonable that satellite galaxies would preferentially reside in late-forming halos.
Formation time is strongly correlated with halo concentration, with early forming halos having higher concentrations \citep[e.g.,][]{Wechsler2002, Zhao2003}, and thus fewer satellite galaxies.
This also explains why satellite galaxies preferentially reside in halos with low-density environments: satellites residing in high-density environments are more vulnerable to mergers, and thus host halos in high-density environments will ultimately contain fewer satellites.

Meanwhile, among Milky Way-sized halos, galaxies residing in higher concentration halos tend to be more luminous \citep{Zentner2019}. 
This is because at fixed mass, halos with higher concentrations have deeper potential wells, allowing gravity to more strongly bind the stellar and gas contents of these halos, possibly leading to more rapid star formation (or less vulnerability to processes that suppress star formation).
Thus, we can understand why central galaxies residing in Milky Way-sized halos seem to preferentially reside in high concentration halos: the deep potential well of the host halo ultimately leads to a more luminous central galaxy, which is more likely to pass our luminosity cut than a central galaxy living in a shallow potential well.
The same logic can be used to understand why central galaxies prefer to reside in halos with high-density environments - such environments are more conducive to merging, leading to a more luminous central galaxy in the end.

The lack of assembly bias signature among high-luminosity galaxies has a few possible explanations.
It has been found that among Milky Way-sized halos, the number of subhalos in a given host halo depends heavily on host halo environment, but among cluster-mass halos, the abundance of subhalos exhibits no such environmental-dependence \citep[e.g.,][]{Zentner2019}.
This explains why when using halo environment as our secondary halo property, we detect satellite assembly bias among low-luminosity galaxies but not among high-luminosity galaxies.
Another possible explanation is that satellite galaxies in the -21 sample are all recent additions to the halo, and so they have not had enough time to be destroyed via mergers; thus, whether the host halo is high- or low-concentration (or lives in a high- or low-density environment) makes no difference for the presence of satellite galaxies in this sample. 
This could also explain why we detect satellite galaxy velocity bias among low-luminosity galaxies but not among high-luminosity galaxies: satellite galaxies in the -19 sample have likely been slowed down via dynamical friction over time, whereas satellite galaxies in the -21 sample are all more recent acquisitions to the halo and thus have not had time to be significantly affected by dynamical friction.

While several physical explanations are reasonable, the fact remains that among low-luminosity galaxies the model that includes both assembly bias and satellite velocity bias exhibits minimal tension; in other words, the model is in good agreement with the clustering of $-19$ SDSS galaxies. 
This is consistent with our expectation for low-luminosity galaxies, and with the minimal tension found in previous studies of this nature for low-luminosity galaxies.
By contrast, the model is \textit{not} in good agreement with the clustering of $-21$ SDSS galaxies. 
This high degree of tension for high-luminosity galaxies is in contrast with these previous studies, which did not find any significant tension with SDSS. The larger constraining power of our results is likely, in part, due to the large set of optimal clustering statistics that we use.

This tension that we find among the higher luminosity sample could be indicative of several things.
For example, it is possible that central galaxies do indeed exhibit significant velocity bias \citep[e.g.,][]{Guo2015,Guo2015b}, and including this in the model would lead to better agreement with SDSS. However, we have examined the impact on our observables after adding central velocity bias to our best-fit model at the level found in previous works, and found that this has a negligible effect on our clustering measurements. Thus, if central velocity bias is indeed present in SDSS, it is not currently detectable with our clustering measurements given our uncertainty due to cosmic variance.
It is also possible that galaxies do not trace the spatial distribution of dark matter within halos \citep[i.e., there is spatial bias][]{Watson2012, Piscionere2015}.
Additionally, the standard HOD model assumes that the number of satellite galaxies in each halo is governed by a Poisson distribution, but recent results indicate that this is probably not the case \citep{Boylan-Kolchin2010, Mao2015, Jimenez2019}.

It is also possible that a change in halo definition or halo finder could alter our results. S18 repeated their analysis twice, once using $M_{200b}$ halos and again using $M_{vir}$ halos, and found similar results, with $M_{vir}$ halos producing slightly tighter constraints. For this reason we have proceeded using only $M_{vir}$ halos. Because a small change in halo definition simply leads to a small change in mass for all halos, we think it likely that the HOD parameters can compensate for any small change in halo definition; in this case, our best-fit parameter values would change slightly, but our overall conclusions about the presence of assembly bias or the goodness-of-fit of our model would not change. However, a significant change in halo definition or halo finder could potentially lead to changes in our conclusions.
For example, several works have found that the proper treatment of splashback halos could lead to a reduction in the assembly bias signature for low-luminosity galaxies \citep{Villarreal2017, Mansfield2020}. 

In future work, it is worth investigating whether accounting for these possibilities leads to improved agreement between our model and the observed clustering of high-luminosity galaxies.
However, we think it unlikely that accounting for these affects would be enough to explain the amount of tension we are finding.
In fact, in hydrodynamic simulations, the standard HOD model proved to be a good fit to the clustering of high-luminosity galaxies, provided that the model was applied to a DMO simulation with the same cosmological model as the hydrodynamic simulation in question \citep{Beltz-Mohrmann2020}.
Thus, the fact that we find such significant tension in our analysis between our best-fit HOD model and the clustering of high-luminosity galaxies leads us to believe that there may be an issue with our cosmological model.
It is possible that our clustering statistics are able to detect such an issue for our high-luminosity sample, but are not sensitive enough to pick up on a cosmological discrepancy among low-luminosity galaxies.
Such a result would be consistent with the findings of several other recent analyses \citep[e.g.,][]{Chapman2022, Lange2022, Wibking2020, Zhai2022}.
For example, \citet{Lange2022} used an HOD model with both assembly bias and velocity bias parameters to obtain cosmological constraints from the BOSS LOWZ sample. Using $V_\mathrm{max}$ as their assembly bias property, they did not find significant evidence of either central or satellite galaxy assembly bias, and found only minimal evidence for central velocity bias and no evidence of satellite velocity bias. 
However, they found that their best cosmological constraints were slightly inconsistent with the Planck observations.
Similarly, \citet{Zhai2022} used the Aemulus suite of cosmological N-body simulations to model the clustering of BOSS galaxies, using an HOD model with both assembly bias (based on environment) and velocity bias. They found some evidence for positive galaxy assembly bias but no evidence for satellite galaxy velocity bias. Additionally, they found that their cosmological constraints exhibited some tension with the Planck observations.

In future work, we intend to explore whether a change in cosmological parameters could be the key to alleviating this tension that we are finding.
It is worth noting that the Dark Energy Spectroscopic Instrument \citep[DESI,][]{DESI2016} will have better precision than the SDSS due to its larger volume, allowing it to potentially detect even smaller differences in clustering measurements.
Applying our model to upcoming DESI data could allow us to gain better constraints on our halo model parameters, differentiate between different baryonic feedback implementations, and ultimately constrain cosmology using small-scale galaxy clustering.

\begin{acknowledgments}
We would like to sincerely thank our anonymous referee for helpful comments which improved the quality of this paper.
This project has been supported by the National Science Foundation (NSF) through Award (AST-1909631), and has made use of NASA's Astrophysics Data System; \textsc{matplotlib}, a Python library for publication quality graphics \citep{Hunter:2007}; \textsc{scipy} \citep{Virtanen_2020}; the \textsc{ipython} package \citep{PER-GRA:2007}; \textsc{astropy}, a community-developed core Python package for Astronomy \citep{Astropy, 2013A&A...558A..33A}; \textsc{numpy} \citep{harris2020array}; \textsc{pandas} \citep{McKinney_2010, McKinney_2011}, and \textsc{chainconsumer} \citep{Hinton2016}.
Funding for the SDSS and SDSS-II has been provided by the Alfred P. Sloan Foundation, the Participating Institutions, the National Science Foundation, the U.S. Department of Energy, the National Aeronautics and Space Administration, the Japanese Monbukagakusho, the Max Planck Society, and the Higher Education Funding Council for England. The SDSS Web Site is \url{http://www.sdss.org/}. 
The SDSS is managed by the Astrophysical Research Consortium for the Participating Institutions. 
The Participating Institutions are the American Museum of Natural History, Astrophysical Institute Potsdam, University of Basel, University of Cambridge, Case Western Reserve University, University of Chicago, Drexel University, Fermilab, the Institute for Advanced Study, the Japan Participation Group, Johns Hopkins University, the Joint Institute for Nuclear Astrophysics, the Kavli Institute for Particle Astrophysics and Cosmology, the Korean Scientist Group, the Chinese Academy of Sciences (LAMOST), Los Alamos National Laboratory, the Max-Planck-Institute for Astronomy (MPIA), the Max-Planck-Institute for Astrophysics (MPA), New Mexico State University, Ohio State University, University of Pittsburgh, University of Portsmouth, Princeton University, the United States Naval Observatory, and the University of Washington.
The mock catalogues used in this paper were produced by the LasDamas project (\url{http://lss.phy.vanderbilt.edu/lasdamas/}); we thank NSF XSEDE for providing the computational resources for LasDamas.
The MCMCs in this work were run on the Texas Advanced Computing Center's Stampede2 supercomputer.
Some of the computational facilities used in this project were provided by the Vanderbilt Advanced Computing Center for Research and Education (ACCRE). Parts of this research were conducted by the Australian Research Council Centre of Excellence for All Sky Astrophysics in 3 Dimensions (ASTRO 3D), through project number CE170100013.
These acknowledgements were compiled using the Astronomy Acknowledgement Generator (\url{http://astrofrog.github.io/acknowledgment-generator/}).
\end{acknowledgments}

\bibliography{assembly_bias}

\begin{thebibliography}{}
\expandafter\ifx\csname natexlab\endcsname\relax\def\natexlab#1{#1}\fi
\providecommand{\url}[1]{\href{#1}{#1}}
\providecommand{\dodoi}[1]{doi:~\href{http://doi.org/#1}{\nolinkurl{#1}}}
\providecommand{\doeprint}[1]{\href{http://ascl.net/#1}{\nolinkurl{http://ascl.net/#1}}}
\providecommand{\doarXiv}[1]{\href{https://arxiv.org/abs/#1}{\nolinkurl{https://arxiv.org/abs/#1}}}

\bibitem[{{Abazajian} {et~al.}(2009){Abazajian}, {Adelman-McCarthy},
  {Ag{\"u}eros}, {Allam}, {Allende Prieto}, {An}, {Anderson}, {Anderson},
  {Annis}, {Bahcall}, \& et~al.}]{Abazajian2009}
{Abazajian}, K.~N., {Adelman-McCarthy}, J.~K., {Ag{\"u}eros}, M.~A., {et~al.}
  2009, \apjs, 182, 543, \dodoi{10.1088/0067-0049/182/2/543}

\bibitem[{{Artale} {et~al.}(2018){Artale}, {Zehavi}, {Contreras}, \&
  {Norberg}}]{Artale2018}
{Artale}, M.~C., {Zehavi}, I., {Contreras}, S., \& {Norberg}, P. 2018, \mnras,
  480, 3978, \dodoi{10.1093/mnras/sty2110}

\bibitem[{{Astropy Collaboration} {et~al.}(2013){Astropy Collaboration},
  {Robitaille}, {Tollerud}, {Greenfield}, {Droettboom}, {Bray}, {Aldcroft},
  {Davis}, {Ginsburg}, {Price-Whelan}, {Kerzendorf}, {Conley}, {Crighton},
  {Barbary}, {Muna}, {Ferguson}, {Grollier}, {Parikh}, {Nair}, {Unther},
  {Deil}, {Woillez}, {Conseil}, {Kramer}, {Turner}, {Singer}, {Fox}, {Weaver},
  {Zabalza}, {Edwards}, {Azalee Bostroem}, {Burke}, {Casey}, {Crawford},
  {Dencheva}, {Ely}, {Jenness}, {Labrie}, {Lim}, {Pierfederici}, {Pontzen},
  {Ptak}, {Refsdal}, {Servillat}, \& {Streicher}}]{2013A&A...558A..33A}
{Astropy Collaboration}, {Robitaille}, T.~P., {Tollerud}, E.~J., {et~al.} 2013,
  \aap, 558, A33, \dodoi{10.1051/0004-6361/201322068}

\bibitem[{{Astropy Collaboration} {et~al.}(2018){Astropy Collaboration},
  {Price-Whelan}, {Sip{\H{o}}cz}, {G{\"u}nther}, {Lim}, {Crawford}, {Conseil},
  {Shupe}, {Craig}, {Dencheva}, {Ginsburg}, {VanderPlas}, {Bradley},
  {P{\'e}rez-Su{\'a}rez}, {de Val-Borro}, {Aldcroft}, {Cruz}, {Robitaille},
  {Tollerud}, {Ardelean}, {Babej}, {Bach}, {Bachetti}, {Bakanov}, {Bamford},
  {Barentsen}, {Barmby}, {Baumbach}, {Berry}, {Biscani}, {Boquien}, {Bostroem},
  {Bouma}, {Brammer}, {Bray}, {Breytenbach}, {Buddelmeijer}, {Burke},
  {Calderone}, {Cano Rodr{\'\i}guez}, {Cara}, {Cardoso}, {Cheedella}, {Copin},
  {Corrales}, {Crichton}, {D'Avella}, {Deil}, {Depagne}, {Dietrich}, {Donath},
  {Droettboom}, {Earl}, {Erben}, {Fabbro}, {Ferreira}, {Finethy}, {Fox},
  {Garrison}, {Gibbons}, {Goldstein}, {Gommers}, {Greco}, {Greenfield},
  {Groener}, {Grollier}, {Hagen}, {Hirst}, {Homeier}, {Horton}, {Hosseinzadeh},
  {Hu}, {Hunkeler}, {Ivezi{\'c}}, {Jain}, {Jenness}, {Kanarek}, {Kendrew},
  {Kern}, {Kerzendorf}, {Khvalko}, {King}, {Kirkby}, {Kulkarni}, {Kumar},
  {Lee}, {Lenz}, {Littlefair}, {Ma}, {Macleod}, {Mastropietro}, {McCully},
  {Montagnac}, {Morris}, {Mueller}, {Mumford}, {Muna}, {Murphy}, {Nelson},
  {Nguyen}, {Ninan}, {N{\"o}the}, {Ogaz}, {Oh}, {Parejko}, {Parley}, {Pascual},
  {Patil}, {Patil}, {Plunkett}, {Prochaska}, {Rastogi}, {Reddy Janga},
  {Sabater}, {Sakurikar}, {Seifert}, {Sherbert}, {Sherwood-Taylor}, {Shih},
  {Sick}, {Silbiger}, {Singanamalla}, {Singer}, {Sladen}, {Sooley},
  {Sornarajah}, {Streicher}, {Teuben}, {Thomas}, {Tremblay}, {Turner},
  {Terr{\'o}n}, {van Kerkwijk}, {de la Vega}, {Watkins}, {Weaver}, {Whitmore},
  {Woillez}, {Zabalza}, \& {Astropy Contributors}}]{Astropy}
{Astropy Collaboration}, {Price-Whelan}, A.~M., {Sip{\H{o}}cz}, B.~M., {et~al.}
  2018, \aj, 156, 123, \dodoi{10.3847/1538-3881/aabc4f}

\bibitem[{{Baugh} {et~al.}(1999){Baugh}, {Benson}, {Cole}, {Frenk}, \&
  {Lacey}}]{Baugh1999}
{Baugh}, C.~M., {Benson}, A.~J., {Cole}, S., {Frenk}, C.~S., \& {Lacey}, C.~G.
  1999, \mnras, 305, L21, \dodoi{10.1046/j.1365-8711.1999.02590.x}

\bibitem[{{Behroozi} {et~al.}(2022){Behroozi}, {Hearin}, \&
  {Moster}}]{Behroozi2022}
{Behroozi}, P., {Hearin}, A., \& {Moster}, B.~P. 2022, \mnras, 509, 2800,
  \dodoi{10.1093/mnras/stab3193}

\bibitem[{{Behroozi} {et~al.}(2013){Behroozi}, {Wechsler}, \&
  {Wu}}]{Behroozi2013}
{Behroozi}, P.~S., {Wechsler}, R.~H., \& {Wu}, H.-Y. 2013, \apj, 762, 109,
  \dodoi{10.1088/0004-637X/762/2/109}

\bibitem[{{Beltz-Mohrmann} \& {Berlind}(2021)}]{Beltz-Mohrmann2021}
{Beltz-Mohrmann}, G.~D., \& {Berlind}, A.~A. 2021, arXiv e-prints,
  arXiv:2103.05076.
\newblock \doarXiv{2103.05076}

\bibitem[{{Beltz-Mohrmann} {et~al.}(2020){Beltz-Mohrmann}, {Berlind}, \&
  {Szewciw}}]{Beltz-Mohrmann2020}
{Beltz-Mohrmann}, G.~D., {Berlind}, A.~A., \& {Szewciw}, A.~O. 2020, \mnras,
  491, 5771, \dodoi{10.1093/mnras/stz3442}

\bibitem[{{Benson} {et~al.}(2000){Benson}, {Cole}, {Frenk}, {Baugh}, \&
  {Lacey}}]{Benson2000b}
{Benson}, A.~J., {Cole}, S., {Frenk}, C.~S., {Baugh}, C.~M., \& {Lacey}, C.~G.
  2000, \mnras, 311, 793, \dodoi{10.1046/j.1365-8711.2000.03101.x}

\bibitem[{{Berlind} \& {Weinberg}(2002)}]{Berlind2002}
{Berlind}, A.~A., \& {Weinberg}, D.~H. 2002, \apj, 575, 587,
  \dodoi{10.1086/341469}

\bibitem[{{Berlind} {et~al.}(2003){Berlind}, {Weinberg}, {Benson}, {Baugh},
  {Cole}, {Dav{\'e}}, {Frenk}, {Jenkins}, {Katz}, \& {Lacey}}]{Berlind2003}
{Berlind}, A.~A., {Weinberg}, D.~H., {Benson}, A.~J., {et~al.} 2003, \apj, 593,
  1, \dodoi{10.1086/376517}

\bibitem[{{Berlind} {et~al.}(2006){Berlind}, {Frieman}, {Weinberg}, {Blanton},
  {Warren}, {Abazajian}, {Scranton}, {Hogg}, {Scoccimarro}, {Bahcall},
  {Brinkmann}, {Gott}, {Kleinman}, {Krzesinski}, {Lee}, {Miller}, {Nitta},
  {Schneider}, {Tucker}, {Zehavi}, \& {SDSS Collaboration}}]{Berlind2006a}
{Berlind}, A.~A., {Frieman}, J., {Weinberg}, D.~H., {et~al.} 2006, \apjs, 167,
  1, \dodoi{10.1086/508170}

\bibitem[{{Blanton} {et~al.}(2005){Blanton}, {Schlegel}, {Strauss},
  {Brinkmann}, {Finkbeiner}, {Fukugita}, {Gunn}, {Hogg}, {Ivezi{\'c}}, {Knapp},
  {Lupton}, {Munn}, {Schneider}, {Tegmark}, \& {Zehavi}}]{Blanton2005}
{Blanton}, M.~R., {Schlegel}, D.~J., {Strauss}, M.~A., {et~al.} 2005, \aj, 129,
  2562, \dodoi{10.1086/429803}

\bibitem[{{Bose} {et~al.}(2019){Bose}, {Eisenstein}, {Hernquist}, {Pillepich},
  {Nelson}, {Marinacci}, {Springel}, \& {Vogelsberger}}]{Bose2019}
{Bose}, S., {Eisenstein}, D.~J., {Hernquist}, L., {et~al.} 2019, \mnras, 490,
  5693, \dodoi{10.1093/mnras/stz2546}

\bibitem[{{Boylan-Kolchin} {et~al.}(2010){Boylan-Kolchin}, {Springel}, {White},
  \& {Jenkins}}]{Boylan-Kolchin2010}
{Boylan-Kolchin}, M., {Springel}, V., {White}, S.~D.~M., \& {Jenkins}, A. 2010,
  \mnras, 406, 896, \dodoi{10.1111/j.1365-2966.2010.16774.x}

\bibitem[{{Bryan} \& {Norman}(1998)}]{Bryan1998}
{Bryan}, G.~L., \& {Norman}, M.~L. 1998, \apj, 495, 80, \dodoi{10.1086/305262}

\bibitem[{{Chapman} {et~al.}(2022){Chapman}, {Mohammad}, {Zhai}, {Percival},
  {Tinker}, {Bautista}, {Brownstein}, {Burtin}, {Dawson}, {Gil-Mar{\'\i}n}, {de
  la Macorra}, {Ross}, {Rossi}, {Schneider}, \& {Zhao}}]{Chapman2022}
{Chapman}, M.~J., {Mohammad}, F.~G., {Zhai}, Z., {et~al.} 2022, \mnras, 516,
  617, \dodoi{10.1093/mnras/stac1923}

\bibitem[{{Contreras} {et~al.}(2023){Contreras}, {Angulo}, {Chaves-Montero},
  {White}, \& {Aric{\`o}}}]{Contreras2023}
{Contreras}, S., {Angulo}, R.~E., {Chaves-Montero}, J., {White}, S. D.~M., \&
  {Aric{\`o}}, G. 2023, \mnras, 520, 489, \dodoi{10.1093/mnras/stad122}

\bibitem[{{Contreras} {et~al.}(2021){Contreras}, {Angulo}, \&
  {Zennaro}}]{Contreras2021}
{Contreras}, S., {Angulo}, R.~E., \& {Zennaro}, M. 2021, \mnras, 504, 5205,
  \dodoi{10.1093/mnras/stab1170}

\bibitem[{{Contreras} {et~al.}(2019){Contreras}, {Zehavi}, {Padilla}, {Baugh},
  {Jim{\'e}nez}, \& {Lacerna}}]{Contreras2019}
{Contreras}, S., {Zehavi}, I., {Padilla}, N., {et~al.} 2019, \mnras, 484, 1133,
  \dodoi{10.1093/mnras/stz018}

\bibitem[{{Cooray} \& {Sheth}(2002)}]{Cooray2002}
{Cooray}, A., \& {Sheth}, R. 2002, \physrep, 372, 1,
  \dodoi{10.1016/S0370-1573(02)00276-4}

\bibitem[{{Coupon} {et~al.}(2015){Coupon}, {Arnouts}, {van Waerbeke},
  {Moutard}, {Ilbert}, {van Uitert}, {Erben}, {Garilli}, {Guzzo}, {Heymans},
  {Hildebrandt}, {Hoekstra}, {Kilbinger}, {Kitching}, {Mellier}, {Miller},
  {Scodeggio}, {Bonnett}, {Branchini}, {Davidzon}, {De Lucia}, {Fritz}, {Fu},
  {Hudelot}, {Hudson}, {Kuijken}, {Leauthaud}, {Le F{\`e}vre}, {McCracken},
  {Moscardini}, {Rowe}, {Schrabback}, {Semboloni}, \& {Velander}}]{Coupon2015}
{Coupon}, J., {Arnouts}, S., {van Waerbeke}, L., {et~al.} 2015, \mnras, 449,
  1352, \dodoi{10.1093/mnras/stv276}

\bibitem[{{Crocce} {et~al.}(2006){Crocce}, {Pueblas}, \&
  {Scoccimarro}}]{Crocce2006}
{Crocce}, M., {Pueblas}, S., \& {Scoccimarro}, R. 2006, \mnras, 373, 369,
  \dodoi{10.1111/j.1365-2966.2006.11040.x}

\bibitem[{{Crocce} {et~al.}(2012){Crocce}, {Pueblas}, \&
  {Scoccimarro}}]{Crocce2012}
---. 2012, {2LPTIC: 2nd-order Lagrangian Perturbation Theory Initial
  Conditions}, Astrophysics Source Code Library.
\newblock \doeprint{1201.005}

\bibitem[{{Croton} {et~al.}(2007){Croton}, {Gao}, \& {White}}]{Croton2007}
{Croton}, D.~J., {Gao}, L., \& {White}, S.~D.~M. 2007, \mnras, 374, 1303,
  \dodoi{10.1111/j.1365-2966.2006.11230.x}

\bibitem[{{DESI Collaboration} {et~al.}(2016){DESI Collaboration}, {Aghamousa},
  {Aguilar}, {Ahlen}, {Alam}, {Allen}, {Allende Prieto}, {Annis}, {Bailey},
  {Balland}, \& et~al.}]{DESI2016}
{DESI Collaboration}, {Aghamousa}, A., {Aguilar}, J., {et~al.} 2016, arXiv
  e-prints.
\newblock \doarXiv{1611.00036}

\bibitem[{{Dong-P{\'a}ez} {et~al.}(2022){Dong-P{\'a}ez}, {Smith}, {Szewciw},
  {Ereza}, {Abdullah}, {Hern{\'a}ndez-Aguayo}, {Trusov}, {Prada}, {Klypin},
  {Ishiyama}, {Berlind}, {Zarrouk}, {L{\'o}pez Cacheiro}, \&
  {Ruedas}}]{Uchuu2022}
{Dong-P{\'a}ez}, C.~A., {Smith}, A., {Szewciw}, A.~O., {et~al.} 2022, arXiv
  e-prints, arXiv:2208.00540.
\newblock \doarXiv{2208.00540}

\bibitem[{{Foreman-Mackey} {et~al.}(2013){Foreman-Mackey}, {Hogg}, {Lang}, \&
  {Goodman}}]{EMCEE2013}
{Foreman-Mackey}, D., {Hogg}, D.~W., {Lang}, D., \& {Goodman}, J. 2013, \pasp,
  125, 306, \dodoi{10.1086/670067}

\bibitem[{{Gao} {et~al.}(2005){Gao}, {Springel}, \& {White}}]{Gao2005}
{Gao}, L., {Springel}, V., \& {White}, S.~D.~M. 2005, \mnras, 363, L66,
  \dodoi{10.1111/j.1745-3933.2005.00084.x}

\bibitem[{{Gao} \& {White}(2007)}]{Gao2007}
{Gao}, L., \& {White}, S. D.~M. 2007, \mnras, 377, L5,
  \dodoi{10.1111/j.1745-3933.2007.00292.x}

\bibitem[{{Guo} {et~al.}(2015{\natexlab{a}}){Guo}, {Zheng}, {Zehavi}, {Dawson},
  {Skibba}, {Tinker}, {Weinberg}, {White}, \& {Schneider}}]{Guo2015}
{Guo}, H., {Zheng}, Z., {Zehavi}, I., {et~al.} 2015{\natexlab{a}}, \mnras, 446,
  578, \dodoi{10.1093/mnras/stu2120}

\bibitem[{{Guo} {et~al.}(2015{\natexlab{b}}){Guo}, {Zheng}, {Zehavi},
  {Behroozi}, {Chuang}, {Comparat}, {Favole}, {Gottloeber}, {Klypin}, {Prada},
  {Weinberg}, \& {Yepes}}]{Guo2015b}
---. 2015{\natexlab{b}}, \mnras, 453, 4368, \dodoi{10.1093/mnras/stv1966}

\bibitem[{{Guo} {et~al.}(2016){Guo}, {Zheng}, {Behroozi}, {Zehavi}, {Chuang},
  {Comparat}, {Favole}, {Gottloeber}, {Klypin}, {Prada},
  {Rodr{\'{\i}}guez-Torres}, {Weinberg}, \& {Yepes}}]{Guo2016}
{Guo}, H., {Zheng}, Z., {Behroozi}, P.~S., {et~al.} 2016, \mnras, 459, 3040,
  \dodoi{10.1093/mnras/stw845}

\bibitem[{{Hadzhiyska} {et~al.}(2021{\natexlab{a}}){Hadzhiyska}, {Bose},
  {Eisenstein}, \& {Hernquist}}]{Hadzhiyska2021}
{Hadzhiyska}, B., {Bose}, S., {Eisenstein}, D., \& {Hernquist}, L.
  2021{\natexlab{a}}, \mnras, 501, 1603, \dodoi{10.1093/mnras/staa3776}

\bibitem[{{Hadzhiyska} {et~al.}(2020){Hadzhiyska}, {Bose}, {Eisenstein},
  {Hernquist}, \& {Spergel}}]{Hadzhiyska2020}
{Hadzhiyska}, B., {Bose}, S., {Eisenstein}, D., {Hernquist}, L., \& {Spergel},
  D.~N. 2020, \mnras, 493, 5506, \dodoi{10.1093/mnras/staa623}

\bibitem[{{Hadzhiyska} {et~al.}(2021{\natexlab{b}}){Hadzhiyska}, {Liu},
  {Somerville}, {Gabrielpillai}, {Bose}, {Eisenstein}, \&
  {Hernquist}}]{Hadzhiyska2021c}
{Hadzhiyska}, B., {Liu}, S., {Somerville}, R.~S., {et~al.} 2021{\natexlab{b}},
  \mnras, 508, 698, \dodoi{10.1093/mnras/stab2564}

\bibitem[{{Hadzhiyska} {et~al.}(2021{\natexlab{c}}){Hadzhiyska}, {Tacchella},
  {Bose}, \& {Eisenstein}}]{Hadzhiyska2021b}
{Hadzhiyska}, B., {Tacchella}, S., {Bose}, S., \& {Eisenstein}, D.~J.
  2021{\natexlab{c}}, \mnras, 502, 3599, \dodoi{10.1093/mnras/stab243}

\bibitem[{{Hahn} {et~al.}(2019){Hahn}, {Beutler}, {Sinha}, {Berlind}, {Ho}, \&
  {Hogg}}]{Hahn2019}
{Hahn}, C., {Beutler}, F., {Sinha}, M., {et~al.} 2019, \mnras, 485, 2956,
  \dodoi{10.1093/mnras/stz558}

\bibitem[{Harris {et~al.}(2020)Harris, Millman, van~der Walt, Gommers,
  Virtanen, Cournapeau, Wieser, Taylor, Berg, Smith, Kern, Picus, Hoyer, van
  Kerkwijk, Brett, Haldane, del R{'{\i}}o, Wiebe, Peterson,
  G{'{e}}rard-Marchant, Sheppard, Reddy, Weckesser, Abbasi, Gohlke, \&
  Oliphant}]{harris2020array}
Harris, C.~R., Millman, K.~J., van~der Walt, S.~J., {et~al.} 2020, Nature, 585,
  357, \dodoi{10.1038/s41586-020-2649-2}

\bibitem[{{Hearin} {et~al.}(2016){Hearin}, {Zentner}, {van den Bosch},
  {Campbell}, \& {Tollerud}}]{Hearin2016}
{Hearin}, A.~P., {Zentner}, A.~R., {van den Bosch}, F.~C., {Campbell}, D., \&
  {Tollerud}, E. 2016, \mnras, 460, 2552, \dodoi{10.1093/mnras/stw840}

\bibitem[{{Hinton}(2016)}]{Hinton2016}
{Hinton}, S.~R. 2016, The Journal of Open Source Software, 1, 00045,
  \dodoi{10.21105/joss.00045}

\bibitem[{Hunter(2007)}]{Hunter:2007}
Hunter, J.~D. 2007, Computing In Science \& Engineering, 9, 90

\bibitem[{{Jim{\'e}nez} {et~al.}(2019){Jim{\'e}nez}, {Contreras}, {Padilla},
  {Zehavi}, {Baugh}, \& {Gonzalez-Perez}}]{Jimenez2019}
{Jim{\'e}nez}, E., {Contreras}, S., {Padilla}, N., {et~al.} 2019, \mnras, 490,
  3532, \dodoi{10.1093/mnras/stz2790}

\bibitem[{{Jing} {et~al.}(1998){Jing}, {Mo}, \& {B{\"o}rner}}]{Jing1998}
{Jing}, Y.~P., {Mo}, H.~J., \& {B{\"o}rner}, G. 1998, \apj, 494, 1,
  \dodoi{10.1086/305209}

\bibitem[{{Kauffmann} {et~al.}(1999){Kauffmann}, {Colberg}, {Diaferio}, \&
  {White}}]{Kauffmann1999}
{Kauffmann}, G., {Colberg}, J.~M., {Diaferio}, A., \& {White}, S.~D.~M. 1999,
  \mnras, 303, 188, \dodoi{10.1046/j.1365-8711.1999.02202.x}

\bibitem[{{Kauffmann} {et~al.}(1997){Kauffmann}, {Nusser}, \&
  {Steinmetz}}]{Kauffmann1997}
{Kauffmann}, G., {Nusser}, A., \& {Steinmetz}, M. 1997, \mnras, 286, 795,
  \dodoi{10.1093/mnras/286.4.795}

\bibitem[{{Klypin} {et~al.}(2011){Klypin}, {Trujillo-Gomez}, \&
  {Primack}}]{Klypin2011}
{Klypin}, A.~A., {Trujillo-Gomez}, S., \& {Primack}, J. 2011, \apj, 740, 102,
  \dodoi{10.1088/0004-637X/740/2/102}

\bibitem[{{Kravtsov} {et~al.}(2004){Kravtsov}, {Berlind}, {Wechsler}, {Klypin},
  {Gottl{\"o}ber}, {Allgood}, \& {Primack}}]{Kravtsov2004}
{Kravtsov}, A.~V., {Berlind}, A.~A., {Wechsler}, R.~H., {et~al.} 2004, \apj,
  609, 35, \dodoi{10.1086/420959}

\bibitem[{{Lacey} \& {Cole}(1994)}]{Lacey1994}
{Lacey}, C., \& {Cole}, S. 1994, \mnras, 271, 676,
  \dodoi{10.1093/mnras/271.3.676}

\bibitem[{{Lange} {et~al.}(2022){Lange}, {Hearin}, {Leauthaud}, {van den
  Bosch}, {Guo}, \& {DeRose}}]{Lange2022}
{Lange}, J.~U., {Hearin}, A.~P., {Leauthaud}, A., {et~al.} 2022, \mnras, 509,
  1779, \dodoi{10.1093/mnras/stab3111}

\bibitem[{{Lange} {et~al.}(2019){Lange}, {van den Bosch}, {Zentner}, {Wang},
  {Hearin}, \& {Guo}}]{Lange2019c}
{Lange}, J.~U., {van den Bosch}, F.~C., {Zentner}, A.~R., {et~al.} 2019,
  Monthly Notices of the Royal Astronomical Society, 490, 1870,
  \dodoi{10.1093/mnras/stz2664}

\bibitem[{{Leauthaud} {et~al.}(2012){Leauthaud}, {Tinker}, {Bundy}, {Behroozi},
  {Massey}, {Rhodes}, {George}, {Kneib}, {Benson}, {Wechsler}, {Busha},
  {Capak}, {Cort{\^e}s}, {Ilbert}, {Koekemoer}, {Le F{\`e}vre}, {Lilly},
  {McCracken}, {Salvato}, {Schrabback}, {Scoville}, {Smith}, \&
  {Taylor}}]{Leauthaud2012}
{Leauthaud}, A., {Tinker}, J., {Bundy}, K., {et~al.} 2012, \apj, 744, 159,
  \dodoi{10.1088/0004-637X/744/2/159}

\bibitem[{{Lehmann} {et~al.}(2017){Lehmann}, {Mao}, {Becker}, {Skillman}, \&
  {Wechsler}}]{Lehmann2017}
{Lehmann}, B.~V., {Mao}, Y.-Y., {Becker}, M.~R., {Skillman}, S.~W., \&
  {Wechsler}, R.~H. 2017, \apj, 834, 37, \dodoi{10.3847/1538-4357/834/1/37}

\bibitem[{{Ma} \& {Fry}(2000)}]{Ma2000}
{Ma}, C.-P., \& {Fry}, J.~N. 2000, \apj, 543, 503, \dodoi{10.1086/317146}

\bibitem[{{Mansfield} \& {Kravtsov}(2020)}]{Mansfield2020}
{Mansfield}, P., \& {Kravtsov}, A.~V. 2020, \mnras, 493, 4763,
  \dodoi{10.1093/mnras/staa430}

\bibitem[{{Mao} {et~al.}(2015){Mao}, {Williamson}, \& {Wechsler}}]{Mao2015}
{Mao}, Y.-Y., {Williamson}, M., \& {Wechsler}, R.~H. 2015, \apj, 810, 21,
  \dodoi{10.1088/0004-637X/810/1/21}

\bibitem[{{Mao} {et~al.}(2018){Mao}, {Zentner}, \& {Wechsler}}]{Mao2018}
{Mao}, Y.-Y., {Zentner}, A.~R., \& {Wechsler}, R.~H. 2018, \mnras, 474, 5143,
  \dodoi{10.1093/mnras/stx3111}

\bibitem[{{McBride} {et~al.}(2009){McBride}, {Berlind}, {Scoccimarro},
  {Wechsler}, {Busha}, {Gardner}, \& {van den Bosch}}]{McBride2009}
{McBride}, C., {Berlind}, A., {Scoccimarro}, R., {et~al.} 2009, in Bulletin of
  the American Astronomical Society, Vol.~41, American Astronomical Society
  Meeting Abstracts \#213, 253

\bibitem[{{McCarthy} {et~al.}(2019){McCarthy}, {Zheng}, \&
  {Guo}}]{McCarthy2019}
{McCarthy}, K.~S., {Zheng}, Z., \& {Guo}, H. 2019, \mnras, 487, 2424,
  \dodoi{10.1093/mnras/stz1461}

\bibitem[{{McCarthy} {et~al.}(2022){McCarthy}, {Zheng}, {Guo}, {Luo}, \&
  {Lin}}]{McCarthy2022}
{McCarthy}, K.~S., {Zheng}, Z., {Guo}, H., {Luo}, W., \& {Lin}, Y.-T. 2022,
  \mnras, 509, 380, \dodoi{10.1093/mnras/stab2602}

\bibitem[{{McClelland} \& {Silk}(1977)}]{McClelland1977}
{McClelland}, J., \& {Silk}, J. 1977, \apj, 217, 331, \dodoi{10.1086/155583}

\bibitem[{{McCullagh} {et~al.}(2017){McCullagh}, {Norberg}, {Cole},
  {Gonzalez-Perez}, {Baugh}, \& {Helly}}]{McCullagh2017}
{McCullagh}, N., {Norberg}, P., {Cole}, S., {et~al.} 2017, arXiv e-prints.
\newblock \doarXiv{1705.01988}

\bibitem[{McKinney(2010)}]{McKinney_2010}
McKinney, W. 2010, in Proceedings of the 9th Python in Science Conference, Vol.
  445, Austin, TX, 51--56

\bibitem[{McKinney(2011)}]{McKinney_2011}
McKinney, W. 2011, Python for High Performance and Scientific Computing, 14

\bibitem[{{Navarro} {et~al.}(1997){Navarro}, {Frenk}, \& {White}}]{Navarro1997}
{Navarro}, J.~F., {Frenk}, C.~S., \& {White}, S.~D.~M. 1997, \apj, 490, 493,
  \dodoi{10.1086/304888}

\bibitem[{{Neyman} \& {Scott}(1952)}]{Neyman1952}
{Neyman}, J., \& {Scott}, E.~L. 1952, \apj, 116, 144, \dodoi{10.1086/145599}

\bibitem[{{Norberg} {et~al.}(2009){Norberg}, {Baugh}, {Gazta{\~n}aga}, \&
  {Croton}}]{Norberg2009}
{Norberg}, P., {Baugh}, C.~M., {Gazta{\~n}aga}, E., \& {Croton}, D.~J. 2009,
  \mnras, 396, 19, \dodoi{10.1111/j.1365-2966.2009.14389.x}

\bibitem[{{Padilla} {et~al.}(2019){Padilla}, {Contreras}, {Zehavi}, {Baugh}, \&
  {Norberg}}]{Padilla2019}
{Padilla}, N., {Contreras}, S., {Zehavi}, I., {Baugh}, C.~M., \& {Norberg}, P.
  2019, \mnras, 486, 582, \dodoi{10.1093/mnras/stz824}

\bibitem[{{Parejko} {et~al.}(2013){Parejko}, {Sunayama}, {Padmanabhan}, {Wake},
  {Berlind}, {Bizyaev}, {Blanton}, {Bolton}, {van den Bosch}, {Brinkmann},
  {Brownstein}, {da Costa}, {Eisenstein}, {Guo}, {Kazin}, {Maia},
  {Malanushenko}, {Maraston}, {McBride}, {Nichol}, {Oravetz}, {Pan},
  {Percival}, {Prada}, {Ross}, {Ross}, {Schlegel}, {Schneider}, {Simmons},
  {Skibba}, {Tinker}, {Tojeiro}, {Weaver}, {Wetzel}, {White}, {Weinberg},
  {Thomas}, {Zehavi}, \& {Zheng}}]{Parejko2013}
{Parejko}, J.~K., {Sunayama}, T., {Padmanabhan}, N., {et~al.} 2013, \mnras,
  429, 98, \dodoi{10.1093/mnras/sts314}

\bibitem[{{Peacock} \& {Smith}(2000)}]{Peacock2000}
{Peacock}, J.~A., \& {Smith}, R.~E. 2000, \mnras, 318, 1144,
  \dodoi{10.1046/j.1365-8711.2000.03779.x}

\bibitem[{{Peebles}(1974)}]{Peebles1974}
{Peebles}, P.~J.~E. 1974, \aap, 32, 197

\bibitem[{P\'erez \& Granger(2007)}]{PER-GRA:2007}
P\'erez, F., \& Granger, B.~E. 2007, Computing in Science and Engineering, 9,
  21, \dodoi{10.1109/MCSE.2007.53}

\bibitem[{{Perez} {et~al.}(2021){Perez}, {Malhotra}, {Rhoads}, \&
  {Tilvi}}]{Perez2021}
{Perez}, L.~A., {Malhotra}, S., {Rhoads}, J.~E., \& {Tilvi}, V. 2021, \apj,
  906, 58, \dodoi{10.3847/1538-4357/abc88b}

\bibitem[{{Piscionere} {et~al.}(2015){Piscionere}, {Berlind}, {McBride}, \&
  {Scoccimarro}}]{Piscionere2015}
{Piscionere}, J.~A., {Berlind}, A.~A., {McBride}, C.~K., \& {Scoccimarro}, R.
  2015, \apj, 806, 125, \dodoi{10.1088/0004-637X/806/1/125}

\bibitem[{{Planck Collaboration} {et~al.}(2014){Planck Collaboration}, {Ade},
  {Aghanim}, {Armitage-Caplan}, {Arnaud}, {Ashdown}, {Atrio-Barand ela},
  {Aumont}, {Baccigalupi}, {Banday}, {Barreiro}, {Bartlett}, {Battaner},
  {Benabed}, {Beno{\^\i}t}, {Benoit-L{\'e}vy}, {Bernard}, {Bersanelli},
  {Bielewicz}, {Bobin}, {Bock}, {Bonaldi}, {Bond}, {Borrill}, {Bouchet},
  {Bridges}, {Bucher}, {Burigana}, {Butler}, {Calabrese}, {Cappellini},
  {Cardoso}, {Catalano}, {Challinor}, {Chamballu}, {Chary}, {Chen}, {Chiang},
  {Chiang}, {Christensen}, {Church}, {Clements}, {Colombi}, {Colombo},
  {Couchot}, {Coulais}, {Crill}, {Curto}, {Cuttaia}, {Danese}, {Davies},
  {Davis}, {de Bernardis}, {de Rosa}, {de Zotti}, {Delabrouille}, {Delouis},
  {D{\'e}sert}, {Dickinson}, {Diego}, {Dolag}, {Dole}, {Donzelli}, {Dor{\'e}},
  {Douspis}, {Dunkley}, {Dupac}, {Efstathiou}, {Elsner}, {En{\ss}lin},
  {Eriksen}, {Finelli}, {Forni}, {Frailis}, {Fraisse}, {Franceschi}, {Gaier},
  {Galeotta}, {Galli}, {Ganga}, {Giard}, {Giardino}, {Giraud-H{\'e}raud},
  {Gjerl{\o}w}, {Gonz{\'a}lez-Nuevo}, {G{\'o}rski}, {Gratton}, {Gregorio},
  {Gruppuso}, {Gudmundsson}, {Haissinski}, {Hamann}, {Hansen}, {Hanson},
  {Harrison}, {Henrot-Versill{\'e}}, {Hern{\'a}ndez-Monteagudo}, {Herranz},
  {Hildebrand t}, {Hivon}, {Hobson}, {Holmes}, {Hornstrup}, {Hou}, {Hovest},
  {Huffenberger}, {Jaffe}, {Jaffe}, {Jewell}, {Jones}, {Juvela},
  {Keih{\"a}nen}, {Keskitalo}, {Kisner}, {Kneissl}, {Knoche}, {Knox}, {Kunz},
  {Kurki-Suonio}, {Lagache}, {L{\"a}hteenm{\"a}ki}, {Lamarre}, {Lasenby},
  {Lattanzi}, {Laureijs}, {Lawrence}, {Leach}, {Leahy}, {Leonardi},
  {Le{\'o}n-Tavares}, {Lesgourgues}, {Lewis}, {Liguori}, {Lilje},
  {Linden-V{\o}rnle}, {L{\'o}pez-Caniego}, {Lubin}, {Mac{\'\i}as-P{\'e}rez},
  {Maffei}, {Maino}, {Mand olesi}, {Maris}, {Marshall}, {Martin},
  {Mart{\'\i}nez-Gonz{\'a}lez}, {Masi}, {Massardi}, {Matarrese}, {Matthai},
  {Mazzotta}, {Meinhold}, {Melchiorri}, {Melin}, {Mendes}, {Menegoni},
  {Mennella}, {Migliaccio}, {Millea}, {Mitra}, {Miville-Desch{\^e}nes},
  {Moneti}, {Montier}, {Morgante}, {Mortlock}, {Moss}, {Munshi}, {Murphy},
  {Naselsky}, {Nati}, {Natoli}, {Netterfield}, {N{\o}rgaard-Nielsen},
  {Noviello}, {Novikov}, {Novikov}, {O'Dwyer}, {Osborne}, {Oxborrow}, {Paci},
  {Pagano}, {Pajot}, {Paladini}, {Paoletti}, {Partridge}, {Pasian},
  {Patanchon}, {Pearson}, {Pearson}, {Peiris}, {Perdereau}, {Perotto},
  {Perrotta}, {Pettorino}, {Piacentini}, {Piat}, {Pierpaoli}, {Pietrobon},
  {Plaszczynski}, {Platania}, {Pointecouteau}, {Polenta}, {Ponthieu}, {Popa},
  {Poutanen}, {Pratt}, {Pr{\'e}zeau}, {Prunet}, {Puget}, {Rachen}, {Reach},
  {Rebolo}, {Reinecke}, {Remazeilles}, {Renault}, {Ricciardi}, {Riller},
  {Ristorcelli}, {Rocha}, {Rosset}, {Roudier}, {Rowan-Robinson},
  {Rubi{\~n}o-Mart{\'\i}n}, {Rusholme}, {Sandri}, {Santos}, {Savelainen},
  {Savini}, {Scott}, {Seiffert}, {Shellard}, {Spencer}, {Starck}, {Stolyarov},
  {Stompor}, {Sudiwala}, {Sunyaev}, {Sureau}, {Sutton}, {Suur-Uski}, {Sygnet},
  {Tauber}, {Tavagnacco}, {Terenzi}, {Toffolatti}, {Tomasi}, {Tristram},
  {Tucci}, {Tuovinen}, {T{\"u}rler}, {Umana}, {Valenziano}, {Valiviita}, {Van
  Tent}, {Vielva}, {Villa}, {Vittorio}, {Wade}, {Wandelt}, {Wehus}, {White},
  {White}, {Wilkinson}, {Yvon}, {Zacchei}, \& {Zonca}}]{Planck2014}
{Planck Collaboration}, {Ade}, P.~A.~R., {Aghanim}, N., {et~al.} 2014, \aap,
  571, A16, \dodoi{10.1051/0004-6361/201321591}

\bibitem[{{Pujol} {et~al.}(2017){Pujol}, {Hoffmann}, {Jim{\'e}nez}, \&
  {Gazta{\~n}aga}}]{Pujol2017}
{Pujol}, A., {Hoffmann}, K., {Jim{\'e}nez}, N., \& {Gazta{\~n}aga}, E. 2017,
  \aap, 598, A103, \dodoi{10.1051/0004-6361/201629121}

\bibitem[{{Salcedo} {et~al.}(2018){Salcedo}, {Maller}, {Berlind}, {Sinha},
  {McBride}, {Behroozi}, {Wechsler}, \& {Weinberg}}]{Salcedo2018}
{Salcedo}, A.~N., {Maller}, A.~H., {Berlind}, A.~A., {et~al.} 2018, \mnras,
  475, 4411, \dodoi{10.1093/mnras/sty109}

\bibitem[{{Salcedo} {et~al.}(2022){Salcedo}, {Zu}, {Zhang}, {Wang}, {Yang},
  {Wu}, {Jing}, {Mo}, \& {Weinberg}}]{Salcedo2022}
{Salcedo}, A.~N., {Zu}, Y., {Zhang}, Y., {et~al.} 2022, Science China Physics,
  Mechanics, and Astronomy, 65, 109811, \dodoi{10.1007/s11433-022-1955-7}

\bibitem[{{Scherrer} \& {Bertschinger}(1991)}]{Scherrer1991}
{Scherrer}, R.~J., \& {Bertschinger}, E. 1991, \apj, 381, 349,
  \dodoi{10.1086/170658}

\bibitem[{{Scoccimarro}(1998)}]{Scoccimarro1998}
{Scoccimarro}, R. 1998, \mnras, 299, 1097,
  \dodoi{10.1046/j.1365-8711.1998.01845.x}

\bibitem[{{Scoccimarro} {et~al.}(2001){Scoccimarro}, {Sheth}, {Hui}, \&
  {Jain}}]{Scoccimarro2001}
{Scoccimarro}, R., {Sheth}, R.~K., {Hui}, L., \& {Jain}, B. 2001, \apj, 546,
  20, \dodoi{10.1086/318261}

\bibitem[{{Seljak}(2000)}]{Seljak2000}
{Seljak}, U. 2000, \mnras, 318, 203, \dodoi{10.1046/j.1365-8711.2000.03715.x}

\bibitem[{{Seljak} \& {Zaldarriaga}(1996)}]{Seljak1996}
{Seljak}, U., \& {Zaldarriaga}, M. 1996, \apj, 469, 437, \dodoi{10.1086/177793}

\bibitem[{{Sheth} {et~al.}(2001){Sheth}, {Hui}, {Diaferio}, \&
  {Scoccimarro}}]{Sheth2001}
{Sheth}, R.~K., {Hui}, L., {Diaferio}, A., \& {Scoccimarro}, R. 2001, \mnras,
  325, 1288, \dodoi{10.1046/j.1365-8711.2001.04222.x}

\bibitem[{{Sheth} \& {Tormen}(2004)}]{Sheth2004}
{Sheth}, R.~K., \& {Tormen}, G. 2004, \mnras, 350, 1385,
  \dodoi{10.1111/j.1365-2966.2004.07733.x}

\bibitem[{{Shi} \& {Sheth}(2018)}]{Shi2018}
{Shi}, J., \& {Sheth}, R.~K. 2018, \mnras, 473, 2486,
  \dodoi{10.1093/mnras/stx2277}

\bibitem[{{Sinha} {et~al.}(2018){Sinha}, {Berlind}, {McBride}, {Scoccimarro},
  {Piscionere}, \& {Wibking}}]{Sinha2018}
{Sinha}, M., {Berlind}, A.~A., {McBride}, C.~K., {et~al.} 2018, \mnras, 478,
  1042, \dodoi{10.1093/mnras/sty967}

\bibitem[{{Sinha} \& {Garrison}(2019)}]{Sinha2019}
{Sinha}, M., \& {Garrison}, L.~H. 2019, in Software Challenges to Exascale
  Computing. Second Workshop, 3--20, \dodoi{10.1007/978-981-13-7729-7_1}

\bibitem[{{Sinha} \& {Garrison}(2020)}]{Sinha2020}
{Sinha}, M., \& {Garrison}, L.~H. 2020, \mnras, 491, 3022,
  \dodoi{10.1093/mnras/stz3157}

\bibitem[{{Springel}(2005)}]{Springel2005}
{Springel}, V. 2005, \mnras, 364, 1105,
  \dodoi{10.1111/j.1365-2966.2005.09655.x}

\bibitem[{{Storey-Fisher} {et~al.}(2022){Storey-Fisher}, {Tinker}, {Zhai},
  {DeRose}, {Wechsler}, \& {Banerjee}}]{Storey-Fisher2022}
{Storey-Fisher}, K., {Tinker}, J., {Zhai}, Z., {et~al.} 2022, arXiv e-prints,
  arXiv:2210.03203.
\newblock \doarXiv{2210.03203}

\bibitem[{{Szewciw} {et~al.}(2022){Szewciw}, {Beltz-Mohrmann}, {Berlind}, \&
  {Sinha}}]{Szewciw2022}
{Szewciw}, A.~O., {Beltz-Mohrmann}, G.~D., {Berlind}, A.~A., \& {Sinha}, M.
  2022, \apj, 926, 15, \dodoi{10.3847/1538-4357/ac3a7c}

\bibitem[{{Tinker} {et~al.}(2008){Tinker}, {Conroy}, {Norberg}, {Patiri},
  {Weinberg}, \& {Warren}}]{Tinker2008}
{Tinker}, J.~L., {Conroy}, C., {Norberg}, P., {et~al.} 2008, \apj, 686, 53,
  \dodoi{10.1086/589983}

\bibitem[{{Tinker} {et~al.}(2006{\natexlab{a}}){Tinker}, {Weinberg}, \&
  {Warren}}]{Tinker2006a}
{Tinker}, J.~L., {Weinberg}, D.~H., \& {Warren}, M.~S. 2006{\natexlab{a}},
  \apj, 647, 737, \dodoi{10.1086/504795}

\bibitem[{{Tinker} {et~al.}(2006{\natexlab{b}}){Tinker}, {Weinberg}, \&
  {Zheng}}]{Tinker2006b}
{Tinker}, J.~L., {Weinberg}, D.~H., \& {Zheng}, Z. 2006{\natexlab{b}}, \mnras,
  368, 85, \dodoi{10.1111/j.1365-2966.2006.10114.x}

\bibitem[{{Tonegawa} {et~al.}(2020){Tonegawa}, {Park}, {Zheng}, {Park}, {Hong},
  {Hwang}, \& {Kim}}]{Tonegawa2020}
{Tonegawa}, M., {Park}, C., {Zheng}, Y., {et~al.} 2020, \apj, 897, 17,
  \dodoi{10.3847/1538-4357/ab95ff}

\bibitem[{{Vakili} \& {Hahn}(2019)}]{Vakili2019}
{Vakili}, M., \& {Hahn}, C. 2019, \apj, 872, 115,
  \dodoi{10.3847/1538-4357/aaf1a1}

\bibitem[{{Villarreal} {et~al.}(2017){Villarreal}, {Zentner}, {Mao}, {Purcell},
  {van den Bosch}, {Diemer}, {Lange}, {Wang}, \& {Campbell}}]{Villarreal2017}
{Villarreal}, A.~S., {Zentner}, A.~R., {Mao}, Y.-Y., {et~al.} 2017, \mnras,
  472, 1088, \dodoi{10.1093/mnras/stx2045}

\bibitem[{{Virtanen} {et~al.}(2020){Virtanen}, {Gommers}, {Oliphant},
  {Haberland}, {Reddy}, {Cournapeau}, {Burovski}, {Peterson}, {Weckesser},
  {Bright}, {van der Walt}, {Brett}, {Wilson}, {Jarrod Millman}, {Mayorov},
  {Nelson}, {Jones}, {Kern}, {Larson}, {Carey}, {Polat}, {Feng}, {Moore}, {Vand
  erPlas}, {Laxalde}, {Perktold}, {Cimrman}, {Henriksen}, {Quintero}, {Harris},
  {Archibald}, {Ribeiro}, {Pedregosa}, {van Mulbregt}, \&
  {Contributors}}]{Virtanen_2020}
{Virtanen}, P., {Gommers}, R., {Oliphant}, T.~E., {et~al.} 2020, Nature
  Methods, 17, 261, \dodoi{https://doi.org/10.1038/s41592-019-0686-2}

\bibitem[{{Walsh} \& {Tinker}(2019)}]{Walsh2019}
{Walsh}, K., \& {Tinker}, J. 2019, \mnras, \dodoi{10.1093/mnras/stz1351}

\bibitem[{{Wang} {et~al.}(2022){Wang}, {Mao}, {Zentner}, {Guo}, {Lange}, {van
  den Bosch}, \& {Mezini}}]{Wang2022}
{Wang}, K., {Mao}, Y.-Y., {Zentner}, A.~R., {et~al.} 2022, arXiv e-prints,
  arXiv:2204.05332.
\newblock \doarXiv{2204.05332}

\bibitem[{{Wang} {et~al.}(2019){Wang}, {Mao}, {Zentner}, {van den Bosch},
  {Lange}, {Schafer}, {Villarreal}, {Hearin}, \& {Campbell}}]{Wang2019}
---. 2019, \mnras, 488, 3541, \dodoi{10.1093/mnras/stz1733}

\bibitem[{{Watson} {et~al.}(2012){Watson}, {Berlind}, {McBride}, {Hogg}, \&
  {Jiang}}]{Watson2012}
{Watson}, D.~F., {Berlind}, A.~A., {McBride}, C.~K., {Hogg}, D.~W., \& {Jiang},
  T. 2012, \apj, 749, 83, \dodoi{10.1088/0004-637X/749/1/83}

\bibitem[{{Wechsler} {et~al.}(2002){Wechsler}, {Bullock}, {Primack},
  {Kravtsov}, \& {Dekel}}]{Wechsler2002}
{Wechsler}, R.~H., {Bullock}, J.~S., {Primack}, J.~R., {Kravtsov}, A.~V., \&
  {Dekel}, A. 2002, \apj, 568, 52, \dodoi{10.1086/338765}

\bibitem[{{Wechsler} \& {Tinker}(2018)}]{Wechsler2018}
{Wechsler}, R.~H., \& {Tinker}, J.~L. 2018, \araa, 56, 435,
  \dodoi{10.1146/annurev-astro-081817-051756}

\bibitem[{{Wechsler} {et~al.}(2006){Wechsler}, {Zentner}, {Bullock},
  {Kravtsov}, \& {Allgood}}]{Wechsler2006}
{Wechsler}, R.~H., {Zentner}, A.~R., {Bullock}, J.~S., {Kravtsov}, A.~V., \&
  {Allgood}, B. 2006, \apj, 652, 71, \dodoi{10.1086/507120}

\bibitem[{{White} {et~al.}(2001){White}, {Hernquist}, \&
  {Springel}}]{White2001}
{White}, M., {Hernquist}, L., \& {Springel}, V. 2001, \apjl, 550, L129,
  \dodoi{10.1086/319644}

\bibitem[{{Wibking} {et~al.}(2020){Wibking}, {Weinberg}, {Salcedo}, {Wu},
  {Singh}, {Rodr{\'\i}guez-Torres}, {Garrison}, \& {Eisenstein}}]{Wibking2020}
{Wibking}, B.~D., {Weinberg}, D.~H., {Salcedo}, A.~N., {et~al.} 2020, \mnras,
  492, 2872, \dodoi{10.1093/mnras/stz3423}

\bibitem[{{Xu} \& {Zheng}(2020)}]{Xu2020}
{Xu}, X., \& {Zheng}, Z. 2020, \mnras, 492, 2739, \dodoi{10.1093/mnras/staa009}

\bibitem[{{York} {et~al.}(2000){York}, {Adelman}, {Anderson}, {Anderson},
  {Annis}, {Bahcall}, {Bakken}, {Barkhouser}, {Bastian}, {Berman}, {Boroski},
  {Bracker}, {Briegel}, {Briggs}, {Brinkmann}, {Brunner}, {Burles}, {Carey},
  {Carr}, {Castander}, {Chen}, {Colestock}, {Connolly}, {Crocker}, {Csabai},
  {Czarapata}, {Davis}, {Doi}, {Dombeck}, {Eisenstein}, {Ellman}, {Elms},
  {Evans}, {Fan}, {Federwitz}, {Fiscelli}, {Friedman}, {Frieman}, {Fukugita},
  {Gillespie}, {Gunn}, {Gurbani}, {de Haas}, {Haldeman}, {Harris}, {Hayes},
  {Heckman}, {Hennessy}, {Hindsley}, {Holm}, {Holmgren}, {Huang}, {Hull},
  {Husby}, {Ichikawa}, {Ichikawa}, {Ivezi{\'c}}, {Kent}, {Kim}, {Kinney},
  {Klaene}, {Kleinman}, {Kleinman}, {Knapp}, {Korienek}, {Kron}, {Kunszt},
  {Lamb}, {Lee}, {Leger}, {Limmongkol}, {Lindenmeyer}, {Long}, {Loomis},
  {Loveday}, {Lucinio}, {Lupton}, {MacKinnon}, {Mannery}, {Mantsch}, {Margon},
  {McGehee}, {McKay}, {Meiksin}, {Merelli}, {Monet}, {Munn}, {Narayanan},
  {Nash}, {Neilsen}, {Neswold}, {Newberg}, {Nichol}, {Nicinski}, {Nonino},
  {Okada}, {Okamura}, {Ostriker}, {Owen}, {Pauls}, {Peoples}, {Peterson},
  {Petravick}, {Pier}, {Pope}, {Pordes}, {Prosapio}, {Rechenmacher}, {Quinn},
  {Richards}, {Richmond}, {Rivetta}, {Rockosi}, {Ruthmansdorfer}, {Sandford},
  {Schlegel}, {Schneider}, {Sekiguchi}, {Sergey}, {Shimasaku}, {Siegmund},
  {Smee}, {Smith}, {Snedden}, {Stone}, {Stoughton}, {Strauss}, {Stubbs},
  {SubbaRao}, {Szalay}, {Szapudi}, {Szokoly}, {Thakar}, {Tremonti}, {Tucker},
  {Uomoto}, {Vanden Berk}, {Vogeley}, {Waddell}, {Wang}, {Watanabe},
  {Weinberg}, {Yanny}, {Yasuda}, \& {SDSS Collaboration}}]{York2000}
{York}, D.~G., {Adelman}, J., {Anderson}, Jr., J.~E., {et~al.} 2000, \aj, 120,
  1579, \dodoi{10.1086/301513}

\bibitem[{{Yuan} {et~al.}(2021){Yuan}, {Garrison}, {Hadzhiyska}, {Bose}, \&
  {Eisenstein}}]{Yuan2021}
{Yuan}, S., {Garrison}, L.~H., {Hadzhiyska}, B., {Bose}, S., \& {Eisenstein},
  D.~J. 2021, \mnras, \dodoi{10.1093/mnras/stab3355}

\bibitem[{{Zaldarriaga} \& {Seljak}(2000)}]{Zaldarriaga2000}
{Zaldarriaga}, M., \& {Seljak}, U. 2000, \apjs, 129, 431,
  \dodoi{10.1086/313423}

\bibitem[{{Zaldarriaga} {et~al.}(1998){Zaldarriaga}, {Seljak}, \&
  {Bertschinger}}]{Zaldarriaga1998}
{Zaldarriaga}, M., {Seljak}, U., \& {Bertschinger}, E. 1998, \apj, 494, 491,
  \dodoi{10.1086/305223}

\bibitem[{{Zehavi} {et~al.}(2018){Zehavi}, {Contreras}, {Padilla}, {Smith},
  {Baugh}, \& {Norberg}}]{Zehavi2018}
{Zehavi}, I., {Contreras}, S., {Padilla}, N., {et~al.} 2018, \apj, 853, 84,
  \dodoi{10.3847/1538-4357/aaa54a}

\bibitem[{{Zehavi} {et~al.}(2002){Zehavi}, {Blanton}, {Frieman}, {Weinberg},
  {Mo}, {Strauss}, {Anderson}, {Annis}, {Bahcall}, {Bernardi}, {Briggs},
  {Brinkmann}, {Burles}, {Carey}, {Castander}, {Connolly}, {Csabai},
  {Dalcanton}, {Dodelson}, {Doi}, {Eisenstein}, {Evans}, {Finkbeiner},
  {Friedman}, {Fukugita}, {Gunn}, {Hennessy}, {Hindsley}, {Ivezi{\'c}}, {Kent},
  {Knapp}, {Kron}, {Kunszt}, {Lamb}, {Leger}, {Long}, {Loveday}, {Lupton},
  {McKay}, {Meiksin}, {Merrelli}, {Munn}, {Narayanan}, {Newcomb}, {Nichol},
  {Owen}, {Peoples}, {Pope}, {Rockosi}, {Schlegel}, {Schneider}, {Scoccimarro},
  {Sheth}, {Siegmund}, {Smee}, {Snir}, {Stebbins}, {Stoughton}, {SubbaRao},
  {Szalay}, {Szapudi}, {Tegmark}, {Tucker}, {Uomoto}, {Vanden Berk}, {Vogeley},
  {Waddell}, {Yanny}, \& {York}}]{Zehavi2002}
{Zehavi}, I., {Blanton}, M.~R., {Frieman}, J.~A., {et~al.} 2002, \apj, 571,
  172, \dodoi{10.1086/339893}

\bibitem[{{Zehavi} {et~al.}(2004){Zehavi}, {Weinberg}, {Zheng}, {Berlind},
  {Frieman}, {Scoccimarro}, {Sheth}, {Blanton}, {Tegmark}, {Mo}, {Bahcall},
  {Brinkmann}, {Burles}, {Csabai}, {Fukugita}, {Gunn}, {Lamb}, {Loveday},
  {Lupton}, {Meiksin}, {Munn}, {Nichol}, {Schlegel}, {Schneider}, {SubbaRao},
  {Szalay}, {Uomoto}, {York}, \& {SDSS Collaboration}}]{Zehavi2004}
{Zehavi}, I., {Weinberg}, D.~H., {Zheng}, Z., {et~al.} 2004, \apj, 608, 16,
  \dodoi{10.1086/386535}

\bibitem[{{Zehavi} {et~al.}(2005){Zehavi}, {Zheng}, {Weinberg}, {Frieman},
  {Berlind}, {Blanton}, {Scoccimarro}, {Sheth}, {Strauss}, {Kayo}, {Suto},
  {Fukugita}, {Nakamura}, {Bahcall}, {Brinkmann}, {Gunn}, {Hennessy},
  {Ivezi{\'c}}, {Knapp}, {Loveday}, {Meiksin}, {Schlegel}, {Schneider},
  {Szapudi}, {Tegmark}, {Vogeley}, {York}, \& {SDSS
  Collaboration}}]{Zehavi2005}
{Zehavi}, I., {Zheng}, Z., {Weinberg}, D.~H., {et~al.} 2005, \apj, 630, 1,
  \dodoi{10.1086/431891}

\bibitem[{{Zehavi} {et~al.}(2011){Zehavi}, {Zheng}, {Weinberg}, {Blanton},
  {Bahcall}, {Berlind}, {Brinkmann}, {Frieman}, {Gunn}, {Lupton}, {Nichol},
  {Percival}, {Schneider}, {Skibba}, {Strauss}, {Tegmark}, \&
  {York}}]{Zehavi2011}
---. 2011, \apj, 736, 59, \dodoi{10.1088/0004-637X/736/1/59}

\bibitem[{{Zentner} {et~al.}(2005){Zentner}, {Berlind}, {Bullock}, {Kravtsov},
  \& {Wechsler}}]{Zentner2005}
{Zentner}, A.~R., {Berlind}, A.~A., {Bullock}, J.~S., {Kravtsov}, A.~V., \&
  {Wechsler}, R.~H. 2005, \apj, 624, 505, \dodoi{10.1086/428898}

\bibitem[{{Zentner} {et~al.}(2019){Zentner}, {Hearin}, {van den Bosch},
  {Lange}, \& {Villarreal}}]{Zentner2019}
{Zentner}, A.~R., {Hearin}, A., {van den Bosch}, F.~C., {Lange}, J.~U., \&
  {Villarreal}, A. 2019, \mnras, 485, 1196, \dodoi{10.1093/mnras/stz470}

\bibitem[{{Zentner} {et~al.}(2014){Zentner}, {Hearin}, \& {van den
  Bosch}}]{Zentner2014}
{Zentner}, A.~R., {Hearin}, A.~P., \& {van den Bosch}, F.~C. 2014, \mnras, 443,
  3044, \dodoi{10.1093/mnras/stu1383}

\bibitem[{{Zhai} {et~al.}(2022){Zhai}, {Tinker}, {Banerjee}, {DeRose}, {Guo},
  {Mao}, {McLaughlin}, {Storey-Fisher}, \& {Wechsler}}]{Zhai2022}
{Zhai}, Z., {Tinker}, J.~L., {Banerjee}, A., {et~al.} 2022, arXiv e-prints,
  arXiv:2203.08999.
\newblock \doarXiv{2203.08999}

\bibitem[{{Zhao} {et~al.}(2003){Zhao}, {Mo}, {Jing}, \&
  {B{\"o}rner}}]{Zhao2003}
{Zhao}, D.~H., {Mo}, H.~J., {Jing}, Y.~P., \& {B{\"o}rner}, G. 2003, \mnras,
  339, 12, \dodoi{10.1046/j.1365-8711.2003.06135.x}

\bibitem[{{Zheng}(2004)}]{Zheng2004}
{Zheng}, Z. 2004, \apj, 610, 61, \dodoi{10.1086/421542}

\bibitem[{{Zheng} {et~al.}(2007){Zheng}, {Coil}, \& {Zehavi}}]{Zheng2007a}
{Zheng}, Z., {Coil}, A.~L., \& {Zehavi}, I. 2007, \apj, 667, 760,
  \dodoi{10.1086/521074}

\bibitem[{{Zheng} \& {Weinberg}(2007)}]{Zheng2007b}
{Zheng}, Z., \& {Weinberg}, D.~H. 2007, \apj, 659, 1, \dodoi{10.1086/512151}

\bibitem[{{Zheng} {et~al.}(2005){Zheng}, {Berlind}, {Weinberg}, {Benson},
  {Baugh}, {Cole}, {Dav{\'e}}, {Frenk}, {Katz}, \& {Lacey}}]{Zheng2005}
{Zheng}, Z., {Berlind}, A.~A., {Weinberg}, D.~H., {et~al.} 2005, \apj, 633,
  791, \dodoi{10.1086/466510}

\bibitem[{{Zu} \& {Mandelbaum}(2018)}]{Zu2018}
{Zu}, Y., \& {Mandelbaum}, R. 2018, \mnras, 476, 1637,
  \dodoi{10.1093/mnras/sty279}

\end{thebibliography}
\bibliographystyle{aasjournal}

\end{document}